\documentclass[12pt,reqno]{article}
\pdfoutput=1
\usepackage[a4paper,width=170mm,top=25mm,bottom=25mm]{geometry}
\usepackage{hyperref}
\hypersetup{
    colorlinks=false,                       
    linkcolor=red,                          
    citecolor=blue                          
}
\usepackage{setspace}
\usepackage{fancyhdr}
\usepackage{amsmath,amsthm,amsfonts,amssymb,braket,mathtools,mathrsfs}
\usepackage{url}
\usepackage{bbm, dsfont}
\usepackage{bbold}
\usepackage{enumerate}
\usepackage{multirow}
\usepackage[all]{xy}
\usepackage{color}
\usepackage{graphicx}
\usepackage{caption}
\usepackage{subcaption}

\numberwithin{equation}{section}
\theoremstyle{plain}
\newtheorem{theorem}{Theorem}[section]
\newtheorem{corollary}[theorem]{Corollary}
\newtheorem{lemma}[theorem]{Lemma}
\newtheorem{proposition}[theorem]{Proposition}
\newtheorem{main theorem}{Main Theorem}
\newtheorem{thmalpha}{Theorem}

\theoremstyle{definition}

\newtheorem{remark}[theorem]{Remark}

\newtheorem{definition}[theorem]{Definition}

\newtheorem*{acknowledgements}{Acknowledgements}
\newtheorem*{conflictofinterest}{Conflict of interest statement}
\newtheorem*{dataavailability}{Data availability statement}

\usepackage{tikz}
\usepackage{xy}
\usepackage{tikz-cd}
\usepackage{xcolor}
\usetikzlibrary{decorations.pathmorphing}

\tikzset{snake/.style={decorate, decoration=snake}}
\usetikzlibrary {arrows.meta,bending,positioning,shapes.geometric,decorations.text,decorations.markings}
\newcommand{\VeryThinLine}{\draw[line width=0.5pt]}

\tikzset{block/.style={rectangle, draw, fill=blue!20, text width=5em, 
    text centered, rounded corners, minimum height=4em},
    cloud/.style={draw, ellipse,fill=red!20, node distance=3cm, minimum height=2em},
    line/.style={draw, -latex'},
    Cbox/.style = {circle, draw, thick, fill=white, opaque}, 
    }

\newcommand{\cM}{\mathcal{M}}

\newcommand{\coev}{\textbf{coev}}

\newcommand{\ev}{\textbf{ev}}
\newcommand{\Hom}{\text{Hom}}

\newcommand{\Tr}{\mathrm{Tr}}
\newcommand{\tr}{\mathrm{tr}}
\newcommand{\range}{\mathrm{Range}}
\newcommand{\Irr}{\mathrm{Irr}}
\newcommand{\ONB}{\mathrm{ONB}}
\NewDocumentCommand{\tens}{t_}
 {%
  \IfBooleanTF{#1}
   {\tensop}
   {\otimes}%
 }
\NewDocumentCommand{\tensop}{m}
 {%
  \mathbin{\mathop{\otimes}\displaylimits_{#1}}%
 }

\begin{document}
\title{Topological Orders from Reflection Positive Frustration-free Hamiltonians}
\author{
  Zishuo Zhao$^{2}$\thanks{zishuozhao0602@gmail.com}, Zhengwei Liu$^{1,2,3}$\thanks{liuzhengwei@mail.tsinghua.edu.cn}\\
  \\
  \small $^{1}$Department of Mathematics, Tsinghua University, Beijing 100084, China \\
  \small $^{2}$Yau Mathematical Sciences Center, Tsinghua University, Beijing 100084, China \\
  \small $^{3}$Yanqi Lake Beijing Institute of Mathematical Sciences and Applications, Beijing 100407, China
}
\date{\today}
\maketitle
\begin{abstract}
    We establish a framework based on reflection positivity for analyzing topologically ordered quantum spin systems and reconstructing their boundary algebras. 
    For any reflection positive frustration-free Hamiltonian, we prove that the local topological quantum order (LTQO) condition of ground states on a disk holds, if and only if the ground state on the sphere obtained by gluing the disk with its reflection is nondegenerate. 
    Furthermore, we show that Osterwalder-Schrader reconstruction produces the local net of boundary operator algebras from the local ground states, offering a constructive approach to topological holography through spatial reflection positivity. 
\end{abstract}
\section{Introduction}
The fractional quantum Hall effect exhibits phases of matter beyond Landau's symmetry-breaking paradigm, in which phases are classified by patterns of spontaneous symmetry breaking. 
Such phases have been studied as topologically ordered phases since the pioneering works of Wen \cite{Wen1990GroundState,LevinWen2005} and Kitaev \cite{Kitaev2003}.  
Topologically ordered systems may exhibit ground-state degeneracies that depend on the topology of the underlying spatial manifold. 
These macroscopic, topology-dependent degeneracies can be described by $(2+1)$-dimensional topological quantum field theories (TQFTs) \cite{Atiyah1988}. 

From the viewpoint of many-body physics, an alternative way to study topological orders microscopically is through local Hamiltonians and their local ground states. 
Local topological quantum order (LTQO) is a key hypothesis in stability theorems for suitable gapped local Hamiltonians \cite{BravyiHastingsMichalakis2010,MichalakisZwolak2013stability}. 
A fundamental problem is to find model-independent criteria ensuring LTQO for a given Hamiltonian. 
For commuting projector Hamiltonians, such a criterion based on the uniqueness of the infinite-volume ground state is given in \cite{ogata2023BoundaryStates}. 
For frustration-free Hamiltonians, it is shown that the uniqueness of the infinite-volume ground state is equivalent to approximate LTQO in the strong-operator topology \cite{OjitoProdanStoiber2025}. 

In \cite{Liu2024}, the second author proposed a functional integral approach to characterize an $(n+1)$-dimensional topologically ordered ground state wave function through \emph{reflection positivity} (RP), a fundamental condition in constructive quantum field theory \cite{OS1973,OS1975}. 
In this paper, we use reflection positivity to characterize topological order in frustration-free Hamiltonians. 
Our first result is a criterion for local indistinguishability in terms of the nondegeneracy of the ground state on the sphere: 

\begin{thmalpha}[Theorem \ref{thm:: local-indistinguishability and ground state degeneracy}, informal]
    For a quantum spin system with a frustration-free, reflection positive, local Hamiltonian $H$ on the sphere $\mathbb{S}^n = \mathbb{D}^n_-\cup \mathbb{D}^n_+$, the following statements are equivalent:
    \begin{enumerate}
        \item The ground state of $H$ on the sphere is nondegenerate;
        \item Ground states of the local Hamiltonian in a neighborhood of $\mathbb{D}^n_+$ are indistinguishable in $\mathbb{D}^n_+$. 
    \end{enumerate}
\end{thmalpha}

Recently, it has been proposed that topological orders should be understood holographically \cite{JiWeni2019NoninvertibleAnomalies}. 
In \cite{kong2020AlgebraicHigherSymmetrya}, it was proposed that the holographic dual of the bulk topological order is captured by the local symmetric operator algebras. 
These are subalgebras of the local operators fixed by the generalized symmetry actions. 
Subsequent studies have demonstrated this viewpoint for exactly solvable models \cite{ChatterjeeWen2023,InamuraWen2023}. 
Building on the LTQO condition, an axiomatic approach to boundary algebras was developed in \cite{JNPW2025,chuah2024boundary,jones2025HolographyBulkboundaryLocal}. 
However, a constructive approach to boundary algebras from the ground state of a frustration-free Hamiltonian is still lacking. 

In this paper, we show that the Osterwalder-Schrader (OS) reconstruction \cite{OS1973,OS1975} provides a constructive approach to the boundary algebra from the ground state of a frustration-free, reflection positive, local Hamiltonian. 
Our analysis relies on the connection between RP and complete positivity. 
This connection has been explored to understand the ground states of QFTs under perturbations, with an emphasis on uniqueness \cite{GlimmJaffe1970phi4}. 
By contrast, the ground states of a local Hamiltonian are often degenerate because of local symmetries. 
Inspired by the study of the algebraic structure of the Perron--Frobenius eigenspace in \cite{HJLW2023}, we relate these degenerate ground states to local symmetries of the Hamiltonian. 
Specifically, a reflection positive Hamiltonian acting on $\mathcal{H}_-\otimes \mathcal{H}_+$ always admits a canonical ground state $\ket{\phi_{\text{PF}}}$. 
The range projection $\widehat{\Pi}$ of the reduced density matrix associated with $\ket{\phi_{\text{PF}}}$ defines the symmetric sub-Hilbert space $\widehat{\Pi}\mathcal{H}_+$. 
\begin{thmalpha}[Theorem \ref{theorem:: ground state of H}]
    For any ground state $\ket{\phi}$ of a Hamiltonian $H$ on $\mathcal{H}_-\otimes \mathcal{H}_+$ with RP, there is a unique operator $W$ on $\mathcal{H}_{+}$ with $[\mathrm{I}\otimes W,H]=0$ and $W\widehat{\Pi}=W$ such that 
    \begin{align*}
        \ket{\phi} = (\mathrm{I}\otimes W)\ket{\phi_{\text{PF}}} = (\Theta(W^\dagger)\otimes \mathrm{I})\ket{\phi_{\text{PF}}}. 
    \end{align*}
\end{thmalpha}
\noindent 
This theorem provides a bijection between the ground state degeneracy of the Hamiltonian and local symmetries acting on the symmetric sub-Hilbert space. 

Based on the above result, we adopt OS reconstruction with respect to the ground state correlation functions. 
In this case, OS reconstruction implements the following two-step truncation: in the first step, operators localized in the upper half-space are compressed to the symmetric sub-Hilbert space; the second step selects the operators that commute with the local symmetries. 
The frustration-free condition ensures that the local symmetric operators can be organized into a quasilocal algebra:  
\begin{thmalpha}[Theorem \ref{thm:: local net of field algebras}]
    Let $\{\Pi(X)\}_{X\in \mathcal{X}}$ be an extendable family of local ground state projections of a frustration-free, reflection positive interaction. 
    Denote by $\cM_X$ the $C^*$-algebra produced by OS reconstruction applied to $\Pi(X)$. 
    Then for any $X\subseteq Y$ in $\mathcal{X}$, there exists an injective $*$-homomorphism $\iota_{Y,X}: \cM_X\rightarrow \cM_Y$ such that the following properties hold: 
    \begin{enumerate}
        \item[(a)] For $X\subseteq Y\subseteq Z$ in $\mathcal{X}$, we have $\iota_{Z,Y}\circ \iota_{Y,X} = \iota_{Z,X}$.
        \item[(b)] For $X,Y\subseteq Z$ with $X\cap Y = \emptyset$, $\iota_{Z,X}(\cM_{X})$ and $\iota_{Z,Y}(\cM_{Y})$ commute. 
    \end{enumerate}
\end{thmalpha}
Furthermore, we find that the modular automorphism groups of the ground states on each local algebra $\cM_X$ extend consistently to the inductive limit (see Theorem \ref{thm:: consistent dynamics in the inductive limit}). 

When the interaction has finite range, we prove that the local net of operator algebras reduces to an $(n-1)$-dimensional boundary algebra; namely, the embedding $\iota_{Y,X}$ becomes an isomorphism if $Y\backslash X$ contains no sites within the interaction range of the reflection hyperplane: 
\begin{thmalpha}[Theorem \ref{thm:: boundary algebra of a local RP frustration-free Hamiltonian}]
    Let $\varPhi$ be a reflection positive frustration-free interaction with interaction range $R$, such that the family of local ground state projections $\{\Pi(X)\}_{X\in \mathcal{X}}$ is extendable. 
    Let $X,Y\in \mathcal{X}$ be such that $X\subseteq Y$. 
    Suppose that $Y\backslash X$ contains no sites within distance $R$ of the reflection hyperplane. 
    Then the inclusion $\iota_{Y,X}:\cM_X\rightarrow \cM_Y$ is an isomorphism. 
\end{thmalpha}

We explicitly compute this boundary algebra for exactly solvable models, including the toric code model \cite{Kitaev2003} and the Levin--Wen string-net models \cite{LevinWen2005}. 
These models are known to have RP \cite{JL2020}, and their boundary theories have been studied from the categorical point of view \cite{kitaev_kong_2012,Kong2014Universal}. 
In our framework, we assume RP and frustration-freeness, which are arguably easier to verify for concrete models, and we derive local indistinguishability from the nondegeneracy of the ground state. 
We do not know if other LTO axioms in \cite{JNPW2025} can be derived for ground state projections of frustration-free Hamiltonians based on RP. 
It is also interesting to connect our framework to the operator-algebraic approach to boundary theory \cite{Wallick2023ToricCodeBoundary,ogata2023BoundaryStates}, given recent study of invertible states through reflection positivity \cite{sopenko2025ReflectionPositivityRefined}. 
There, one works directly in the infinite-volume limit using the DHR superselection theory, assuming approximate Haag duality of the ground state. 
Recently, a large class of models with commuting projector Hamiltonians have been shown to satisfy this condition \cite{ogata2025HaagDuality}. 
Their proof uses certain corner algebras of the cone von Neumann algebras, which are closely related to (the completion of) both the boundary algebras derived from OS reconstruction and those derived from the LTO axioms. 
It is therefore natural to ask under what conditions Haag duality of the ground state follows from the factoriality of the boundary algebra. 
Finally, it is interesting to ask whether the boundary algebra constructed in this paper can be used to infer the DHR theory of ground states of frustration-free Hamiltonians, demonstrating topological holography for those systems. 
We believe that the theory of DHR bimodules developed in \cite{Jones2024DHR} could be useful in this direction. 

The rest of the paper is organized as follows: in Section \ref{sec:: reflection positivity}, we review the abstract framework of RP. 
In Section \ref{sec:: ground states of a reflection positive Hamiltonian}, we study the ground state structure of a reflection positive Hamiltonian using Perron--Frobenius theory of completely positive maps. 
Then, in Section \ref{section:: FF interaction with RP}, we derive the criterion for local indistinguishability of ground states in terms of global nondegeneracy for a frustration-free, reflection positive interaction. 
In Section \ref{sec:: OS reconstruction and local operator algebras}, we construct the local net of boundary algebras using OS reconstruction.  
In Section \ref{sec:: examples of boundary algebras}, we compute the boundary net for the toric code and Levin--Wen string-net models. 

\begin{acknowledgements}
The authors thank Arthur Jaffe for his warm hospitality during their visit to Harvard University, where part of this work was carried out. 
The authors thank Zixuan Feng, Arthur Jaffe, and Xiaogang Wen for valuable discussions. 
Zhengwei Liu was supported by Beijing Natural Science Foundation Key Program (Grant No. Z220002). 
All authors were supported by Beijing Natural Science Foundation (Grant No. Z221100002722017).
\end{acknowledgements}

\section{Reflection positivity}\label{sec:: reflection positivity}
In this section, we set up reflection positivity and establish the connection to complete positivity. 
Historically, RP has played an essential role in the construction of quantum field theories \cite{OS1973,OS1975,GlimmJaffe1987}. 
For any observable $A$ localized in the upper half-space, RP of the partition function is expressed as 
\begin{equation}
    \langle \vartheta(A)A\rangle\geq0, 
\end{equation}
where $\vartheta(A)$ is the observable obtained by reflecting $A$ about the time-zero slice and taking its complex conjugate. 
RP represents the notion of \emph{unitarity} of a Euclidean quantum field theory.  
After its introduction in constructive QFT, RP has found applications in the study of phase transitions in field theories, as well as in classical and quantum lattice models \cite{Glimm1975Phase,FSS1976infrared,FILS1978}. 
We refer the reader to \cite{Frohlich2023,Frohlich2024} for a historical review. 
There, RP takes the form 
\begin{equation}
    \langle \Theta(A)\otimes A\rangle_{\tau}\geq 0,
\end{equation}
where $\Theta$ is the reflection about a spatial hyperplane and $\langle \cdot\rangle_{\tau}$ denotes the thermal expectation at inverse temperature $\tau$. 
This is the notion that we adopt in the paper. 

\begin{definition}[Reflection Positivity]\label{def:reflection-positivity}
    Let $\mathfrak{A}_{+}$ and $\mathfrak{A}_{-}$ be $\mathrm{C}^*$-algebras, and let $\Theta:\mathfrak{A}_{+}\rightarrow \mathfrak{A}_{-}$ be an anti-linear $*$-isomorphism.
    A linear functional $Z$ on the $*$-algebra $\mathfrak{A}_{-}\otimes \mathfrak{A}_{+}$ is said to be \textbf{reflection positive} with respect to $\Theta$ if for any $A \in \mathfrak{A}_{+}$, we have
    \begin{equation}
        Z(\Theta(A)\otimes A) \geq 0.
    \end{equation}
    When the choice of $\Theta$ is clear from the context, we simply call $Z$ reflection positive.
\end{definition}

In the ensuing discussion, we set $\mathfrak{A}_+ = \mathcal{B}(\mathcal{H}_+)$ and $\mathfrak{A}_- = \mathcal{B}(\mathcal{H}_-)$, where $\mathcal{H}_+$ is a finite-dimensional Hilbert space.
In this case, every anti-linear $*$-isomorphism $\Theta$ from $\mathfrak{A}_{+}$ to $\mathfrak{A}_{-}$ is implemented by an anti-unitary operator (unique up to a phase factor) $\widehat{\theta}:\mathcal{H}_{+} \to \mathcal{H}_{-}$ via
\begin{equation}
    \Theta(X) = \widehat{\theta} X \widehat{\theta}^{-1},\quad X \in \mathfrak{A}_{+}.
\end{equation}
In specific models, $\mathcal{H}_+$ represents the state space of the system within a certain region, while $\mathcal{H}_-$ represents the state space in the mirror-reflected region.
The operator $\widehat{\theta}$ corresponds to the action induced on the state spaces by spatial reflection and time reversal.

The reflection positivity of an operator on the Hilbert space $\mathcal{H}_{-}\otimes \mathcal{H}_+$ is defined through the induced linear functional:
\begin{definition}\label{def:RP-operator}
    Let $\mathcal{H}_+, \mathcal{H}_-$ be defined as above, and $\Theta: \mathcal{B}(\mathcal{H}_+) \to \mathcal{B}(\mathcal{H}_-)$ be an anti-linear $*$-isomorphism.
    An operator $T$ on $\mathcal{H}_- \otimes \mathcal{H}_+$ is reflection positive with respect to $\Theta$ if for any $A \in \mathcal{B}(\mathcal{H}_+)$, we have
    \begin{equation}
        \Tr(T(\Theta(A)\otimes A))\geq 0,\quad A \in \mathcal{B}(\mathcal{H}_+).
    \end{equation}
    We say that a vector state $\ket{\psi}\in \mathcal{H}_- \otimes \mathcal{H}_+$ is reflection positive if the projection operator $\ket{\psi}\bra{\psi}$ is reflection positive.
\end{definition}

We now discuss the connection between operator reflection positivity and complete positivity.
There is a natural isomorphism from $\mathcal{H}^*_{+}\otimes \mathcal{H}_{+}$ to $\mathcal{B}(\mathcal{H}_+)$. 
Specifying an anti-unitary operator $\widehat{\theta}:\mathcal{H}_+ \to \mathcal{H}_-$ is equivalent to defining a unitary operator from $\mathcal{H}^*_{+}$ to $\mathcal{H}_{-}$.
Therefore, we obtain the following isomorphism:
\begin{definition}\label{def:: state-operator correspondence}
    For $\ket{\xi},\ket{\eta}\in \mathcal{H}_{+}$, define an operator in $\mathcal{B}(\mathcal{H}_{+})$ by
    \begin{equation}
        \mathcal{O}(\widehat{\theta}\ket{\eta}\otimes \ket{\xi}) = \ket{\xi}\bra{\eta}.
    \end{equation}
    This map uniquely extends to an isomorphism from $\mathcal{H}_{-}\otimes \mathcal{H}_{+}$ to $\mathcal{B}(\mathcal{H}_{+})$. 
\end{definition}
To avoid overly cumbersome notation, from now on, we abbreviate $\mathcal{O}(\ket{\psi})$ as $\mathcal{O}(\psi)$.

\begin{lemma}\label{lemma:: O is an isometry}
    For any $\ket{\psi},\ket{\phi}\in \mathcal{H}_{-}\otimes \mathcal{H}_{+}$, we have
    \begin{equation}
        \braket{\phi,\psi} = \Tr(\mathcal{O}(\phi)^\dagger \mathcal{O}(\psi)).
    \end{equation}
    In particular, for $\ket{\xi},\ket{\eta} \in \mathcal{H}_{+}$ and $\ket{\zeta}\in\mathcal{H}_{-}\otimes \mathcal{H}_{+}$, we have
    \begin{equation}
        \bra{\xi}\mathcal{O}(\zeta)\ket{\eta} = \braket{\widehat{\theta} \eta\otimes  \xi|\zeta}.
    \end{equation}
\end{lemma}
\begin{proof}
    By sesquilinearity, it suffices to verify this for $\ket{\psi} = \widehat{\theta}\ket{\psi_1}\otimes \ket{\psi_2}$ and $\ket{\phi} = \widehat{\theta}\ket{\phi_1}\otimes \ket{\phi_2}$, where $\ket{\psi_i},\ket{\phi_i}\in \mathcal{H}_{+}$ for $i=1,2$.
    By definition,
    \begin{equation}
        \Tr(\mathcal{O}(\phi)^\dagger \mathcal{O}(\psi)) = \Tr(\ket{\phi_1}\bra{\phi_2}\ket{\psi_2}\bra{\psi_1}) = \braket{\phi_2|\psi_2}\braket{\psi_1|\phi_1} = \braket{\phi|\psi},
    \end{equation}
    where the final step uses the anti-linearity of $\widehat{\theta}$.
    For the second equation, simply substitute $\ket{\psi} = \ket{\zeta}$ and $\ket{\phi} = \widehat{\theta}\ket{\eta}\otimes \ket{\xi}$.
\end{proof}
\begin{remark}
    By Lemma \ref{lemma:: O is an isometry}, we can identify $\mathcal{H}_{-}\otimes \mathcal{H}_{+}$ with $\mathcal{B}(\mathcal{H}_{+})$ endowed with the Hilbert--Schmidt inner product.
\end{remark}

\begin{lemma}\label{lemma:: s-to-o map}
    For any $X,Y\in \mathcal{B}(\mathcal{H}_{+})$, we have
    \begin{equation}
        \mathcal{O}(\mathrm{I}_{\mathcal{H}_{-}}\otimes X)\mathcal{O}^{-1}(Y) = XY,\quad \mathcal{O}(\Theta(X)\otimes \mathrm{I}_{\mathcal{H}_{+}})\mathcal{O}^{-1}(Y) = YX^\dagger. 
    \end{equation}
\end{lemma}
\begin{proof}
    By linearity, it is sufficient to prove the lemma for $Y = \ket{\xi}\bra{\eta}$ with $\xi,\eta\in\mathcal{H}_+$.
    For the first equation, we have 
    \begin{equation}
        \mathcal{O}( \mathrm{I}_{\mathcal{H}_{-}}\otimes X )\mathcal{O}^{-1}(\ket{\xi}\bra{\eta}) = \mathcal{O}(\widehat{\theta}\eta\otimes X\xi) = X\ket{\xi}\bra{\eta}.
    \end{equation}
    For the second equation, we have 
    \begin{equation}
        (\Theta(X)\otimes \mathrm{I}_{\mathcal{H}_{+}})\mathcal{O}^{-1}(\ket{\xi}\bra{\eta}) = \widehat{\theta} X\widehat{\theta}^{-1}\widehat{\theta}\ket{\eta}\otimes \ket{\xi} = \widehat{\theta} X\ket{\eta}\otimes \ket{\xi},
    \end{equation}
    and thus 
    \begin{equation}
        \mathcal{O}(\Theta(X)\otimes \mathrm{I}_{\mathcal{H}_{+}})\mathcal{O}^{-1}(\ket{\xi}\bra{\eta}) = \mathcal{O}(\widehat{\theta} X\eta\otimes \xi) = \ket{\xi}\bra{\eta} X^{\dagger}.
    \end{equation}
\end{proof}

Through the isomorphism $\mathcal{O}$, we can translate the condition of reflection positivity into complete positivity.
\begin{theorem}\label{thm:: charactorization of RP based on s-to-o map}
    An operator $T$ on the Hilbert space $\mathcal{H}_{-}\otimes \mathcal{H}_{+}$ is reflection positive with respect to $\Theta$ if and only if the map $\mathcal{O}T\mathcal{O}^{-1}: \mathcal{B}(\mathcal{H}_{+})\rightarrow \mathcal{B}(\mathcal{H}_{+})$ is completely positive.
\end{theorem}
\begin{proof}
    Consider rank-one operators $Y = \ket{\xi}\bra{\eta}$ and $Y' = \ket{\xi'}\bra{\eta'}$ in $\mathcal{B}(\mathcal{H}_{+})$.
    Let $\{\ket{i}\}_{i\in \mathcal{I}}$ be an orthonormal basis of $\mathcal{H}_{+}$.
    Then by Lemma \ref{lemma:: O is an isometry} and Lemma \ref{lemma:: s-to-o map}, we have
    \begin{equation}
    \begin{aligned}
        \Tr (T(\Theta(Y')\otimes Y)) &= \sum_{i,j\in \mathcal{I}} \bra{\widehat{\theta}j\otimes i} T(\Theta(Y')\otimes Y) \ket{\widehat{\theta}j\otimes i}\\
        &= \sum_{i,j\in \mathcal{I}}\Tr (\mathcal{O}T\mathcal{O}^{-1}(Y\ket{i}\bra{j} Y'^\dagger)\ket{j}\bra{i})\\
        &= \sum_{i,j\in \mathcal{I}}\bra{i} \mathcal{O}T\mathcal{O}^{-1}(Y\ket{i}\bra{j} Y'^\dagger)\ket{j} = \bra{\eta} \mathcal{O}T\mathcal{O}^{-1}(\ket{\xi}\bra{\xi'})\ket{\eta'}.
    \end{aligned}
    \end{equation}
    Now, for any positive integer $n$ and any operator $X = \sum^n_{k=1}\ket{\xi_k}\bra{\eta_k}$ in $\mathcal{B}(\mathcal{H}_{+})$ with rank at most $n$, we obtain
    \begin{equation}
    \begin{aligned}
        \Tr (T(\Theta(X)\otimes X)) &= \sum^n_{k,k'=1} \bra{\eta_k} \mathcal{O}T\mathcal{O}^{-1}(\ket{\xi_k} \bra{\xi_{k'}})\ket{\eta_{k'}}.
    \end{aligned}
    \end{equation}
    Therefore, $T$ is reflection positive with respect to $\Theta$ if and only if for any positive integer $n$ and any operator $X\in \mathcal{B}(\mathcal{H}_{+})$ of rank at most $n$, we have
    \begin{equation}
        \sum^n_{k,k'=1} \bra{\eta_k} \mathcal{O}T\mathcal{O}^{-1}(\ket{\xi_k} \bra{\xi_{k'}})\ket{\eta_{k'}} = \braket{(\eta_{k})_k\vert [\mathcal{O}T\mathcal{O}^{-1}(\ket{\xi_k} \bra{\xi_{k'}})]_{k,k'}\vert (\eta_{k'})_{k'}}\geq 0,
    \end{equation}
    which means that the block matrix $(\mathcal{O}T\mathcal{O}^{-1}\otimes \mathrm{id}_n)([\ket{\xi_k} \bra{\xi_{k'}}]_{k,k'})$ is positive semidefinite.
    By \cite[Lemma 3.13]{Paulsen2003}, every positive element in $\mathcal{B}(\mathcal{H}_+)\otimes \mathrm{M}_n(\mathbb{C})$ can be written as a nonnegative linear combination of elements of the form $[\ket{\xi_k} \bra{\xi_{k'}}]_{k,k'}$.
    Since $n$ is arbitrary, the above inequality holds if and only if $\mathcal{O}T\mathcal{O}^{-1}$ is a completely positive map.
\end{proof}

\begin{definition}
    Let the self-adjoint operator $H$ be a Hamiltonian on $\mathcal{H}_- \otimes \mathcal{H}_+$.
    We say that $H$ is reflection positive with respect to $\Theta$ if, for any $\tau \geq 0$, the operator $e^{-\tau H}$ is reflection positive with respect to $\Theta$; that is,
    \begin{equation}
        \Tr(e^{-\tau H} (\Theta(A) \otimes A)) \geq 0,\quad A \in \mathcal{B}(\mathcal{H}_+),\tau\geq 0.
    \end{equation}
\end{definition}

Next, we record a sufficient condition for a Hamiltonian to be reflection positive. 
Similar criteria were given previously in \cite{FILS1978,JB2017,JL2017}. 

\begin{theorem}\label{thm:: structure of RP Hamiltonian}
    Let $H$ be a self-adjoint operator on $\mathcal{H}_- \otimes \mathcal{H}_+$.
    Suppose that there exist a self-adjoint operator $H_+$ on $\mathcal{H}_{+}$ and a finite set $\{O_j\}^m_{j=1}$ in $\mathcal{B}(\mathcal{H}_{+})$ such that
    \begin{equation}
        H = \mathrm{I}_{\mathcal{H}_{-}}\otimes H_+ + \Theta(H_+)\otimes \mathrm{I}_{\mathcal{H}_{+}} - \sum^m_{j=1}\Theta(O_j)\otimes O_j.
    \end{equation}
    Then $H$ is reflection positive with respect to $\Theta$.
\end{theorem}
\begin{proof}
    Define the completely positive map $\Phi$ and the linear map $\mathcal{A}$ on $\mathcal{B}(\mathcal{H}_+)$ by
    \begin{equation}
        \Phi(X)=\sum_{j=1}^m O_jXO_j^\dagger,\qquad \mathcal{A}(X)=H_+X+XH_+.
    \end{equation}
    By Lemma \ref{lemma:: s-to-o map}, $\mathcal{L}:=\mathcal{O}H\mathcal{O}^{-1}=\mathcal{A}-\Phi$.
    Since $H_+$ is self-adjoint, the map
    \begin{equation}
        e^{-\tau\mathcal{A}}(X)=e^{-\tau H_+}Xe^{-\tau H_+}
    \end{equation}
    is completely positive for every $\tau\geq0$.
    Moreover, $e^{\tau\Phi}=\sum_{k=0}^{\infty}\tau^k\Phi^k/k!$ is completely positive for every $\tau\geq0$.
    The Lie--Trotter product formula gives
    \begin{equation}
        e^{-\tau\mathcal{L}}=\lim_{n\to\infty}\left(e^{-\tau\mathcal{A}/n} e^{\tau\Phi/n}\right)^n.
    \end{equation}
    Hence $e^{-\tau\mathcal{L}}$ is completely positive because compositions and norm limits of completely positive maps remain completely positive.
    Finally, $\mathcal{O}e^{-\tau H}\mathcal{O}^{-1}=e^{-\tau\mathcal{L}}$, so Theorem \ref{thm:: charactorization of RP based on s-to-o map} implies that $e^{-\tau H}$ is reflection positive for every $\tau\geq0$.
    Therefore, $H$ is reflection positive with respect to $\Theta$.
\end{proof}

\begin{proposition}\label{prop:: reduced density matrix in terms of O}
    For a vector $\ket{\Omega}\in \mathcal{H}_-\otimes \mathcal{H}_+$, the following statements hold:
    \begin{itemize}
        \item[(1)] The reduced density matrix of $\ket{\Omega}$ on $\mathcal{H}_{+}$ is given by
        \begin{equation}
        \Tr_{\mathcal{H}_-}\ket{\Omega}\bra{\Omega} = \mathcal{O}(\Omega) \mathcal{O}(\Omega)^\dagger. 
        \end{equation}
        \item[(2)] $\ket{\Omega}$ is reflection positive if and only if there exists $\theta_0\in[0,2\pi)$ such that $e^{i\theta_0}\mathcal{O}(\Omega)\geq 0$, or, equivalently,
        \begin{equation}
            e^{-i\theta_0}\braket{\Omega|\widehat{\theta}\xi\otimes \xi}\geq 0,\quad \ket{\xi}\in \mathcal{H}_{+}.
        \end{equation}
    \end{itemize}
\end{proposition}
\begin{proof}
    (1) Let $\{\ket{i}\}_{i\in I}$ be an orthonormal basis of $\mathcal{H}_{+}$.
    Then, for any $\ket{\xi},\ket{\eta} \in \mathcal{H}_+$,
    \begin{equation}
    \begin{aligned}
        \Tr (\ket{\xi}\bra{\eta} \mathcal{O}(\Omega)\mathcal{O}(\Omega)^\dagger) &= \Tr (\mathcal{O}(\Omega)^\dagger \ket{\xi}\bra{\eta}\mathcal{O}(\Omega))\\
        &= \sum_{i}\bra{i} \mathcal{O}(\Omega)^\dagger \ket{\xi}\bra{\eta}\mathcal{O}(\Omega) \ket{i} \\
        &= \sum_{i}\braket{\widehat{\theta} i\otimes \eta| \Omega} \braket{\Omega |\widehat{\theta} i\otimes \xi}\\
        &= \sum_i \Tr (\ket{\Omega} \bra{\Omega} \ket{\widehat{\theta} i\otimes \xi} \bra{\widehat{\theta} i\otimes \eta})\\
        &=  \bra{\Omega}(\mathrm{I}_{\mathcal{H}_{-}}\otimes \ket{\xi}\bra{\eta}) \ket{\Omega}. 
    \end{aligned}
      \end{equation}
      This proves that $\mathcal{O}(\Omega)\mathcal{O}(\Omega)^\dagger = \Tr_{\mathcal{H}_-}\ket{\Omega}\bra{\Omega}$.

      (2) According to Theorem \ref{thm:: charactorization of RP based on s-to-o map}, the reflection positivity of $\ket{\Omega}$ is equivalent to the complete positivity of the rank-one map
      \begin{equation}
          \mathcal{L}_{\Omega}(X)=\mathcal{O}(\Omega)\Tr(X\mathcal{O}(\Omega)^\dagger).
      \end{equation}
      If $\mathcal{O}(\Omega)=0$, the claim is immediate.
      Otherwise, if $\mathcal{L}_{\Omega}$ is completely positive, then, since positive operators span $\mathcal{B}(\mathcal{H}_+)$, there exists $X\geq 0$ such that $\mathcal{L}_{\Omega}(X)=c\mathcal{O}(\Omega)\neq 0$; hence $c\mathcal{O}(\Omega)\geq0$, so $\mathcal{O}(\Omega)$ is a scalar multiple of a positive operator.
      Conversely, if $\mathcal{O}(\Omega)=cP$ with $P\geq0$, then $\mathcal{L}_{\Omega}(X)=\lvert c\rvert^2P\Tr(XP)$ is completely positive.
      Finally, if $e^{i\theta_0}\mathcal{O}(\Omega)\geq0$, Lemma \ref{lemma:: O is an isometry} gives
      \begin{equation}
          e^{i\theta_0}\braket{\widehat{\theta}\xi\otimes\xi|\Omega}=e^{i\theta_0}\bra{\xi}\mathcal{O}(\Omega)\ket{\xi}\geq0,
      \end{equation}
      whose complex conjugate is the stated equivalent condition.
\end{proof}
\section{Ground States of a Reflection Positive Hamiltonian}\label{sec:: ground states of a reflection positive Hamiltonian}
In this section, we study the ground state structure of a reflection positive Hamiltonian using Perron--Frobenius theory of completely positive maps. 
Consider a self-adjoint, reflection positive operator $H$ on $\mathcal{H} = \mathcal{H}_-\otimes \mathcal{H}_+$. 
By definition, the operator $e^{-\tau H}$ is reflection positive for all $\tau\geq 0$. 
By Theorem \ref{thm:: charactorization of RP based on s-to-o map}, we then obtain a semigroup of completely positive maps $\Phi_{\tau} = \mathcal{O}e^{-\tau H}\mathcal{O}^{-1}$ on $\mathfrak{A}_{+}$. 
By Lemma \ref{lemma:: O is an isometry}, $\mathcal{O}$ is a unitary map from $\mathcal{H}$ to $\mathfrak{A}_{+}$ equipped with the Hilbert--Schmidt inner product.
Thus $\Phi_{\tau}$ is self-adjoint with respect to the Hilbert--Schmidt inner product.
Since $\Phi_\tau$ preserves adjoint, for all $X,Y\in \mathfrak{A}_{+}$ we have
\begin{equation}
    \Tr(\Phi_{\tau}(X)Y)=\Tr(\Phi_{\tau}(X^\dagger)^\dagger Y)=\Tr((X^\dagger)^\dagger \Phi_{\tau}(Y))=\Tr(X\Phi_{\tau}(Y)).
\end{equation}
Thus, $\Phi_{\tau}$ is \emph{symmetric}. 
In particular, ground states of a reflection positive Hamiltonian correspond to eigenvectors of symmetric completely positive maps with largest eigenvalue. 

\subsection{Perron--Frobenius theory}
Let $\mathcal{K}$ be a finite-dimensional Hilbert space and consider a symmetric completely positive map $\Psi$ on $\mathcal{B}(\mathcal{K})$. 
The Perron--Frobenius theorem for (possibly reducible) positive maps on an ordered Banach space \cite[Lemma 3.5]{GluckMironchenko2021} states that the spectral radius $\rho(\Psi)$ is an eigenvalue of $\Psi$ with a nonzero positive eigenvector; irreducibility is not required for this existence statement. 
Denote by $\mathcal{E}$ the eigenspace of $\Psi$ with eigenvalue $\rho(\Psi)$, and call it the Perron--Frobenius eigenspace. 
We call $X\in \mathcal{E}$ a Perron--Frobenius (PF) eigenvector of $\Psi$ if $X\geq 0$. 

Perron--Frobenius theory further guarantees the existence of a PF eigenvector $\Xi\in\mathcal{B}(\mathcal{K})$ of $\Psi$ such that, for every PF eigenvector $X$ of $\Psi$, there exists $\lambda>0$ such that $X \leq\lambda \Xi$. 
In other words, $\Xi$ is a PF eigenvector of $\Psi$ with maximal range projection. 
Define $p_{\text{max}} = \range(\Xi)$; it is known that $p_{\text{max}}$ depends only on $\Psi$. 
Suppose $\rho(\Psi)>0$, and set $T=\Psi/\rho(\Psi)$. 
Since $\Psi$ is symmetric, $T$ is self-adjoint with respect to the Hilbert--Schmidt inner product and $\lVert T\rVert_2=1$. 
The spectral decomposition of $T$ therefore shows that the following Cesàro limit exists: 
\begin{equation}\label{eqn:: canonical PF eigenvector}
    \Xi=\lim_{N\to\infty}\frac{1}{N}\sum_{n=0}^{N-1}T^n(\mathrm{I}).
\end{equation}
The limit $\Xi$ is a positive fixed point of $T$.
Moreover, if $X$ is a PF eigenvector normalized so that $X\leq\mathrm{I}$, positivity gives $X\leq\Xi$, so $\Xi$ has maximal range projection (compare \cite[Lemma 3.6]{HJLW2025phase_group}).
We will refer to this choice of $\Xi$ as the \emph{canonical PF eigenvector} of $\Psi$. 

\begin{lemma}\label{lemma:: maximal range projection commute with Phi}
     Let $\Phi$ be a symmetric completely positive map on $\mathcal{B}(\mathcal{K})$.
     Then we have
    \begin{equation}
        \Phi(X p_{\text{max}}) = \Phi(X )p_{\text{max}},\quad X\in\mathcal{B}(\mathcal{K}). 
    \end{equation}
    Taking the adjoint, we also obtain $\Phi(p_{\text{max}}X) = p_{\text{max}}\Phi(X )$.
\end{lemma}
\begin{proof}
 Set $p=p_{\text{max}}$.
 Since $p$ is the range projection of $\Xi$, for any positive operator $X\in \mathcal{B}(\mathcal{K})$, there exists $\lambda>0$ such that $pXp\leq \lambda \Xi$.
 Since $\Phi$ is a positive map,
 \begin{equation}
    \Phi(pXp)\leq \lambda\Phi(\Xi) = \lambda \rho(\Phi) \Xi,
 \end{equation}
 which implies $\Phi(pXp) = p\Phi(pXp)p$.
 By polarization, every operator in $\mathcal{B}(\mathcal{K})$ can be decomposed into a linear combination of four positive operators, which shows that
 \begin{equation}
    \Phi(pXp) = p\Phi(pXp)p,\quad  X\in\mathcal{B}(\mathcal{K}).
 \end{equation}
 Now, since $\Phi$ is symmetric, for all $X,Y\in \mathcal{B}(\mathcal{K})$, we have
\begin{equation}
\begin{aligned}
    \Tr   (Yp\Phi(X)p) &= \Tr   (\Phi(pYp)X )\\
    &= \Tr  (p\Phi(pYp)pX )\\
    &= \Tr   (Yp\Phi(pXp)p).
\end{aligned}
\end{equation}
Therefore, we obtain 
\begin{equation}
    p\Phi(X)p = p\Phi(pXp)p = \Phi(pXp).
\end{equation}
Let $\{K_i\}_i$ be a set of Kraus operators for $\Phi$.
Then
\begin{equation}
    \sum_i K_i p K^\dagger_i = \Phi(p) = p\Phi(\mathrm{I}_{\mathcal{K}})p \leq \Vert \Phi(\mathrm{I}_{\mathcal{K}})\Vert p. 
\end{equation}
Each term on the left-hand side is positive, so $K_i p K^\dagger_i\leq \Vert \Phi(\mathrm{I}_{\mathcal{K}})\Vert p$ for every $i$.
It follows that $(\mathrm{I}_{\mathcal{K}}-p)K_i p=0$, and hence $K_i p=pK_i p$.
On the other hand, by the trace symmetry of $\Phi$, we have $\Phi(Y)=\sum_iK_i^\dagger YK_i$ for all $Y\in \mathcal{B}(\mathcal{K})$, so $\{K_i^\dagger\}_i$ is also a set of Kraus operators for $\Phi$. 
Applying the preceding argument to this Kraus representation gives $K_i^\dagger p=pK_i^\dagger p$; taking adjoints yields $pK_i=pK_i p$.
Consequently, $K_i p=pK_i$ for every $i$, and therefore
\begin{equation}
    \Phi(Xp)=\sum_iK_iXpK_i^\dagger=\sum_iK_iXK_i^\dagger p=\Phi(X)p,\quad X\in\mathcal{B}(\mathcal{K}).
\end{equation}
\end{proof}

\begin{corollary}\label{corollary:: canonical PF eigenvector for truncated map}
    Let $\Phi$ be a non-zero symmetric completely positive map on $\mathcal{B}(\mathcal{K})$, and let $\Xi$ be its canonical Perron--Frobenius eigenvector.
    Then we have
    \begin{equation}
        \Xi = \lim_{n\to \infty} \frac{1}{n} \sum^{n-1}_{k=0}\left( \frac{\Phi}{\rho(\Phi)} \right)^k(p_{\text{max}}).
    \end{equation}
\end{corollary}
\begin{proof}
    By Lemma \ref{lemma:: maximal range projection commute with Phi}, for all positive integers $k$, we have
    \begin{equation}
        \Phi^k(p_{\text{max}}) = p_{\text{max}} \Phi^k(\mathrm{I}_{\mathcal{K}}) p_{\text{max}}.
    \end{equation}
    Since $p_{\text{max}}$ is the range projection of $\Xi$, it follows that
    \begin{equation}
        \Xi = p_{\text{max}} \Xi p_{\text{max}} = \lim_{n\to \infty} \frac{1}{n} \sum^{n-1}_{k=0}p_{\text{max}}\left( \frac{\Phi}{\rho(\Phi)} \right)^k(\mathrm{I}_{\mathcal{K}}) p_{\text{max}} = \lim_{n\to \infty} \frac{1}{n} \sum^{n-1}_{k=0} \left( \frac{\Phi}{\rho(\Phi)} \right)^k(p_{\text{max}}).
    \end{equation}
\end{proof}

\begin{proposition}\label{prop:: PF eigenspace of truncated map}
    Let $\Phi$ be a non-zero symmetric completely positive map on $\mathcal{B}(\mathcal{K})$, and let $\mathcal{E}$ be its Perron--Frobenius eigenspace.
    Consider the truncated map 
    \begin{equation}\label{eqn:: truncated map}
        \Phi_0(Y) = \Phi(p_{\text{max}}Y p_{\text{max}}),\quad Y\in \mathcal{B}(\mathcal{K}).
    \end{equation}
    Then $\mathcal{E}$ is also the Perron--Frobenius eigenspace of $\Phi_0$.
\end{proposition}
\begin{proof}
    Define the idempotent map $\mathcal{P}(Y) = p_{\text{max}}Y p_{\text{max}}$ on $\mathcal{B}(\mathcal{K})$.
    By Lemma \ref{lemma:: maximal range projection commute with Phi}, we have $\Phi_0 = \mathcal{P}\Phi = \Phi\mathcal{P}$.
    Thus, it suffices to show that the PF eigenspace of $\Phi$ is contained in the range of $\mathcal{P}$, i.e., 
    \begin{equation}
        \mathcal{E}\subseteq p_{\text{max}}\mathcal{B}(\mathcal{K})p_{\text{max}}.
    \end{equation}
    Let $X\in \mathcal{E}$.
    If $X=0$, the conclusion is immediate, so assume that $X\neq 0$.
    Since $\Phi$ is adjoint-preserving, by taking real and imaginary parts we can assume that $X^\dagger = X$. 
    Let $X_+$ and $X_-$ be the positive and negative parts of $X$.
    We show that both $X_+$ and $X_-$ are Perron--Frobenius eigenvectors of $\Phi$. 
    Since $\Phi$ is a positive map, by $|X| = X_+ + X_-$ and $X = X_+ - X_-$ we have
    \begin{equation}
    \begin{aligned}
        \Tr   (\Phi(X)X) &= \Tr   (\Phi(X_+)X_+) + \Tr   (\Phi(X_-)X_-) - \Tr   (\Phi(X_+)X_-) - \Tr   (\Phi(X_-)X_+)\\
        &= \Tr   (\Phi(\vert X\vert)\vert X\vert) -2\Tr  (\Phi(X_+)X_-) - 2\Tr  (\Phi(X_-)X_+)\\
        &\leq \Tr   (\Phi(\vert X\vert)\vert X\vert).
    \end{aligned}
    \end{equation}
    Since $X_+X_- = 0$, we have $\Tr   (\vert X\vert^2) = \Tr   (X^2)$.
    Therefore, we obtain
    \begin{equation}
        \frac{\Tr   (\Phi(\vert X\vert)\vert X\vert)}{\Tr   (\vert X\vert^2)}\geq \frac{\Tr   (\Phi(X)X)}{\Tr   (X^2)} = \rho(\Phi).
    \end{equation}
    Since $\Phi$ is self-adjoint with respect to the Hilbert--Schmidt inner product and $\rho(\Phi)$ is its maximum eigenvalue, the Rayleigh quotient on the left-hand side is at most $\rho(\Phi)$.
    Hence, equality holds.
    Equivalently,
    \begin{equation}
        \langle \vert X\vert,(\rho(\Phi)\mathrm{I}-\Phi)(\vert X\vert)\rangle_{\mathrm{HS}}=0.
    \end{equation}
    Since $\rho(\Phi)\mathrm{I}-\Phi$ is positive semidefinite, this implies $(\rho(\Phi)\mathrm{I}-\Phi)(\vert X\vert)=0$, and therefore $\Phi(\vert X\vert)=\rho(\Phi)\vert X\vert$.
    Since $X_\pm=(\vert X\vert\pm X)/2$, the identities $\Phi(\vert X\vert)=\rho(\Phi)\vert X\vert$ and $\Phi(X)=\rho(\Phi)X$ imply that $\Phi(X_\pm)=\rho(\Phi)X_\pm$; hence, $X_+$ and $X_-$ are Perron--Frobenius eigenvectors of $\Phi$ and belong to $\mathcal{E}$.
    By the maximality of $p_{\text{max}}$, it follows that $X_{-},X_{+}\in p_{\text{max}}\mathcal{B}(\mathcal{K})p_{\text{max}}$.
    Consequently, we have $X = X_{+} - X_{-}\in p_{\text{max}}\mathcal{B}(\mathcal{K})p_{\text{max}}$.
\end{proof}

The following $C^*$-algebra plays an important role in characterizing the PF eigenspace:

\begin{definition}\label{def:: strong symmetry}
    Let $\mathcal{A}$ be a $C^*$-algebra, and let $\Phi: \mathcal{A}\rightarrow \mathcal{A}$ be a linear map.
    We define
    \begin{equation}
        \mathrm{Sym}(\Phi) = \{X\in \mathcal{A}\vert \Phi(XY) = X\Phi(Y),\Phi(X^\dagger Y) = X^\dagger\Phi(Y),Y\in \mathcal{A}\}. 
    \end{equation}
    Then $\mathrm{Sym}(\Phi)$ forms a $*$-subalgebra of $\mathcal{A}$.
\end{definition}
When $\Phi$ is adjoint-preserving, we have 
\begin{equation}
    \Phi(XY) = X\Phi(Y)\quad \forall Y\in \mathcal{A} \iff \Phi(Y X^\dagger) = \Phi(Y)X^\dagger \quad \forall Y\in \mathcal{A} .
\end{equation}
Replace $X$ by $X^\dagger$, we see that for such maps, 
\begin{equation}
    \mathrm{Sym}(\Phi) = \{X\in \mathcal{A}\vert \Phi(YX) = \Phi(Y)X \text{ and }\Phi(XY) = X\Phi(Y), Y\in \mathcal{A}\}.
\end{equation}
That is, $\mathrm{Sym}(\Phi)$ is the largest $*$-subalgebra of $\mathcal{A}$ such that $\Phi$ becomes a bimodule map over it.
By the bimodule property of $\Phi$, for any $X\in \mathrm{Sym}(\Phi)$ and any positive integer $n$, we have 
\begin{equation}
    X\Phi^n(\mathrm{I}_{\mathcal{K}}) = \Phi^n(X\mathrm{I}_{\mathcal{K}}) = \Phi^n(\mathrm{I}_{\mathcal{K}}X) = \Phi^n(\mathrm{I}_{\mathcal{K}})X,
\end{equation}
Thus, by Equation \eqref{eqn:: canonical PF eigenvector}, the canonical Perron--Frobenius eigenvector $\Xi$ satisfies
\begin{equation}\label{eqn:: PF eigenvector commute with strong symmetry}
    [\Xi,\mathrm{Sym}(\Phi)] = 0.
\end{equation}
A direct corollary is that $\mathrm{Sym}(\Phi)\Xi \subseteq \mathcal{E}$.

\begin{lemma}\label{lemma:: strong symmetry of a symmetric CP map}
    Let $\Phi$ be a symmetric completely positive map on $\mathcal{B}(\mathcal{K})$, and let $\{K_i\}_i$ be a set of its Kraus operators.
    Then
    \begin{equation}
        \mathrm{Sym}(\Phi) = \{K_i,K^\dagger_i\}'_{i}. 
    \end{equation}
\end{lemma}
\begin{proof}
    If $X\in  \{K_i,K^\dagger_i\}'_{i}$, then the Kraus decomposition $\Phi(\cdot) = \sum_i K_i(\cdot)K^\dagger_i$ implies that $X\in \mathrm{Sym}(\Phi)$.

    Without loss of generality, we choose $\{K_i\}_i$ to be a linearly independent set of Kraus operators.
    Let $\{L_i\}_i$ be operators such that $\Tr(K_j L^\dagger_i) = \delta_{ij}$.
    Let $\{E_{jk}\}_{j,k}$ be a set of matrix units for $\mathcal{B}(\mathcal{K})$.
    Then for any $X\in \mathrm{Sym}(\Phi)$,
    \begin{equation}
    \begin{aligned}
        \sum_{j,k}E_{kj}\Phi(L^\dagger_i E_{jk}X) &= \sum_{i'} \sum_{j,k}\left( E_{kj}K_{i'}L^\dagger_iE_{jk} \right) XK^\dagger_{i'} \\
        &= \sum_{i'}\Tr(K_{i'}L^\dagger_i) XK^\dagger_{i'} = X K^\dagger_i.
    \end{aligned}
    \end{equation}
    Similarly, one obtains
    \begin{equation}
        \sum_{j,k}E_{kj}\Phi(L^\dagger_i E_{jk})X = K^\dagger_i X.
    \end{equation}
    Therefore, $XK^\dagger_i = K^\dagger_i X$ holds for all $i$.
    Because $\Phi$ is symmetric, $\{K^\dagger_i\}_i$ also serves as a set of Kraus operators for $\Phi$, so the same argument shows that $XK_i = K_i X$ holds for all $i$.
\end{proof}

\begin{proposition}\label{prop:: B(Psi) vs B(Psi_0)}
    Let $\Phi$ be a symmetric completely positive map on $\mathcal{B}(\mathcal{K})$, and let $\Phi_0$ be the truncated map defined in Proposition \ref{prop:: PF eigenspace of truncated map}.
    Then the following statements hold:
    \begin{itemize}
        \item[(1)] $\mathrm{Sym}(\Phi)\subseteq \mathrm{Sym}(\Phi_0)$;
        \item[(2)] $p_{\text{max}}\in \mathcal{Z}(\mathrm{Sym}(\Phi_0))\cap \mathcal{Z}(\mathrm{Sym}(\Phi))$;
        \item[(3)] $\mathrm{Sym}(\Phi)p_{\text{max}} = \mathrm{Sym}(\Phi_0)p_{\text{max}}$.
    \end{itemize}
\end{proposition}
\begin{proof}
    (1) Take $X\in \mathrm{Sym}(\Phi)$.
    Then $[X,\Xi] = [X^\dagger,\Xi] = 0$.
    Since the range projection of $\Xi$ is $p_{\text{max}}$, it follows that $[X,p_{\text{max}}] = [X^\dagger,p_{\text{max}}] = 0$.
    Therefore, for any operator $Y\in \mathcal{B}(\mathcal{K})$, we have
    \begin{equation}
        \Phi_0(XY) = \Phi(p_{\text{max}}X Y p_{\text{max}}) = \Phi(X p_{\text{max}}Y p_{\text{max}}) = X\Phi_0(Y).
    \end{equation}
    Similarly, one can show that $\Phi_0(X^\dagger Y) = X^\dagger \Phi_0(Y)$, and hence $X\in \mathrm{Sym}(\Phi_0)$.\\
    (2) From Corollary \ref{corollary:: canonical PF eigenvector for truncated map}, we know that $\Xi$ is the common canonical Perron--Frobenius eigenvector for both $\Phi$ and $\Phi_0$.
    Therefore, by Lemma \ref{lemma:: maximal range projection commute with Phi}, $p_{\text{max}}\in \mathrm{Sym}(\Phi_0)\cap \mathrm{Sym}(\Phi)$.
    As seen in the proof of (1), $p_{\text{max}}$ commutes with any element in $\mathrm{Sym}(\Phi)$.
    Thus, $p_{\text{max}}\in \mathcal{Z}(\mathrm{Sym}(\Phi))$.
    Similarly, $p_{\text{max}}$ commutes with every element of $\mathrm{Sym}(\Phi_0)$.
    Thus, $p_{\text{max}}\in \mathcal{Z}(\mathrm{Sym}(\Phi_0))$.\\
    (3) From (1) and (2), we immediately have $\mathrm{Sym}(\Phi)p_{\text{max}} \subseteq \mathrm{Sym}(\Phi_0)p_{\text{max}}$.
    Conversely, take $X\in \mathrm{Sym}(\Phi_0)p_{\text{max}}$, and let $\{K_i\}_i$ be a set of Kraus operators for $\Phi$.
    By Lemma \ref{lemma:: maximal range projection commute with Phi}, $p_{\text{max}}$ commutes with every $K_i$, and $\{K_i p_{\text{max}}\}_i$ is a set of Kraus operators for $\Phi_0$.
    Therefore, Lemma \ref{lemma:: strong symmetry of a symmetric CP map}, applied to $\Phi_0$, implies that $X$ commutes with $K_i p_{\text{max}}$ for every $i$.
    Since $X=Xp_{\text{max}}=p_{\text{max}}X$, we have 
    \begin{equation}
        XK_i = Xp_{\text{max}}K_i = X K_i p_{\text{max}} = K_i p_{\text{max}}X = K_i X,
    \end{equation}
    which implies $X\in \mathrm{Sym}(\Phi)$, completing the proof of (3).
\end{proof}

\begin{lemma}\label{lemma:: strong symmetry of UCP map}
    Let $\Psi$ be a UCP map on a $C^*$-algebra $\mathcal{A}$, and suppose $\Psi$ has a faithful invariant state $\omega$.
    Then we have
    \begin{equation}
        \mathrm{Sym}(\Psi) = \mathrm{Fix}(\Psi). 
    \end{equation}
\end{lemma}
\begin{proof}
    For $x\in \mathrm{Sym}(\Psi)$, we have $\Psi(x) = \Psi(\mathrm{I}_{\mathcal{A}}x) = \Psi(\mathrm{I}_{\mathcal{A}})x = x$.
    Thus, $\mathrm{Sym}(\Psi)\subseteq \mathrm{Fix}(\Psi)$.
    Conversely, take $x\in \mathrm{Fix}(\Psi)$. 
    By the Kadison--Schwarz inequality,
    \begin{equation}
        \Psi(x^*x)-\Psi(x)^*\Psi(x)\geq 0.
    \end{equation}
    Since $\omega$ is invariant and $\Psi(x)=x$, we have
    \begin{equation}
        \omega\bigl(\Psi(x^*x)-\Psi(x)^*\Psi(x)\bigr)=\omega(x^*x)-\omega(x^*x)=0.
    \end{equation}
    The faithfulness of $\omega$ therefore implies that $\Psi(x^*x)=\Psi(x)^*\Psi(x)$. 
    Applying the same argument to $x^*$ and using Choi's theorem \cite[Theorem 3.1]{Choi1974}, we conclude that $x$ belongs to the multiplicative domain of $\Psi$. 
    Therefore, for any $y\in \mathcal{A}$,
    \begin{equation}
        \Psi(xy) = \Psi(x)\Psi(y) = x\Psi(y),\quad \Psi(yx) = \Psi(y)\Psi(x) = \Psi(y)x.
    \end{equation}
    This demonstrates that $x\in \mathrm{Sym}(\Psi)$, proving the lemma.
\end{proof}

\begin{theorem}\label{thm:: structure of maximum eigenspace of a symmetric CP map}
    Let $\Phi:\mathcal{B}(\mathcal{K})\rightarrow \mathcal{B}(\mathcal{K})$ be a non-zero symmetric completely positive map.
    Let $\Xi$ be the canonical PF eigenvector of $\Phi$.
    Then for any operator $X\in p_{\text{max}}\mathcal{B}(\mathcal{K})p_{\text{max}}$, we have $\Xi^{1/2} X \Xi^{1/2}\in \mathcal{E}$ if and only if $X\in \mathrm{Sym}(\Phi)p_{\text{max}}$.
    In particular,
    \begin{equation}
        \mathcal{E} = \mathrm{Sym}(\Phi)\Xi.
    \end{equation}
\end{theorem}
\begin{proof}
    Let $X\in \mathrm{Sym}(\Phi)p_{\text{max}}$.
    By the fact that $p_{\text{max}}$ is the range projection of $\Xi$: 
    \begin{equation}
        \Phi(\Xi^{1/2}X\Xi^{1/2}) = \Phi(X\Xi) = X\Phi(\Xi) = \rho(\Phi) \Xi^{1/2}X\Xi^{1/2},
    \end{equation}
    so $\Xi^{1/2}X\Xi^{1/2}\in\mathcal{E}$, which proves the necessity.\\
    Next, we prove the sufficiency.
    Define a UCP map $\widetilde{\Phi}_0$ on $p_{\text{max}}\mathcal{B}(\mathcal{K})p_{\text{max}}$ by
    \begin{equation}
        \widetilde{\Phi}_0(X) = \frac{1}{\rho(\Phi)}\Xi^{-1/2}\Phi(\Xi^{1/2}X\Xi^{1/2})\Xi^{-1/2},\quad X\in p_{\text{max}}\mathcal{B}(\mathcal{K})p_{\text{max}}.
    \end{equation}
    Denote the Perron--Frobenius eigenspace of $\widetilde{\Phi}_0$ by $\widetilde{\mathcal{E}}_0$.
    For $X\in p_{\text{max}}\mathcal{B}(\mathcal{K})p_{\text{max}}$, we have 
    \begin{equation}
        \Xi^{1/2}\widetilde{\Phi}_0(X) \Xi^{1/2} = \Phi(\Xi^{1/2} X\Xi^{1/2})/\rho(\Phi).
    \end{equation}
    Hence, $\Xi^{1/2} X \Xi^{1/2}\in \mathcal{E}$ if and only if $X\in \widetilde{\mathcal{E}}_0$.
    Because $\Phi$ is symmetric with respect to $\Tr$,
    \begin{equation}
    \begin{aligned}
        \Tr  (\Xi\widetilde{\Phi}_0(X)\Xi) &= \Tr  (\Xi^{1/2}\Phi(\Xi^{1/2}X\Xi^{1/2})\Xi^{1/2})/\rho(\Phi) \\
        &= \Tr  (\Xi \Phi(\Xi^{1/2}X\Xi^{1/2}))/\rho(\Phi) \\
        &= \Tr  (\Phi(\Xi)\Xi^{1/2}X\Xi^{1/2})/\rho(\Phi) \\
        &= \Tr (\Xi X\Xi). 
    \end{aligned}
    \end{equation}
    Therefore, the positive linear functional $\Tr (\Xi\cdot \Xi)/\Tr(\Xi^2)$ on $p_{\text{max}}\mathcal{B}(\mathcal{K})p_{\text{max}}$ is an equilibrium of $\widetilde{\Phi}_0$, which is faithful since $p_{\text{max}}$ is the range projection of $\Xi$. 
    Thus, by Lemma \ref{lemma:: strong symmetry of UCP map}, we have $\widetilde{\mathcal{E}}_0 = \mathrm{Sym}(\widetilde{\Phi}_0)$.
    By Proposition \ref{prop:: B(Psi) vs B(Psi_0)}, $\mathrm{Sym}(\Phi)p_{\text{max}} = \mathrm{Sym}(\Phi_0)p_{\text{max}}$.
    Therefore, it suffices to prove that $\mathrm{Sym}(\widetilde{\Phi}_0) = \mathrm{Sym}(\Phi_0)p_{\text{max}}$.

    By Corollary \ref{corollary:: canonical PF eigenvector for truncated map} and Equation \eqref{eqn:: PF eigenvector commute with strong symmetry}, $\mathrm{Sym}(\Phi_0)p_{\text{max}}$ commutes with $\Xi$.
    Hence, elements of $\mathrm{Sym}(\Phi_0)p_{\text{max}}$ are fixed by $\widetilde{\Phi}_0$ and are in $\mathrm{Sym}(\widetilde{\Phi}_0)$ due to Lemma \ref{lemma:: strong symmetry of UCP map}. 
    For the converse, define the invertible linear map $C$ on $p_{\text{max}}\mathcal{B}(\mathcal{K})p_{\text{max}}$ by $C(Y) = \Xi^{1/2}Y\Xi^{1/2}$. 
    For an operator $X$, let $R_X$ denote the right multiplication map on $\mathcal{B}(\mathcal{K})$.
    Then $X\in \mathrm{Sym}(\widetilde{\Phi}_0)$ implies that
    \begin{equation}
        [C^{-1}\Phi_0 C,R_X] = 0. 
    \end{equation}
    Conjugating both sides by $C$, we obtain $[\Phi_0, CR_{X}C^{-1}] = 0$.
    Since $CR_XC^{-1} = R_{\Xi^{-1/2}X\Xi^{1/2}}$, this means
    \begin{equation}
        \Phi_0(Y\Xi^{-1/2}X\Xi^{1/2}) = \Phi_0(Y)\Xi^{-1/2}X\Xi^{1/2},\quad Y\in \mathcal{B}(\mathcal{K}).
    \end{equation}
    Since $\Phi_0$ is symmetric, by the cyclicity of $\Tr$, for any $Y,Z\in \mathcal{B}(\mathcal{K})$, we have
    \begin{equation}
    \begin{aligned}
        \Tr\left( Z\Phi_0(\Xi^{-1/2}X\Xi^{1/2}Y) \right) &= \Tr\left( \Phi_0(Z)\Xi^{-1/2}X\Xi^{1/2}Y \right) \\
        &= \Tr\left( \Phi_0(Z \Xi^{-1/2}X\Xi^{1/2})Y \right) \\
        &= \Tr\left( Z \Xi^{-1/2}X\Xi^{1/2}\Phi_0(Y) \right).
    \end{aligned}
    \end{equation}
    Therefore, we have $\Phi_0(\Xi^{-1/2}X\Xi^{1/2}Y) = \Xi^{-1/2}X\Xi^{1/2}\Phi_0(Y)$, which implies that  $\Xi^{-1/2}X\Xi^{1/2}$ is in $ \mathrm{Sym}(\Phi_0)p_{\text{max}}$.
    Furthermore, by Corollary \ref{corollary:: canonical PF eigenvector for truncated map} and Equation \eqref{eqn:: PF eigenvector commute with strong symmetry}, we have $[\Xi^{-1/2}X\Xi^{1/2},\Xi] = 0$, which in turn implies $[X,\Xi] = 0$.
    Hence $X = \Xi^{-1/2}X\Xi^{1/2} \in \mathrm{Sym}(\Phi_0)p_{\text{max}}$, and thus we have $\mathrm{Sym}(\widetilde{\Phi}_0) = \mathrm{Sym}(\Phi_0)p_{\text{max}}$.
    Together with Proposition \ref{prop:: B(Psi) vs B(Psi_0)}, this proves the claimed equivalence.
    Finally, since every element of $\mathcal{E}$ is supported on $p_{\text{max}}$ by Proposition \ref{prop:: PF eigenspace of truncated map} and every element of $\mathrm{Sym}(\Phi)p_{\text{max}}$ commutes with $\Xi$, the equivalence yields $\mathcal{E} = \mathrm{Sym}(\Phi)\Xi$.
    This completes the proof of the theorem.
\end{proof}
\begin{remark}
    The map $\widetilde{\Phi}_0$ is KMS-symmetric with respect to $\Tr(\Xi\cdot\Xi)$.  
\end{remark}

\subsection{Ground state subspace of a reflection positive Hamiltonian}

Now we apply the theory developed in the previous section to understand the ground state subspace of a reflection positive Hamiltonian.
Consider a reflection positive Hamiltonian $H$ on the Hilbert space $\mathcal{H}_{-}\otimes \mathcal{H}_{+}$.
In this case, the ground state space of $H$ coincides with the eigenspace corresponding to the maximum eigenvalue of the operator $e^{-\tau H}$.
Consider the semigroup of symmetric completely positive maps on $\mathcal{B}(\mathcal{H}_{+})$ given by
\begin{equation}
    \Phi_{\tau} = \mathcal{O}e^{-\tau H}\mathcal{O}^{-1},\quad \tau\geq 0.
\end{equation}
For $\tau>0$, $\mathcal{O}$ maps the ground state space of $H$ unitarily onto the eigenspace corresponding to the common largest eigenvalue of $\{\Phi_{\tau}\}_{\tau>0}$, denoted by $\mathcal{E}$.

\begin{lemma}
    Let $H$ be a reflection positive Hamiltonian on $\mathcal{H}_{-}\otimes \mathcal{H}_{+}$.
    Define $\Pi$ as the orthogonal projection onto the ground state space of $H$; then $\Pi$ is reflection positive.
    Let $\Xi$ be the canonical Perron--Frobenius eigenvector of $\Phi_{\tau}$.
    Then we have
    \begin{equation}
        \Xi = \mathcal{O}\left( \sum_{i}\Pi \ket{\widehat{\theta}i\otimes i} \right),
    \end{equation}
    where $\{\ket{i}\}_i$ is an orthonormal basis of $\mathcal{H}_{+}$.
\end{lemma}
\begin{proof}
    We have $\Pi = \lim_{\tau\rightarrow \infty}e^{-\tau (H- E_0)}$, where $E_0$ is the ground state energy of $H$.
    Thus, from the reflection positivity of $H$, it follows that $\Pi$ is also reflection positive.
    By the semigroup property of $\Phi_{\tau}$, for any $\tau_0>0$, we have
    \begin{equation}
        \mathcal{O}\Pi \mathcal{O}^{-1} = \lim_{n\rightarrow \infty}\frac{1}{n}\sum^{n-1}_{k=0}\frac{\Phi_{k\tau_0}}{e^{-k\tau_0 E_0}}.
    \end{equation}
    Therefore, we have 
    \begin{equation}
        \Xi = \lim_{n\rightarrow \infty}\frac{1}{n}\sum^{n-1}_{k=0}\frac{\Phi_{k\tau_0}}{e^{-k\tau_0 E_0}}(\mathrm{I}_{\mathcal{H}_{+}}) = \mathcal{O} \Pi(\mathcal{O}^{-1}(\mathrm{I}_{\mathcal{H}_{+}})) = \mathcal{O}\left( \sum_{i}\Pi \ket{\widehat{\theta}i\otimes i} \right).
    \end{equation}
\end{proof}

\begin{definition}[Canonical Perron--Frobenius Ground State]
    Let $H$ be a reflection positive Hamiltonian on $\mathcal{H}_{-}\otimes \mathcal{H}_{+}$. 
    Let $\Pi$ be the orthogonal projection onto the ground state space of $H$.
    Up to normalization, we call 
    \begin{equation}
        \ket{\phi_{\text{PF}}} = \sum_{i} \Pi \ket{\widehat{\theta}i\otimes i}
    \end{equation}
    the canonical PF ground state of $H$.
\end{definition}

\begin{definition}[Entanglement Support]
    Let $\Pi$ be a projection on $\mathcal{H}_{-}\otimes \mathcal{H}_{+}$.
    We define the entanglement support of $\Pi$ as the range projection of the reduced operator $\Tr_{\mathcal{H}_-}(\Pi)$:
    \begin{equation}
        \widehat{\Pi} = \range(\Tr_{\mathcal{H}_-}(\Pi)). 
    \end{equation}
    We define the entanglement support of the Hamiltonian $H$ as the entanglement support of its ground state projection.
\end{definition}

\begin{proposition}\label{prop:: PF eigenvector and entanglement support}
    Let $H$ be a reflection positive Hamiltonian on $\mathcal{H}_{-}\otimes \mathcal{H}_{+}$, and let $\ket{\phi_{\text{PF}}}$ be its canonical PF ground state.
    Then we have
    \begin{equation}
       \range(\Tr_{\mathcal{H}_{-}}\ket{\phi_{\text{PF}}}\bra{\phi_{\text{PF}}}) = \widehat{\Pi}. 
    \end{equation}
\end{proposition}
\begin{proof}
    For $\tau\geq 0$, let $\Xi$ be the canonical PF eigenvector of the map $\Phi_{\tau} = \mathcal{O}e^{-\tau H}\mathcal{O}^{-1}$, and let $p_{\text{max}}$ be the range projection of $\Xi$.
    By Proposition \ref{prop:: reduced density matrix in terms of O}, we know that
    \begin{equation}
        \Tr_{\mathcal{H}_{-}}\ket{\phi_{\text{PF}}}\bra{\phi_{\text{PF}}} = \mathcal{O}(\phi_{\text{PF}})^2 = \Xi^2.
    \end{equation}
    Since the range projection of $\Xi$ is $p_{\text{max}}$,
    \begin{equation}
        \range (\Tr_{\mathcal{H}_{-}}\ket{\phi_{\text{PF}}}\bra{\phi_{\text{PF}}}) = \range(\Xi^2) = p_{\text{max}},
    \end{equation}
    so we have $p_{\text{max}}\leq \widehat{\Pi}$.
    Conversely, suppose that $\ket{\phi}$ is an arbitrary ground state of $H$; then $\mathcal{O}(\phi)\in \mathcal{E}$.
    Therefore, by Proposition \ref{prop:: PF eigenspace of truncated map}, we have $p_{\text{max}}\mathcal{O}(\phi) = \mathcal{O}(\phi)$.
    Thus,
    \begin{equation}
    \begin{aligned}
        p_{\text{max}}\Tr_{\mathcal{H}_-}(\ket{\phi}\bra{\phi}) p_{\text{max}} &= p_{\text{max}}\mathcal{O}(\phi)\mathcal{O}(\phi)^{\dagger}p_{\text{max}} \\
        &= \mathcal{O}(\phi)\mathcal{O}(\phi)^{\dagger} = \Tr_{\mathcal{H}_-}(\ket{\phi}\bra{\phi}). 
    \end{aligned}
    \end{equation}
    This proves that $\widehat{\Pi}\leq p_{\text{max}}$.
\end{proof}

\begin{remark}
    This proposition illustrates that $\ket{\phi_{\text{PF}}}$ possesses the maximal entanglement support among all ground states of $H$.
\end{remark}

\begin{definition}
    Let $T$ be an operator on $\mathcal{H}_{-}\otimes \mathcal{H}_{+}$.
    We define 
    \begin{equation}
        \mathrm{Sym}_{\mathcal{H}_{+}}(T) = \{X\in \mathcal{B}(\mathcal{H}_{+}) : [\mathrm{I}_{\mathcal{H}_{-}} \otimes X, T] = [\mathrm{I}_{\mathcal{H}_{-}} \otimes X^\dagger, T] = 0\}.
    \end{equation}
\end{definition}

If $p$ is a projection on $\mathcal{H}_+$ and $T_Q$ is an operator on $\Theta(p)\mathcal{H}_-\otimes p\mathcal{H}_+$, we denote by $\mathrm{Sym}_{p\mathcal{H}_+}(T_Q)$ the algebra of operators $X\in p\mathcal{B}(\mathcal{H}_+)p$ such that both $\mathrm{I}_{\Theta(p)\mathcal{H}_-}\otimes X$ and $\mathrm{I}_{\Theta(p)\mathcal{H}_-}\otimes X^\dagger$ commute with $T_Q$.

\begin{proposition}\label{prop:: local symmetry v.s. strong symmetry}
    Consider an operator $T$ on $\mathcal{H}_{-}\otimes \mathcal{H}_{+}$.
    Then we have
    \begin{equation}
        \mathrm{Sym}_{\mathcal{H}_{+}}(T) = \mathrm{Sym}(\mathcal{O}T\mathcal{O}^{-1}).
    \end{equation}
\end{proposition}
\begin{proof}
    Set $\Phi_T=\mathcal{O}T\mathcal{O}^{-1}$, and let $L_X$ denote left multiplication by $X$ on $\mathcal{B}(\mathcal{H}_{+})$.
    By Lemma \ref{lemma:: s-to-o map}, $\mathcal{O}(\mathrm{I}_{\mathcal{H}_{-}}\otimes X)\mathcal{O}^{-1}=L_X$.
    Therefore, $[\mathrm{I}_{\mathcal{H}_{-}}\otimes X,T]=0$ if and only if $[L_X,\Phi_T]=0$, or equivalently,
    \begin{equation}
        \Phi_T(XY)=X\Phi_T(Y),\quad Y\in\mathcal{B}(\mathcal{H}_{+}).
    \end{equation}
    Applying the same argument to $X^\dagger$ and using Definition \ref{def:: strong symmetry} proves the claim.
\end{proof}

\begin{proposition}\label{prop: local commutant and maximal support}
    Consider a reflection positive Hamiltonian $H$ on $\mathcal{H}_{-}\otimes \mathcal{H}_{+}$ with ground state projection $\Pi$ and entanglement support $\widehat{\Pi}$.
    Set $Q:=\Theta(\widehat{\Pi})\otimes \widehat{\Pi}$ and $\mathcal{H}_Q:=Q(\mathcal{H}_-\otimes\mathcal{H}_+)$, and let $\widetilde{H}:=H|_{\mathcal{H}_Q}$.
    Then the following statements hold:
    \begin{itemize}
        \item[(1)] The projection $\widehat{\Pi}$, and therefore $Q$, commutes with $H$; 
        \item[(2)] $H$ and $\widetilde{H}$ have the same ground-state subspace;
        \item[(3)] $\widehat{\Pi}\in\mathcal{Z}(\mathrm{Sym}_{\mathcal{H}_+}(H))$; 
        \item[(4)] As $C^*$-subalgebras of $\widehat{\Pi}\mathcal{B}(\mathcal{H}_+)\widehat{\Pi}$,
        \begin{equation}
            \mathrm{Sym}_{\mathcal{H}_+}(\Pi)\widehat{\Pi}
            =\mathrm{Sym}_{\mathcal{H}_+}(H)\widehat{\Pi}
            =\mathrm{Sym}_{\widehat{\Pi}\mathcal{H}_+}(\widetilde{H}).
        \end{equation}
    \end{itemize}
\end{proposition}
\begin{proof}
    Fix $\tau>0$ and set $\Phi_\tau=\mathcal{O}e^{-\tau H}\mathcal{O}^{-1}$.
    Since $x\mapsto e^{-\tau x}$ is injective on $\mathbb{R}$, functional calculus and Proposition \ref{prop:: local symmetry v.s. strong symmetry} give
    \begin{equation}\label{eqn:: local symmetry of H v.s. strong symmetry for Phi_tau}
        \mathrm{Sym}_{\mathcal{H}_+}(H)
        =\mathrm{Sym}_{\mathcal{H}_+}(e^{-\tau H})
        =\mathrm{Sym}(\Phi_\tau).
    \end{equation}
    $(1)$: Proposition \ref{prop:: PF eigenvector and entanglement support} shows that $\widehat{\Pi}$ is the range projection of the canonical PF eigenvector of $\Phi_\tau$.
    Lemma \ref{lemma:: maximal range projection commute with Phi} therefore gives $\widehat{\Pi}\in \mathrm{Sym}(\Phi_{\tau})$. 
    Translating these relations through Lemma \ref{lemma:: s-to-o map} and by Equation \eqref{eqn:: local symmetry of H v.s. strong symmetry for Phi_tau}, 
    \begin{equation}
        [H,\mathrm{I}_{\mathcal{H}_-}\otimes\widehat{\Pi}]
        =[H,\Theta(\widehat{\Pi})\otimes\mathrm{I}_{\mathcal{H}_+}]
        =0.
    \end{equation}
    Hence $Q=\Theta(\widehat{\Pi})\otimes\widehat{\Pi}$ commutes with $H$.

    $(2)$: let $\ket{\phi}$ be a ground state of $H$.
    Then $\mathcal{O}(\phi)$ belongs to the PF eigenspace of $\Phi_\tau$.
    Propositions \ref{prop:: PF eigenspace of truncated map} and \ref{prop:: PF eigenvector and entanglement support} give
    \begin{equation}
        \mathcal{O}(\phi)
        =\widehat{\Pi}\mathcal{O}(\phi)\widehat{\Pi}.
    \end{equation}
    Lemma \ref{lemma:: s-to-o map} therefore gives $Q\ket{\phi}=\ket{\phi}$.
    Thus every ground state of $H$ belongs to $\mathcal{H}_Q$.
    Since $Q$ commutes with $H$, the subspace $\mathcal{H}_Q$ reduces $H$, and therefore $\widetilde{H}$ has exactly the same ground-state subspace as $H$.

    $(3)$: Proposition \ref{prop:: B(Psi) vs B(Psi_0)}(2) gives $\widehat{\Pi}\in\mathcal{Z}(\mathrm{Sym}(\Phi_\tau))$. 
    Equation \eqref{eqn:: local symmetry of H v.s. strong symmetry for Phi_tau} therefore implies
    \begin{equation}
        \widehat{\Pi}\in\mathcal{Z}(\mathrm{Sym}_{\mathcal{H}_+}(H)).
    \end{equation}

    $(4)$: Set $F=\mathcal{O}\Pi\mathcal{O}^{-1}$.
    The PF eigenspace of $\Phi_\tau$ corresponds under $\mathcal{O}$ to the ground-state subspace of $H$, whereas $F$ is the orthogonal projection onto this same subspace.
    Thus $F$ and $\Phi_\tau$ have the same PF eigenspace.
    Moreover, the preceding characterization of $\Xi$ gives
    \begin{equation}
        \Xi=\mathcal{O}\Pi\mathcal{O}^{-1}(\mathrm{I}_{\mathcal{H}_+})=F(\mathrm{I}_{\mathcal{H}_+}),
    \end{equation}
    so $\Xi$ is their common canonical PF eigenvector.
    Consequently, Theorem \ref{thm:: structure of maximum eigenspace of a symmetric CP map} gives $\mathrm{Sym}(F)\widehat{\Pi}
        =\mathrm{Sym}(\Phi_\tau)\widehat{\Pi}$. 
    Proposition \ref{prop:: local symmetry v.s. strong symmetry} and Equation \eqref{eqn:: local symmetry of H v.s. strong symmetry for Phi_tau} now yield the first equalty: 
    \begin{equation}
        \mathrm{Sym}_{\mathcal{H}_+}(\Pi)\widehat{\Pi}
        =\mathrm{Sym}_{\mathcal{H}_+}(H)\widehat{\Pi}.
    \end{equation}

    Under the restriction of $\mathcal{O}$ to $\mathcal{H}_Q$, the operator $e^{-\tau\widetilde{H}}$ corresponds to the restriction of $\Phi_\tau$ to $\widehat{\Pi}\mathcal{B}(\mathcal{H}_+)\widehat{\Pi}$.
    By Proposition \ref{prop:: PF eigenspace of truncated map}, this restricted map and $\Phi_\tau$ have the same PF eigenspace, and Corollary \ref{corollary:: canonical PF eigenvector for truncated map} shows that their canonical PF eigenvector is $\Xi$.
    Theorem \ref{thm:: structure of maximum eigenspace of a symmetric CP map}, Proposition \ref{prop:: local symmetry v.s. strong symmetry}, and Equation \eqref{eqn:: local symmetry of H v.s. strong symmetry for Phi_tau} therefore give
    \begin{equation}
        \mathrm{Sym}_{\mathcal{H}_+}(H)\widehat{\Pi}\Xi
        =\mathrm{Sym}_{\widehat{\Pi}\mathcal{H}_+}(\widetilde{H})\Xi.
    \end{equation}
    Since $\Xi$ is invertible in $\widehat{\Pi}\mathcal{B}(\mathcal{H}_+)\widehat{\Pi}$, right multiplying both sides by $\Xi^{-1}$ gives the second equality:
    \begin{equation}
        \mathrm{Sym}_{\mathcal{H}_+}(H)\widehat{\Pi}
        =\mathrm{Sym}_{\widehat{\Pi}\mathcal{H}_+}(\widetilde{H}).
    \end{equation}
\end{proof}

In the following theorem and corollary, all inverses and negative powers of the canonical PF eigenvector $\Xi$ are taken in the corner $\widehat{\Pi}\mathcal{B}(\mathcal{H}_+)\widehat{\Pi}$.

\begin{theorem}\label{theorem:: ground state of H}
    Let $H$ be a reflection positive Hamiltonian on $\mathcal{H}_{-}\otimes \mathcal{H}_{+}$ with ground state projection $\Pi$.
    Let $\ket{\phi_{\text{PF}}}$ be the canonical PF ground state of $H$.
    Then, for every ground state $\ket{\phi}$ of $H$, there exists a unique $W\in \mathrm{Sym}_{\mathcal{H}_{+}}(H)\widehat{\Pi}$ such that
    \begin{equation}
        \ket{\phi} = (\mathrm{I}\otimes W)\ket{\phi_{\text{PF}}} = (\Theta(W^\dagger)\otimes \mathrm{I})\ket{\phi_{\text{PF}}}.
    \end{equation}
\end{theorem}
\begin{proof}
    Let $\ket{\phi}$ be an arbitrary ground state of $H$.
    Fix $\tau>0$ and set $\Phi_\tau=\mathcal{O}e^{-\tau H}\mathcal{O}^{-1}$.
    Then $\mathcal{O}(\phi)$ belongs to the PF eigenspace $\mathcal{E}$ of $\Phi_\tau$.
    By Proposition \ref{prop:: PF eigenvector and entanglement support} and Proposition \ref{prop:: PF eigenspace of truncated map},
    \begin{equation}
        \mathcal{O}(\phi)=\widehat{\Pi}\mathcal{O}(\phi)\widehat{\Pi}.
    \end{equation}
    Since $\Xi$ is invertible in $\widehat{\Pi}\mathcal{B}(\mathcal{H}_{+})\widehat{\Pi}$, define
    \begin{equation}
        W:=\mathcal{O}(\phi)\Xi^{-1}\in\widehat{\Pi}\mathcal{B}(\mathcal{H}_{+})\widehat{\Pi}.
    \end{equation}
    Then we have $\mathcal{O}(\phi) = W \Xi$. 
    Theorem \ref{thm:: structure of maximum eigenspace of a symmetric CP map} gives $W\in\mathrm{Sym}(\Phi_\tau)\widehat{\Pi}$.
    Proposition \ref{prop:: local symmetry v.s. strong symmetry} and the injectivity of the functional calculus give $W\in\mathrm{Sym}_{\mathcal{H}_+}(H)\widehat{\Pi}$.
    Invertibility of $\Xi$ in the corner proves uniqueness.
    Moreover, by Equation \eqref{eqn:: PF eigenvector commute with strong symmetry}, $W$ commutes with $\Xi$, and Lemma \ref{lemma:: s-to-o map} gives
    \begin{equation}
        \mathcal{O}(\phi) = W\Xi = \mathcal{O}(\mathrm{I}\otimes W\ket{\phi_{\text{PF}}})
        = \Xi W = \mathcal{O}(\Theta(W^\dagger)\otimes \mathrm{I}\ket{\phi_{\text{PF}}}).
    \end{equation}
    Since $\mathcal{O}$ is an isomorphism, the claimed representation follows.
\end{proof}

\begin{corollary}
    Let $H$ be a reflection positive Hamiltonian on $\mathcal{H}_{-}\otimes \mathcal{H}_{+}$, with ground state projection $\Pi$.
    Let $\ket{\phi_{\text{PF}}}$ be the canonical PF ground state of $H$.
    Let $\ket{\psi}$ be a ground state of $H$.
    Then $\ket{\psi}$ is reflection positive if and only if $\ket{\psi} = (\mathrm{I}\otimes W)\ket{\phi_{\text{PF}}}$, where $W\in \mathrm{Sym}_{\mathcal{H}_{+}}( H)\widehat{\Pi}$ is a scalar multiple of a positive operator.
    In this case, $\range (\Tr_{\mathcal{H}_{-}}\ket{\psi}\bra{\psi}) = \range(W)$.
\end{corollary}
\begin{proof}
    Let $\ket{\psi}$ be a ground state of $H$.
    Then, by Theorem \ref{theorem:: ground state of H}, there exists $W\in \mathrm{Sym}_{\mathcal{H}_{+}}( H)\widehat{\Pi}$ such that $\ket{\psi} = (\mathrm{I}\otimes W)\ket{\phi_{\text{PF}}}$.
    By Lemma \ref{lemma:: s-to-o map}, $\mathcal{O}(\psi) = W\Xi$. 
    Since $[W,\Xi]=0$ and $\Xi$ is positive and invertible on $\widehat{\Pi}\mathcal{H}_{+}$, Proposition \ref{prop:: reduced density matrix in terms of O} implies that $\ket{\psi}$ is reflection positive if and only if $W$ is a scalar multiple of a positive operator.
    In this case, write $W=e^{-i\theta_0}P$, where $P\geq0$.
    Then
    \begin{equation}
        \Tr_{\mathcal{H}_{-}}\ket{\psi}\bra{\psi}=W\Xi^2W^\dagger=P\Xi^2P=P^2\Xi^2.
    \end{equation}
    Since $\Xi$ is strictly positive on $\widehat{\Pi}\mathcal{H}_+$ and commutes with $P$, it follows that
    \begin{equation}
        \range(\Tr_{\mathcal{H}_{-}}\ket{\psi}\bra{\psi})=\range(P)=\range(W).
    \end{equation}
\end{proof}


\section{Frustration-free interaction with reflection positivity}\label{section:: FF interaction with RP}
In this section, we analyze the ground state subspace of a reflection positive, frustration-free interaction. 
We prove a criterion for local indistinguishability of the ground states in terms of global nondegeneracy of the ground state.

\subsection{Frustration-free interaction with reflection positivity}
We consider an $n$-dimensional quantum spin system on a lattice $\Gamma$ embedded in an ambient oriented manifold $\Sigma$, and assume that $\Sigma$ is equipped with a distance function $d$ such that $(\Gamma,d)$ forms a metric space. 
Each site $v\in \Gamma$ is assigned a finite-dimensional Hilbert space $\mathcal{H}_v$. 
In this paper, $\Sigma$ will be $\mathbb{R}^n$ or the sphere $\mathbb{S}^n$. 
For simplicity, we assume that all $\mathcal{H}_v$ are isomorphic to the Hilbert space $\mathbb{C}^d$, though our results remain valid in more general situations. 
For each $\ket{\xi}\in \mathbb{C}^d$, we denote by $\ket{\xi_v}$ the corresponding vector in $\mathcal{H}_v$. 
For a finite subset $\Lambda\subset \Gamma$, we define the local Hilbert space $\mathcal{H}_\Lambda = \bigotimes_{v\in \Lambda}\mathcal{H}_v$, and we denote by $\mathrm{I}_\Lambda$ the identity operator on $\mathcal{H}_\Lambda$. 
For a local operator $A$, we denote by $\textbf{supp}(A)$ the support of $A$, that is, the minimal subset $\Lambda\subset \Gamma$ outside which $A$ acts trivially. 
The set of all operators with support contained in $\Lambda$ is denoted by $\mathfrak{A}_\Lambda$, which is a $\mathrm{C}^*$-algebra isomorphic to $\mathcal{B}(\mathcal{H}_\Lambda)$, the set of bounded operators on $\mathcal{H}_\Lambda$. 
For two finite regions $\Lambda_1\subseteq \Lambda_2$, we have a natural inclusion $\mathfrak{A}_{\Lambda_1}\subseteq \mathfrak{A}_{\Lambda_2}$ given by tensoring with $\mathrm{I}_{\Lambda_2\backslash \Lambda_1}$. 
For simplicity, we will often identify $A\otimes \mathrm{I}_{\Lambda_2\backslash \Lambda_1}$ with $A$ for $A\in \mathfrak{A}_{\Lambda_1}$. 
The algebra of local operators is defined as $\mathfrak{A}_{\mathrm{loc}} = \bigcup_{\Lambda\subset \Gamma \text{ finite}} \mathfrak{A}_\Lambda$. 
For a finite set $\Lambda\subseteq \Gamma$, denote by $\overline{\Lambda}$ the complement of $\Lambda$ in $\Gamma$. 
In what follows, we denote the commutant of a subalgebra $\mathcal{A}\subseteq \mathfrak{A}_\Lambda$ in $\mathfrak{A}_\Lambda$ simply by $\mathcal{A}'$. 

Recall that an interaction $\varPhi$ is a map from the set of finite subsets of $\Gamma$ to $\mathfrak{A}_{\text{loc}}$ that satisfies 
\begin{equation}
    \varPhi(\Lambda)\in \mathfrak{A}_\Lambda \quad \text{and} \quad \varPhi(\Lambda)^\dagger = \varPhi(\Lambda). 
\end{equation}
We say that $\varPhi$ is finite-range or \emph{local} if there exists $R>0$ such that $\varPhi(\Lambda) = 0$ whenever the diameter of $\Lambda$ exceeds $R$. 
Given any finite subset $\Lambda\subseteq \Gamma$, define the local Hamiltonian on $\Lambda$ by $H_\Lambda = \sum_{\Lambda_0\subseteq \Lambda}\varPhi(\Lambda_0)$. 

An interaction $\varPhi$ is \emph{frustration-free} if, for every finite subset $\Lambda\subseteq \Gamma$, the ground state space of $H_\Lambda$ is the intersection of the ground state spaces of all local terms $\varPhi(\Lambda_0)$ with $\Lambda_0\subseteq \Lambda$. 
For any finite set $\Lambda\subseteq \Gamma$, define $\Pi(\Lambda)$ as the ground state projection of $H_\Lambda$. 
Then the frustration-free property of the interaction implies the inequality
\begin{equation}
    \Pi(\Lambda_1)\otimes \mathrm{I}_{\Lambda_2\backslash \Lambda_1}\geq \Pi(\Lambda_2)
\end{equation}
holds for all finite subsets $\Lambda_1\subseteq \Lambda_2\subseteq \Gamma$. 
In general, we call a finite set of self-adjoint operators $\{H_i\}^n_{i=1}$ \emph{mutually frustration-free}, if the ground state subspace of $\sum^n_{i=1} H_i$ is the intersection of the ground state subspaces of $H_i$. 

We now set up reflection positivity in the above context in the sense of \cite{FILS1978}. 
Let $\theta:\Sigma\rightarrow \Sigma$ be a continuous involution, and suppose there exist submanifolds $\Sigma_{+}, \Sigma_{-}, \Sigma_{0}$ of $\Sigma$ satisfying
\begin{equation}
    \Sigma = \Sigma_{-} \sqcup \Sigma_{0} \sqcup \Sigma_{+}, \quad \theta(\Sigma_{\pm}) = \Sigma_{\mp}, \quad \theta(\Sigma_{0}) = \Sigma_{0}. 
\end{equation}
For $\Lambda\subseteq \Gamma$, we denote $\Lambda_{\pm} = \Lambda \cap \Sigma_{\pm}$. 
We assume that $\Gamma\cap \Sigma_{0} = \emptyset$, which implies a disjoint union $\Gamma = \Gamma_{-}\sqcup \Gamma_{+}$. 
We further assume that the restriction of $\theta$ to $\Gamma$ is a bijection on $\Gamma$. 
Consequently, we have $\theta(\Gamma_{+}) = \Gamma_{-}$ and $\theta(\Gamma_{-}) = \Gamma_{+}$.

We now consider the quantum spin system $\mathfrak{A}(\Gamma)$ defined on $\Gamma$. 
We select a family of anti-unitary operators 
\begin{equation}
    \widehat{\theta}_i: \mathcal{H}_i \rightarrow \mathcal{H}_{\theta(i)}, \quad i\in \Gamma_{+}.
\end{equation}
For a local observable $X \in \mathfrak{A}_{\text{loc}}(\Gamma_{+})$ supported on $\Gamma_{+}$, we define its reflection (time reversal) in $\Gamma_{-}$ by
\begin{equation}
    \Theta(X) = \widehat{\theta}_{\mathrm{supp} X} X \widehat{\theta}_{\mathrm{supp} X}^{-1},\quad \widehat{\theta}_{\mathrm{supp} X} = \bigotimes_{i\in \mathrm{supp}X} \widehat{\theta}_i.
\end{equation}
Since the operators $\widehat{\theta}_i$ are anti-unitary, the mapping $\Theta$ is norm-preserving and preserves multiplication and adjoints.
Therefore, it extends uniquely to an anti-linear $*$-isomorphism from $\mathfrak{A}(\Gamma_{+})$ to $\mathfrak{A}(\Gamma_{-})$.

\begin{definition}[Reflection Positive Interaction]\label{def:: reflection positive interaction}
    Let $\varPhi$ be an interaction on $\Gamma$. 
    We say that $\varPhi$ is reflection positive with respect to $\Theta$ if the following conditions hold:
    \begin{itemize}
        \item For any finite subset $\Lambda \subseteq \Gamma_{+}$, we have $\Theta(\varPhi(\Lambda)) = \varPhi(\theta(\Lambda))$.
        \item For any finite subset $\Lambda$ such that $\Lambda_{+},\Lambda_{-}\neq\emptyset$, we have $\varPhi(\Lambda)=0$ unless $\theta(\Lambda)=\Lambda$.
        If $\theta(\Lambda)=\Lambda$, there exist a self-adjoint operator $K_{\Lambda}\in \mathfrak{A}_{\Lambda_+}$ and a finite linearly independent set $\{O_{\Lambda,j}\}_{j=1}^{m_{\Lambda}}\subseteq \mathfrak{A}_{\Lambda_+}$ that is invariant under the adjoint operation such that
        \begin{equation}
            \varPhi(\Lambda)=\Theta(K_{\Lambda})\otimes \mathrm{I}_{\Lambda_+}+\mathrm{I}_{\Lambda_-}\otimes K_{\Lambda}-\sum_{j=1}^{m_{\Lambda}}\Theta(O_{\Lambda,j})\otimes O_{\Lambda,j}.
        \end{equation}
    \end{itemize}
\end{definition}

We say that a subset $\Lambda$ of $\Gamma$ is reflection-symmetric if $\theta(\Lambda) = \Lambda$. 
This gives the following proposition:

\begin{proposition}
    Let $\varPhi$ be an interaction on $\Gamma$. 
    If $\varPhi$ is reflection positive, then for any finite reflection-symmetric subset $\Lambda\subseteq \Gamma$, the local Hamiltonian $H_{\Lambda}$ is reflection positive with respect to $\Theta$.
\end{proposition}
\begin{proof}
    Let $\mathcal{C}(\Lambda)$ be the set of reflection-symmetric subsets $X\subseteq\Lambda$ that intersect both $\Gamma_-$ and $\Gamma_+$, and define the self-adjoint operator
    \begin{equation}
        \widetilde{H}_{\Lambda_+}=H_{\Lambda_+}+\sum_{X\in\mathcal{C}(\Lambda)}K_X.
    \end{equation}
    Since $\Lambda$ is reflection-symmetric, the first condition in Definition \ref{def:: reflection positive interaction} gives $H_{\Lambda_-}=\Theta(H_{\Lambda_+})$.
    The second condition then yields
    \begin{equation}
        H_{\Lambda}=\Theta(\widetilde{H}_{\Lambda_+})\otimes \mathrm{I}_{\Lambda_+}+\mathrm{I}_{\Lambda_-}\otimes \widetilde{H}_{\Lambda_+}-\sum_{X\in\mathcal{C}(\Lambda)}\sum_{j=1}^{m_X}\Theta(O_{X,j})\otimes O_{X,j}.
    \end{equation}
    Therefore, Theorem \ref{thm:: structure of RP Hamiltonian} implies that $H_{\Lambda}$ is reflection positive with respect to $\Theta$.
\end{proof}

For a reflection positive interaction $\varPhi$ and a finite reflection-symmetric subset $\Lambda\subseteq \Gamma$, we denote by $\widehat{\Pi}(\Lambda) \in \mathfrak{A}_{\Lambda_+}$ the entanglement support of $\Pi(\Lambda)$. 
In what follows, we deduce the structure theorem for the ground state subspace of a Hamiltonian with a reflection positive, frustration-free decomposition. 

\begin{lemma}\label{lemma:: commutation between H0 and entanglement support}
    Let $H = H_{-}\otimes \mathrm{I} + H_0 + \mathrm{I}\otimes H_{+}$ be a Hamiltonian on $\mathcal{K}_-\otimes \mathcal{K}_+$ satisfying $H_{-} = \Theta(H_{+})$, where $H_0$ is reflection positive. 
    Suppose that $H_{+}$, $H_{-}$, and $H_0$ are mutually frustration-free. 
    Let $\widehat{\Pi}$ be the entanglement support of $H$. 
    Then
    \begin{equation}
        [H_0,\mathrm{I}_{\mathcal{K}_-}\otimes \widehat{\Pi}]
        = [H_0,\Theta(\widehat{\Pi})\otimes\mathrm{I}_{\mathcal{K}_+}] = 0,
    \end{equation}
    and
    \begin{equation}
        [H_0,\Theta(\widehat{\Pi})\otimes\widehat{\Pi}]=0.
    \end{equation}
\end{lemma}
\begin{proof}
    Proposition \ref{prop: local commutant and maximal support} $(1)$ gives
    \begin{equation}
        [H,\mathrm{I}_{\mathcal{K}_-}\otimes\widehat{\Pi}]=0.
    \end{equation}
    Let $\Pi$ denote the ground state projection of $H$, and let $\Pi_+$ denote the ground state projection of $H_+$.
    Mutual frustration-freeness gives $\Pi\leq\mathrm{I}_{\mathcal{K}_-}\otimes\Pi_+$.
    Taking the partial trace over $\mathcal{K}_-$ gives $\widehat{\Pi}\leq\Pi_+$, and hence
    \begin{equation}
        H_+\widehat{\Pi}=E_0\widehat{\Pi},
    \end{equation}
    where $E_0$ is the ground state energy of $H_+$.
    Expanding $[H,\mathrm{I}_{\mathcal{K}_-}\otimes\widehat{\Pi}]=0$ now gives
    \begin{equation}
        [H_0,\mathrm{I}_{\mathcal{K}_-}\otimes\widehat{\Pi}]=0.
    \end{equation}
    Set $\mathcal{L}_0=\mathcal{O}H_0\mathcal{O}^{-1}$.
    Since $H_0$ is reflection positive, $e^{-\tau\mathcal{L}_0}$ is completely positive for every $\tau\geq0$.
    Differentiating at $\tau=0$ shows that $\mathcal{L}_0$ is adjoint-preserving.
    By Lemma \ref{lemma:: s-to-o map}, the preceding commutation relation is equivalent to
    $\mathcal{L}_0(\widehat{\Pi}Y^\dagger)=\widehat{\Pi}\mathcal{L}_0(Y^\dagger)$ for every $Y\in\mathcal{B}(\mathcal{K}_+)$.
    Taking adjoints gives $\mathcal{L}_0(Y\widehat{\Pi})=\mathcal{L}_0(Y)\widehat{\Pi}$.
    The second identity in Lemma \ref{lemma:: s-to-o map} now yields
    \begin{equation}
        [H_0,\Theta(\widehat{\Pi})\otimes\mathrm{I}_{\mathcal{K}_+}]=0.
    \end{equation}
    The two commutation relations imply the final result. 
\end{proof}

\begin{lemma}\label{lemma: compression by entanglement support projection}
    Let $H = H_{-}\otimes \mathrm{I} + H_0 + \mathrm{I}\otimes H_{+}$ be as in Lemma \ref{lemma:: commutation between H0 and entanglement support}. 
    Consider the projection $Q =\Theta(\widehat{\Pi})\otimes \widehat{\Pi}$. 
    Denote by $H_Q$ and $(H_0)_Q$ the restrictions of $H$ and $H_0$ to the range of $Q$, respectively. 
    Then
    \begin{equation}
        \mathrm{Sym}_{\widehat{\Pi}\mathcal{K}_+}(H_Q)=\mathrm{Sym}_{\widehat{\Pi}\mathcal{K}_+}((H_0)_Q).
    \end{equation}
\end{lemma}
\begin{proof}
    Let $\Pi_+$ be the ground state projection of $H_+$.
    By mutual frustration-freeness, we have $\widehat{\Pi}\leq\Pi_+$.
    Since $H_-=\Theta(H_+)$, we have $H_+\widehat{\Pi}=E_0\widehat{\Pi}$ and $H_-\Theta(\widehat{\Pi})=E_0\Theta(\widehat{\Pi})$, where $E_0$ is the ground state energy of $H_+$ and $H_-$.
    Therefore, on $Q(\mathcal{K}_-\otimes\mathcal{K}_+)$,
    \begin{equation}
        H_Q=2E_0\mathrm{I}_{Q(\mathcal{K}_-\otimes\mathcal{K}_+)}+(H_0)_Q.
    \end{equation}
    Since adding a scalar multiple of the identity does not change the local symmetry algebra, the claim holds.
\end{proof}

\begin{theorem}\label{theorem:: local symmetries of a RP frustration-free Hamiltonian}
    Let $H = H_{-}\otimes \mathrm{I} + H_0 + \mathrm{I}\otimes H_{+}$ be a Hamiltonian on $\mathcal{K}_-\otimes \mathcal{K}_+$ such that $H_{-} = \Theta(H_{+})$, where $H_0$ is reflection positive.
    Assume that $H_\pm$ and $H_0$ are mutually frustration-free.
    Let $\widehat{\Pi}$ denote the entanglement support of $H$.
    Then we have
    \begin{equation}
        \mathrm{Sym}_{\mathcal{K}_+}(H)\widehat{\Pi} = \widehat{\Pi}\mathrm{Sym}_{\mathcal{K}_+}(H_0)\widehat{\Pi}. 
    \end{equation}
\end{theorem}
\begin{proof}
    Set $Q:=\Theta(\widehat{\Pi})\otimes\widehat{\Pi}$, and denote by $H_Q$ and $(H_0)_Q$ the restrictions of $H$ and $H_0$ to $Q(\mathcal{K}_-\otimes\mathcal{K}_+)$, respectively.
    Proposition \ref{prop: local commutant and maximal support} and Lemma \ref{lemma: compression by entanglement support projection} give
    \begin{equation}
    \begin{aligned}
        \mathrm{Sym}_{\mathcal{K}_+}(H)\widehat{\Pi} =\mathrm{Sym}_{\widehat{\Pi}\mathcal{K}_+}(H_Q) =\mathrm{Sym}_{\widehat{\Pi}\mathcal{K}_+}((H_0)_Q).
    \end{aligned}
    \end{equation}
    So it remains to prove: 
    \begin{equation}
        \mathrm{Sym}_{\widehat{\Pi}\mathcal{K}_+}((H_0)_Q)=\widehat{\Pi}\mathrm{Sym}_{\mathcal{K}_+}(H_0)\widehat{\Pi}.
    \end{equation}

    Lemma \ref{lemma:: commutation between H0 and entanglement support} gives $[H_0,\mathrm{I}_{\mathcal{K}_-}\otimes\widehat{\Pi}]=[H_0,\Theta(\widehat{\Pi})\otimes\mathrm{I}_{\mathcal{K}_+}]=0$.
    For $\tau\geq0$, define $\Phi_\tau=\mathcal{O}e^{-\tau H_0}\mathcal{O}^{-1}$.
    This is a symmetric completely positive map.
    Let $\{K_i\}_i$ be a set of Kraus operators for $\Phi_\tau$.
    The commutation relations, Proposition \ref{prop:: local symmetry v.s. strong symmetry}, and Lemma \ref{lemma:: strong symmetry of a symmetric CP map} imply that $\widehat{\Pi}$ commutes with every $K_i$ and $K_i^\dagger$.
    Thus $\Phi_\tau$ restricts to a symmetric completely positive map $\Phi_{\tau,Q}$ on $\widehat{\Pi}\mathcal{B}(\mathcal{K}_+)\widehat{\Pi}$ with Kraus operators $K_i\widehat{\Pi}$.
    For the restricted unitary linear isomorphism $\mathcal{O}_Q:Q(\mathcal{K}_-\otimes\mathcal{K}_+)\longrightarrow\widehat{\Pi}\mathcal{B}(\mathcal{K}_+)\widehat{\Pi}$, we have
    \begin{equation}
        \Phi_{\tau,Q}=\mathcal{O}_Qe^{-\tau(H_0)_Q}\mathcal{O}_Q^{-1}.
    \end{equation}

    If $X\in\widehat{\Pi}\mathrm{Sym}_{\mathcal{K}_+}(H_0)\widehat{\Pi}$, then $X$ commutes with every $K_i\widehat{\Pi}$ and $K_i^\dagger\widehat{\Pi}$.
    Lemma \ref{lemma:: strong symmetry of a symmetric CP map} and Proposition \ref{prop:: local symmetry v.s. strong symmetry}, applied in the corner, give $X\in\mathrm{Sym}_{\widehat{\Pi}\mathcal{K}_+}((H_0)_Q)$.
    This proves one inclusion.

    Conversely, take $X\in\mathrm{Sym}_{\widehat{\Pi}\mathcal{K}_+}((H_0)_Q)$.
    Proposition \ref{prop:: local symmetry v.s. strong symmetry} and Lemma \ref{lemma:: strong symmetry of a symmetric CP map}, applied in the corner, give $[X,K_i\widehat{\Pi}]=[X,K_i^\dagger\widehat{\Pi}]=0$ for every $i$.
    Since $X=\widehat{\Pi}X\widehat{\Pi}$ and $\widehat{\Pi}$ commutes with $K_i$ and $K_i^\dagger$, we obtain 
    \begin{equation*}
        [X,K_i]=[X,K_i^\dagger]=0
    \end{equation*}
    for every $i$.
    Applying Lemma \ref{lemma:: strong symmetry of a symmetric CP map} and Proposition \ref{prop:: local symmetry v.s. strong symmetry} again yields $X\in\mathrm{Sym}_{\mathcal{K}_+}(e^{-\tau H_0})$ for $\tau\geq0$.
    Differentiating at $\tau=0$ gives $X\in\mathrm{Sym}_{\mathcal{K}_+}(H_0)$.
    Since $X=\widehat{\Pi}X\widehat{\Pi}$, this proves the reverse inclusion and completes the proof.
\end{proof}

\subsection{Local indistinguishability of ground states}
Consider a quantum spin system defined on a lattice $\Gamma\subset \Sigma$ with a frustration-free interaction $\varPhi$. 
Recall that a disk region $\mathbb{D}^n\subseteq \Sigma$ is said to satisfy the \emph{local topological quantum order} condition if, for any ground states $\ket{\phi}$ and $\ket{\psi}$ of the local Hamiltonian $\sum_{\Lambda\cap \mathbb{D}^n\neq \emptyset}\varPhi(\Lambda)$, we have 
\begin{equation}
    \Tr_{\overline{\mathbb{D}^n}}\ket{\psi}\bra{\psi} = \Tr_{\overline{\mathbb{D}^n}}\ket{\phi}\bra{\phi}, 
\end{equation}
where $\overline{\mathbb{D}^n}$ stands for the complement of $\mathbb{D}^n$. 
In other words, measurements in $\mathbb{D}^n$ cannot distinguish different ground states of the local Hamiltonian consisting of interactions overlapping with $\mathbb{D}^n$. 
Furthermore, the system is said to satisfy the LTQO condition if \emph{every} disk with diameter less than a prescribed radius satisfies this condition. 
The LTQO condition was first proposed in \cite{BravyiHastingsMichalakis2010}, where it was used to establish the stability of ground state degeneracy and the spectral gap of local commuting projector Hamiltonians under small but otherwise arbitrary local perturbations. 
This result was later generalized to frustration-free systems in \cite{MichalakisZwolak2013stability}. 
In this section, we give a criterion for the local indistinguishability of ground states using reflection positivity. 

To simplify the situation, we take the disk region to be a hemisphere. 
That is, we take $\Lambda$ to be a lattice on the sphere $\mathbb{S}^n = \mathbb{D}^n_{-}\cup \mathbb{D}^n_{+}$, which is symmetric with respect to the reflection $\theta$ through the equator. 
We assume that the interaction $\varPhi$ is reflection positive and frustration-free. 
Partitioning the interaction according to its support, the Hamiltonian on the sphere decomposes as  
\begin{equation}
    H_{\mathbb{S}^n} = H_{\mathbb{D}^n_{-}}\otimes \mathrm{I} + \mathrm{I}\otimes H_{\mathbb{D}^n_{+}} + H_0,
\end{equation}
where $H_0 = \sum_{\substack{\Lambda\cap \mathbb{D}^n_{-}\neq \emptyset, \Lambda\cap \mathbb{D}^n_{+}\neq \emptyset}}\varPhi(\Lambda)$
consists of interactions across the equator. 
The local Hamiltonians consisting of interactions within $\mathbb{D}^n_{+}$ or $\mathbb{D}^n_{-}$ and across the equator are given by $\mathrm{I}\otimes H_{\mathbb{D}^n_{+}} + H_0$ and $H_{\mathbb{D}^n_{-}}\otimes \mathrm{I} + H_0$, respectively. 

We would like to understand the structure of the ground state subspace of $\mathrm{I}\otimes H_{\mathbb{D}^n_{+}} + H_0$ using Theorem \ref{theorem:: ground state of H}. 
To do so, we use a dilation argument to restore reflection positivity. 
Define 
\begin{equation}
    \mathcal{K}_{+} = \mathcal{H}^1_{\mathbb{D}^n_{+}}\oplus\mathcal{H}^2_{\mathbb{D}^n_{+}} = \mathcal{H}_{\mathbb{D}^n_{+}}\otimes \mathbb{C}^2. 
\end{equation}
We represent operators on $\mathcal{K}_{+}$ by $\mathfrak{A}_{\mathbb{D}^n_{+}}$-valued $2\times 2$ matrices $(A_{ij})$, with the convention that $A_{ij}:\mathcal{H}^j_{\mathbb{D}^n_{+}}\rightarrow \mathcal{H}^i_{\mathbb{D}^n_{+}}$. 
Define $\mathcal{K}_{-}$ analogously. 
The anti-unitary $\widehat{\theta}$ naturally lifts to $\widehat{\theta}^{(2)}: \mathcal{K}_{+}\to \mathcal{K}_{-}$, mapping $\mathcal{H}^i_{\mathbb{D}^n_{+}}$ to $\mathcal{H}^i_{\mathbb{D}^n_{-}}$ for $i=1,2$. 
Let $\Theta^{(2)}:\mathcal{B}(\mathcal{K}_+)\to\mathcal{B}(\mathcal{K}_-)$ denote the induced anti-linear $*$-isomorphism. 
We now use the decomposition of $H_{\mathbb{S}^n}$ to define a new Hamiltonian on $\mathcal{K}_{-}\otimes \mathcal{K}_{+}$. 
First, define 
    \begin{equation}
       \mathbf{H}_{\pm} = \begin{bmatrix}
            H_{\mathbb{D}^n_{\pm}}-E_0 & 0\\
            0 & 0
        \end{bmatrix},
    \end{equation}
    where $E_0$ is the ground state energy of $H_{\mathbb{D}^n_{\pm}}$. 
    We make the following identification:
    \begin{equation}
    \begin{aligned}
        \mathcal{H}_{\mathbb{D}^n_{-}}\otimes \mathbb{C}^2\otimes \mathcal{H}_{\mathbb{D}^n_{+}}\otimes \mathbb{C}^2 &\longleftrightarrow \left(\mathcal{H}_{\mathbb{D}^n_{-}}\otimes \mathcal{H}_{\mathbb{D}^n_{+}}\right) \otimes M_2(\mathbb{C})\\
        \big( \widehat{\theta}\ket{\eta}\otimes\ket{b} \big) \otimes \big( \ket{\xi} \otimes \ket{a} \big)&\longleftrightarrow \big( \widehat{\theta}\ket{\eta}\otimes \ket{\xi} \big)\otimes \big( \ket{a}\bra{b} \big).
    \end{aligned}
    \end{equation}
    Under this identification, the state--operator correspondence for $\mathcal{K}_{-}\otimes\mathcal{K}_{+}$ is given by $\mathcal{O}\otimes\mathrm{id}_2$, where $\mathrm{id}_2$ denotes the identity linear map on $M_2(\mathbb{C})$.
    Set $\mathcal{L}_0=\mathcal{O}H_0\mathcal{O}^{-1}$.
    We define $\mathbf{H}_0$ by
    \begin{equation}
        (\mathcal{O}\otimes\mathrm{id}_2)\mathbf{H}_0(\mathcal{O}^{-1}\otimes\mathrm{id}_2)=\mathcal{L}_0\otimes\mathrm{id}_2.
    \end{equation}
    Equivalently, for every $\ket{\zeta}\in\mathcal{H}_{\mathbb{D}^n_{-}}\otimes\mathcal{H}_{\mathbb{D}^n_{+}}$ and every matrix unit $E_{ij}\in M_2(\mathbb{C})$, we have
    \begin{equation}
        \mathbf{H}_0(\ket{\zeta}\otimes E_{ij})=H_0\ket{\zeta}\otimes E_{ij}.
    \end{equation}
    Thus, $\mathbf{H}_0$ acts as $H_0$ on each of the four subspaces and is self-adjoint.
    Since $H_0$ is reflection positive, Theorem \ref{thm:: charactorization of RP based on s-to-o map} implies that $e^{-\tau\mathcal{L}_0}$ is completely positive for every $\tau\geq 0$.
    Hence $e^{-\tau(\mathcal{L}_0\otimes\mathrm{id}_2)}=e^{-\tau\mathcal{L}_0}\otimes\mathrm{id}_2$ is completely positive, and therefore $\mathbf{H}_0$ is reflection positive with respect to $\Theta^{(2)}$.
    We define the dilated Hamiltonian as $\mathbf{H} = \mathbf{H}_-\otimes \mathrm{I}_{\mathcal{K}_+} + \mathbf{H}_0 + \mathrm{I}_{\mathcal{K}_-}\otimes \mathbf{H}_+$.
    Moreover, $\mathbf{H}_-=\Theta^{(2)}(\mathbf{H}_+)$, so the Lie--Trotter product formula shows that $\mathbf{H}$ is reflection positive with respect to $\Theta^{(2)}$. 
    Define $P_{ij}(X) = E_{ii}XE_{jj}$ for $X\in M_2(\mathbb{C})$ and $i,j=1,2$.
    Then it is verified that $\mathbf{H}$ decomposes as 
    \begin{equation}\label{eqn:: dilated Hamiltonian decomposition}
        \begin{aligned}
            \mathbf{H} &= \big[ (H_{\mathbb{D}^n_{-}}-E_0)\otimes \mathrm{I} + H_0 + \mathrm{I}\otimes (H_{\mathbb{D}^n_{+}}-E_0)\big]\otimes P_{11} + \big[H_0 + \mathrm{I}\otimes (H_{\mathbb{D}^n_{+}}-E_0)\big]\otimes P_{12} \\
            &+ \big[ (H_{\mathbb{D}^n_{-}}-E_0)\otimes \mathrm{I} + H_0 \big]\otimes P_{21} + H_0 \otimes P_{22}. 
        \end{aligned}
    \end{equation}
\begin{lemma}\label{lemma:: ground states of the dilated Hamiltonian}
    The ground state subspace of $\mathbf{H}$ is spanned by the ground states of the four Hamiltonians in the decomposition above. 
    Moreover, let $\ket{\phi_{\text{PF}}}$ and $\ket{\phi^0_{\text{PF}}}$ be the canonical Perron--Frobenius ground states of $H_{\mathbb{S}^n}$ and $H_0$, respectively, both viewed as Hamiltonians on $\mathcal{H}_{\mathbb{D}^n_{-}}\otimes \mathcal{H}_{\mathbb{D}^n_{+}}$. 
    Then 
    \begin{equation}
        \begin{bmatrix}
        \ket{\phi_{\text{PF}}} & 0\\
        0 & \ket{\phi^0_{\text{PF}}}
    \end{bmatrix}
    \end{equation}
    is the canonical Perron--Frobenius ground state of $\mathbf{H}$. 
\end{lemma}
\begin{proof}
    We show that the four Hamiltonians in the decomposition above have the same ground state energy. 
    Let $\lambda_0$ be the ground state energy of $H_0$. 
    Then, by frustration-freeness, the ground state energy of $\mathrm{I}\otimes (H_{\mathbb{D}^n_{+}}-E_0) + H_0$ is also $\lambda_0$, because each ground state must minimize the energy of both $\mathrm{I}\otimes (H_{\mathbb{D}^n_{+}}-E_0)$ and $H_0$. 
    The same argument shows that the ground state energies of $(H_{\mathbb{D}^n_{-}}-E_0)\otimes \mathrm{I} + H_0$ and $(H_{\mathbb{D}^n_{-}}-E_0)\otimes \mathrm{I} + H_0 + \mathrm{I}\otimes (H_{\mathbb{D}^n_{+}}-E_0)$ are also $\lambda_0$. 
    Therefore, the ground state subspace of $\mathbf{H}$ is spanned by the ground states of the four Hamiltonians in Equation \eqref{eqn:: dilated Hamiltonian decomposition}. 
    To determine the canonical Perron--Frobenius ground state of $\mathbf{H}$, we identify $\left(\mathcal{H}_{\mathbb{D}^n_{-}}\otimes \mathcal{H}_{\mathbb{D}^n_{+}}\right) \otimes M_2(\mathbb{C})$ with $\mathfrak{A}_{\mathbb{D}^n_{+}}\otimes M_2(\mathbb{C})$ via $\mathcal{O}\otimes \mathrm{id}_2$. 
    Under this identification, the canonical Perron--Frobenius ground state of $\mathbf{H}$ is given by
    \begin{equation}
        \mathbf{\Xi} = \lim_{\tau\rightarrow +\infty} e^{\tau \lambda_0}\exp\left[ -\tau (\mathcal{O}\otimes \mathrm{id}_2)\mathbf{H} (\mathcal{O}^{-1}\otimes \mathrm{id}_2) \right]\begin{bmatrix}
            \mathrm{I}_{\mathbb{D}^n_{+}} & 0\\
            0 & \mathrm{I}_{\mathbb{D}^n_{+}}
        \end{bmatrix} = \begin{bmatrix}
        \Xi & 0\\
        0 & \Xi_0
        \end{bmatrix},
    \end{equation}
    where $\Xi = \mathcal{O}\ket{\phi_{\text{PF}}}$ and $\Xi_0 = \mathcal{O}\ket{\phi^{\text{PF}}_0}$ are the canonical Perron--Frobenius eigenvectors of $e^{-\tau \mathcal{O}H_{\mathbb{S}^n}\mathcal{O}^{-1}}$ and $e^{-\tau \mathcal{O}H_0\mathcal{O}^{-1}}$, respectively. 
    The claim follows after applying $\mathcal{O}^{-1}\otimes \mathrm{id}_2$. 
\end{proof}
\noindent In particular, writing $\widehat{\Pi}$ for the entanglement support of $H_{\mathbb{S}^n}$ and $\widehat{\Pi}_0$ for that of $H_0$, the entanglement support of $\mathbf{H}$ is given by $\widehat{\mathbf{\Pi}} = \begin{bmatrix}
        \widehat{\Pi} & 0\\
        0 & \widehat{\Pi}_0
    \end{bmatrix}$. 
\begin{proposition}\label{prop:: ground states of partial sums of a RP Hamiltonian} 
    Let $\varPhi$ be a reflection positive, frustration-free interaction on $\mathbb{S}^n = \mathbb{D}^n_{-}\cup \mathbb{D}^n_{+}$. 
    Define $H_0 = H_{\mathbb{S}^n} - H_{\mathbb{D}^n_{-}} - H_{\mathbb{D}^n_{+}}$, and let $\ket{\phi^{\text{PF}}_0}$ be the canonical Perron--Frobenius ground state of $H_0$ viewed as a Hamiltonian on $\mathcal{H}_{\mathbb{D}^n_{-}}\otimes \mathcal{H}_{\mathbb{D}^n_{+}}$. 
    Set $A=\mathrm{Sym}_{\mathbb{D}^n_+}(H_0)$, $p=\widehat{\Pi}(\mathbb{S}^n)$, and $p_0=\widehat{\Pi}_0$. 
    Then the following statements hold:
    \begin{enumerate}
        \item For any ground state $\ket{\varphi}$ of $H_0 + \mathrm{I}\otimes H_{\mathbb{D}^n_{+}}$, there exists a unique $V\in pAp_0$ such that $\ket{\varphi} = (\mathrm{I}\otimes V)\ket{\phi^{\text{PF}}_0}$;
        \item For any ground state $\ket{\varphi}$ of $H_{\mathbb{D}^n_{-}}\otimes \mathrm{I} + H_0$, there exists a unique $W\in p_0Ap$ such that $\ket{\varphi} = (\Theta(W^\dagger)\otimes \mathrm{I})\ket{\phi^{\text{PF}}_0}$. 
    \end{enumerate}
    Conversely, for any $V\in pAp_0$, $(\mathrm{I}\otimes V)\ket{\phi^{\text{PF}}_0}$ is a ground state of $H_0 + \mathrm{I}\otimes H_{\mathbb{D}^n_{+}}$. 
\end{proposition}
\begin{proof}    
    By Lemma \ref{lemma:: ground states of the dilated Hamiltonian}, the ground state subspace of $\mathbf{H}$ is given by the direct sum of the ground state subspaces of the four Hamiltonians:
    \begin{equation}
        H_{\mathbb{S}^n},\quad H_{\mathbb{D}^n_{-}}\otimes \mathrm{I} + H_0,\quad H_0 + \mathrm{I}\otimes H_{\mathbb{D}^n_{+}},\quad H_0. 
    \end{equation}
    By Theorem \ref{thm:: structure of maximum eigenspace of a symmetric CP map}, the Perron--Frobenius eigenspace of $(\mathcal{O}\otimes\mathrm{id}_2)e^{-\tau \mathbf{H}}(\mathcal{O}^{-1}\otimes\mathrm{id}_2)$ is $\mathrm{Sym}_{\mathcal{K}_{+}}(\mathbf{H})\mathbf{\Xi}$. 
    Since there is no frustration between $\mathbf{H}_{\pm}$ and $\mathbf{H}_0$, Theorem \ref{theorem:: local symmetries of a RP frustration-free Hamiltonian} applies to $\mathbf{H}$, which yields
    \begin{equation}
        \mathrm{Sym}_{\mathcal{K}_{+}}(\mathbf{H})\widehat{\mathbf{\Pi}} = \widehat{\mathbf{\Pi}}\mathrm{Sym}_{\mathcal{K}_{+}}(\mathbf{H}_0)\widehat{\mathbf{\Pi}} = \widehat{\mathbf{\Pi}}\left(\mathrm{Sym}_{\mathbb{D}^n_{+}}(H_0)\otimes M_2(\mathbb{C}) \right)\widehat{\mathbf{\Pi}}. 
    \end{equation}
    Therefore, we obtain 
    \begin{equation}
    \begin{aligned}
        \mathrm{Sym}_{\mathcal{K}_{+}}(\mathbf{H})\widehat{\mathbf{\Pi}} &= \widehat{\mathbf{\Pi}}\left(\mathrm{Sym}_{\mathbb{D}^n_{+}}(H_0)\otimes M_2(\mathbb{C}) \right)\widehat{\mathbf{\Pi}} = \begin{bmatrix}
            pAp & pAp_0\\
            p_0Ap & p_0Ap_0
        \end{bmatrix}.
    \end{aligned}
    \end{equation}
    The claim thus follows by multiplying by $\mathbf{\Xi}$. 
    For the converse, let $V\in pAp_0$. 
    Then 
    \begin{equation}
        \begin{aligned}
            (H_0 + \mathrm{I}\otimes H_{\mathbb{D}^n_{+}})(\mathrm{I}\otimes V)\ket{\phi^{\text{PF}}_0} = (\mathrm{I}\otimes V)H_0\ket{\phi^{\text{PF}}_0} + E_0(\mathrm{I}\otimes V)\ket{\phi^{\text{PF}}_0} = (\lambda_0+E_0)(\mathrm{I}\otimes V)\ket{\phi^{\text{PF}}_0},
        \end{aligned}
    \end{equation}
    where $\lambda_0$ is the ground state energy of $H_0$ and $E_0$ is the ground state energy of $H_{\mathbb{D}^n_{+}}$. 
    In the second equality, we used the facts that $[\mathrm{I}\otimes V,H_0] = 0$ and that the range of $V$ is contained in $\widehat{\Pi}$, which is a subprojection of the ground state space of $H_{\mathbb{D}^n_{+}}$. 
\end{proof}
\begin{theorem}\label{thm:: local-indistinguishability and ground state degeneracy}
    Consider a reflection positive, frustration-free interaction $\varPhi$ on $\mathbb{S}^n = \mathbb{D}^n_{-}\cup \mathbb{D}^n_{+}$. 
    Then the following statements are equivalent: 
    \begin{enumerate}
        \item The ground state of $H_{\mathbb{S}^n}$ is nondegenerate; 
        \item The reduced density matrix $\Tr_{\mathbb{D}^n_{-}}\ket{\varphi}\bra{\varphi}$ is independent of the choice of normalized ground state $\ket{\varphi}$ of $H_0 + \mathrm{I}\otimes H_{\mathbb{D}^n_{+}}$. 
    \end{enumerate}
\end{theorem}
\begin{proof}
    Suppose that the ground state of $H_{\mathbb{S}^n}$ is nondegenerate. 
    Let $\ket{\varphi} = (\mathrm{I}\otimes V)\ket{\phi^{\text{PF}}_0}$ be any ground state of $H_0 + \mathrm{I}\otimes H_{\mathbb{D}^n_+}$ as in Proposition \ref{prop:: ground states of partial sums of a RP Hamiltonian}. 
    Retain the notation $A$, $p$, and $p_0$ from that proposition.
    Then, by Proposition \ref{prop:: reduced density matrix in terms of O}, we have 
    \begin{equation}
        \Tr_{\mathbb{D}^n_{-}}\ket{\varphi}\bra{\varphi} = V\Xi^2_0 V^\dagger = VV^\dagger \Xi^2_0.
    \end{equation}
    Since $V\in pAp_0$, Theorem \ref{theorem:: local symmetries of a RP frustration-free Hamiltonian} gives
    \begin{equation}
        VV^\dagger\in p\mathrm{Sym}_{\mathbb{D}^n_{+}}(H_0)p = \mathrm{Sym}_{\mathbb{D}^n_{+}}(H_{\mathbb{S}^n})p, 
    \end{equation}
    which is one-dimensional since the ground state of $H_{\mathbb{S}^n}$ is nondegenerate. 
    Therefore, $VV^\dagger$ is a scalar multiple of $\widehat{\Pi}$, with normalization fixing the scalar to $1/\Tr(p\Xi_0^2)$. 
    Thus we have 
    \begin{equation}
        \Tr_{\mathbb{D}^n_{-}}\ket{\varphi}\bra{\varphi} = \frac{p\Xi^2_0}{\Tr(p\Xi^2_0)}, 
    \end{equation}
    independent of the choice of $\ket{\varphi}$. 

    Conversely, suppose $\Tr_{\mathbb{D}^n_{-}}\ket{\varphi}\bra{\varphi}$ is independent of the choice of a normalized ground state $\ket{\varphi}$ of $H_0 + \mathrm{I}\otimes H_{\mathbb{D}^n_{+}}$. 
    Since $p\in A$ and $p\leq p_0$, we have $p\in pAp_0$. 
    For nonzero $V\in pAp_0$, Proposition \ref{prop:: ground states of partial sums of a RP Hamiltonian} shows that both $(\mathrm{I}\otimes V)\ket{\phi^{\mathrm{PF}}_0}$ and $(\mathrm{I}\otimes p)\ket{\phi^{\mathrm{PF}}_0}$ are nonzero ground states. 
    After normalization, local indistinguishability gives
    \begin{equation}
        \frac{VV^\dagger\Xi_0^2}{\Tr(VV^\dagger\Xi_0^2)} = \frac{p\Xi_0^2}{\Tr(p\Xi_0^2)}.
    \end{equation}
    Since $V$ and $p$ belong to $A$ and hence commute with $\Xi_0$, and since $\Xi_0$ is invertible on $p_0$, this implies that $VV^\dagger$ is a scalar multiple of $p$. 
    Now take $X=X^\dagger\in pAp$; then both $X$ and $p+X$ belong to $pAp_0$.
    By the preceding conclusion, there exist $\alpha,\beta\geq0$ such that
    \begin{equation}
        X^2=\alpha p,\quad (p+X)^2=\beta p.
    \end{equation}
    Since $p$ is the identity of $pAp$, subtraction gives $2X=(\beta-\alpha-1)p$.
    Thus, every self-adjoint element of $pAp$ is a scalar multiple of $p$, and we conclude that $pAp=\mathbb{C}p$. 
    By Theorem \ref{theorem:: local symmetries of a RP frustration-free Hamiltonian}, this implies that $\mathrm{Sym}_{\mathbb{D}^n_{+}}(H_{\mathbb{S}^n})\widehat{\Pi}(\mathbb{S}^n) = \widehat{\Pi}(\mathbb{S}^n)\mathrm{Sym}_{\mathbb{D}^n_{+}}(H_0)\widehat{\Pi}(\mathbb{S}^n)$ is one-dimensional, and hence the ground state of $H_{\mathbb{S}^n}$ is nondegenerate. 
\end{proof}

\section{Boundary Algebra from Osterwalder-Schrader reconstruction}\label{sec:: OS reconstruction and local operator algebras}
\begin{figure}[htb]
    \centering
    \begingroup%
  \makeatletter%
  \providecommand\color[2][]{%
    \errmessage{(Inkscape) Color is used for the text in Inkscape, but the package 'color.sty' is not loaded}%
    \renewcommand\color[2][]{}%
  }%
  \providecommand\transparent[1]{%
    \errmessage{(Inkscape) Transparency is used (non-zero) for the text in Inkscape, but the package 'transparent.sty' is not loaded}%
    \renewcommand\transparent[1]{}%
  }%
  \providecommand\rotatebox[2]{#2}%
  \newcommand*\fsize{\dimexpr\f@size pt\relax}%
  \newcommand*\lineheight[1]{\fontsize{\fsize}{#1\fsize}\selectfont}%
  \ifx\svgwidth\undefined%
    \setlength{\unitlength}{175.77810333bp}%
    \ifx\svgscale\undefined%
      \relax%
    \else%
      \setlength{\unitlength}{\unitlength * \real{\svgscale}}%
    \fi%
  \else%
    \setlength{\unitlength}{\svgwidth}%
  \fi%
  \global\let\svgwidth\undefined%
  \global\let\svgscale\undefined%
  \makeatother%
  \begin{picture}(1,0.69360297)%
    \lineheight{1}%
    \setlength\tabcolsep{0pt}%
    \put(0.21470428,0.64106868){\makebox(0,0)[lt]{\lineheight{1.25}\smash{\begin{tabular}[t]{l}$\tau$\end{tabular}}}}%
    \put(0,0){\includegraphics[width=\unitlength,page=1]{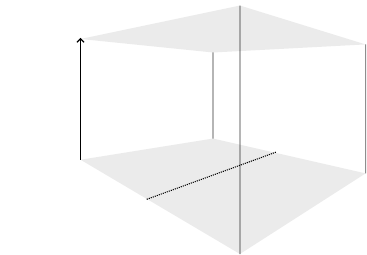}}%
    \put(0.48305456,0.30594358){\makebox(0,0)[lt]{\lineheight{1.25}\smash{\begin{tabular}[t]{l}$\Theta(A)$\end{tabular}}}}%
    \put(0.71172352,0.20485466){\makebox(0,0)[lt]{\lineheight{1.25}\smash{\begin{tabular}[t]{l}$A$\end{tabular}}}}%
    \put(0,0){\includegraphics[width=\unitlength,page=2]{3D_spatial_RP.pdf}}%
    \put(-0.00516679,0.23713871){\makebox(0,0)[lt]{\lineheight{1.25}\smash{\begin{tabular}[t]{l}$\tau = 0$\end{tabular}}}}%
    \put(0,0){\includegraphics[width=\unitlength,page=3]{3D_spatial_RP.pdf}}%
    \put(0.35126185,0.14028707){\makebox(0,0)[lt]{\lineheight{1.25}\smash{\begin{tabular}[t]{l}$\theta$\end{tabular}}}}%
    \put(0,0){\includegraphics[width=\unitlength,page=4]{3D_spatial_RP.pdf}}%
  \end{picture}%
\endgroup%
    \caption{Spatial reflection positivity with periodic boundary conditions in $\tau$.}
    \label{fig:spatial_RP}
\end{figure}

In this section, we turn to the construction of the holographic dual of the bulk ground state. 
We use OS reconstruction to construct the local net of symmetric operator algebras from local ground states of a reflection positive, frustration-free interaction. 
The intuition behind this construction comes from the path integral picture. 
A quantum spin system defined on an $n$-dimensional spatial lattice can be considered as an $(n+1)$-dimensional Euclidean field theory via imaginary-time evolution. 
When the Hamiltonian has reflection positivity, the Euclidean theory has an additional positivity property: the correlation functions are positive with respect to spatial reflection about the chosen hyperplane (see Fig.~\ref{fig:spatial_RP}). 
The OS reconstruction applied to such a system can be thought of as performing a path integral in one spatial direction. 
The resulting Hilbert space then consists of the field configurations at the boundary determined by the spatial reflection. 

\subsection{Osterwalder-Schrader reconstruction from a RP projection}\label{sec:: physical Hilbert space and field algebra}
First, we analyze the case in which OS reconstruction is applied to a reflection positive projection. 
We mainly rely on the results of Section \ref{sec:: ground states of a reflection positive Hamiltonian} and adopt its notation. 
Specifically, let $\Pi$ be a reflection positive projection on $\mathcal{H} = \mathcal{H}_-\otimes \mathcal{H}_+$, viewed as the ground state projection of a reflection positive Hamiltonian. 
Denote by $\widehat{\Pi}$ the range projection of $\Tr_{\mathcal{H}_{-}}\Pi$. 
We then have $\Pi\mathcal{H} = \mathrm{Sym}_{+}( \Pi)\widehat{\Pi}\ket{\phi_{\text{PF}}}$, where $\ket{\phi_{\text{PF}}}$ is the canonical PF state. 
The reflection positivity of $\Pi$ implies that the following sesquilinear form on $\mathfrak{A}_+$ is positive semidefinite: 
\begin{equation}
    \langle Y,X \rangle_0 = \frac{1}{\Tr (\Pi)}\Tr (\Pi \Theta(Y)\otimes X). 
\end{equation}
\begin{definition}\label{def:: physical Hilbert space}
    Denote by $\ker \langle \cdot,\cdot\rangle_0$ the kernel of the sesquilinear form $\langle \cdot,\cdot\rangle_0$. 
    The quotient
    \begin{equation}
        \mathfrak{H} = \mathfrak{A}_+/\ker \langle \cdot,\cdot\rangle_0
    \end{equation}
    equipped with the induced inner product, is a finite-dimensional Hilbert space. 
    For $X\in \mathfrak{A}_+$, we denote by $\psi(X) = X + \ker \langle \cdot,\cdot\rangle_0$ the image of $X$ under the quotient map. 
\end{definition}
Define the linear map $\mathscr{E}: \mathfrak{A}_{+}\rightarrow \mathfrak{A}_{+}$ by
\begin{equation}\label{eq:: the CP idempotent map E}
    \mathscr{E}(X) = \frac{1}{\Tr (\Pi)} \Tr_{\mathcal{H}_{-}}(\Pi\Theta(X^\dagger)\otimes \mathrm{I}),\quad X\in \mathfrak{A}_{+}.
\end{equation}
In terms of $\mathscr{E}$, the sesquilinear form $\langle \cdot,\cdot\rangle_0$ can be expressed as $\langle Y,X\rangle_0 = \Tr (\mathscr{E}(Y^\dagger)X)$. 
\begin{lemma}
    Let $\{\ket{ \phi_i}\}_i$ be an orthonormal basis of $\Pi\mathcal{H}$. 
    Then 
    \begin{equation}
        \mathscr{E}(X) = \frac{1}{\Tr (\Pi)}\sum_i \mathcal{O}(\phi_i) X\mathcal{O}(\phi_i)^\dagger,\quad X\in \mathfrak{A}_{+}.
    \end{equation}
    In particular, $\mathscr{E}$ is completely positive and symmetric, that is, self-adjoint with respect to the Hilbert--Schmidt inner product. 
\end{lemma}
\begin{proof}
     By Lemma \ref{lemma:: s-to-o map}, for any $X,Y\in \mathfrak{A}_{+}$, we have 
    \begin{equation}
    \begin{aligned}
        \Tr (Y\mathscr{E}(X)) &= \frac{1}{\Tr (\Pi)}\Tr (\Pi\Theta(X^\dagger)\otimes Y)= \frac{1}{\Tr (\Pi)}\sum_i \bra{ \phi_i} \Theta(X^\dagger)\otimes Y\ket{ \phi_i}\\
        &= \frac{1}{\Tr (\Pi)} \sum_i \Tr(\mathcal{O}(\phi_i)^\dagger Y\mathcal{O}(\phi_i) X) = \frac{1}{\Tr (\Pi)}\sum_i\Tr (Y \mathcal{O}(\phi_i) X\mathcal{O}(\phi_i)^\dagger). 
    \end{aligned}
    \end{equation}
    This proves that $\mathscr{E}$ is completely positive. 
    Set $F=\mathcal{O}\Pi\mathcal{O}^{-1}$. 
    Since $F$ is completely positive, it is adjoint-preserving, and hence its image is closed under adjoints. 
    Since adjunction preserves the Hilbert--Schmidt inner product, $\{\mathcal{O}(\phi_i)^\dagger\}_i$ is an orthonormal basis of the image of $F$. 
    Equivalently, $\{\mathcal{O}^{-1}(\mathcal{O}(\phi_i)^\dagger)\}_i$ is an orthonormal basis of $\Pi\mathcal{H}$. 
    Since the expression for $\mathscr{E}$ is independent of the choice of orthonormal basis of $\Pi\mathcal{H}$, we have $\displaystyle \mathscr{E}(X) = \frac{1}{\Tr (\Pi)}\sum_i \mathcal{O}(\phi_i)^\dagger X \mathcal{O}(\phi_i)$; thus, $\mathscr{E}$ is symmetric. 
\end{proof}
\begin{remark}
    The complete positivity of $\mathscr{E}$ also follows from the fact that, for any $\{X_i\}^n_{i=1}$ and $\{Y_i\}^n_{i=1}$ in $\mathfrak{A}_{+}$, 
    \begin{equation}
    \begin{aligned}
        \sum^n_{i,j=1}\Tr(Y^\dagger_j\mathscr{E}(X^\dagger_jX_i)Y_i) &= \frac{1}{\Tr (\Pi)}\sum^n_{i,j=1}\Tr (\Pi\Theta(X^\dagger_iX_j)\otimes Y_iY^\dagger_j)\\
        &= \frac{1}{\Tr (\Pi)}\Tr \left(\Pi \left\vert \sum^n_{j=1} \Theta(X_j)\otimes Y^\dagger_j \right\vert^2 \right) \geq 0. 
    \end{aligned}
    \end{equation}
\end{remark}
\begin{proposition}\label{prop:: image of E}
    For any $X\in \mathfrak{A}_{+}$, we have $ \mathscr{E}(X)\in \mathrm{Sym}_{+}(\Pi)'\widehat{\Pi}$, where $\mathrm{Sym}_{+}(\Pi)'$ denotes the commutant in $\mathfrak{A}_+$. 
\end{proposition}
\begin{proof}
    Let $X\in\mathfrak{A}_{+}$. 
    By Theorem \ref{theorem:: ground state of H}, for any $W\in \mathrm{Sym}_{+}(\Pi)$ and $\ket{\phi}\in \Pi\mathcal{H}$, we have $\mathrm{I}\otimes W\ket{\phi}\in \Pi\mathcal{H}$. 
    Since $\{\ket{ \phi_i} \}_i$ is an orthonormal basis of $\Pi\mathcal{H}$, 
    \begin{equation}
        \mathrm{I}\otimes W\ket{ \phi_i} = \sum_j \ket{\phi_j} \bra{ \phi_j} \mathrm{I}\otimes W \ket{ \phi_i}. 
    \end{equation}
    Applying $\mathcal{O}$ to both sides, one obtains
    \begin{equation}
        W\mathcal{O}(\phi_i) = \mathcal{O}\left(\mathrm{I}\otimes W \ket{ \phi_i} \right) = \sum_j \Tr \left(\mathcal{O}(\phi_j)^\dagger W\mathcal{O}(\phi_i) \right) \mathcal{O}(\phi_j).
    \end{equation}
    Taking adjoints on both sides with the indices $i$ and $j$ interchanged, one also obtains
    \begin{equation}
        \mathcal{O}(\phi_j)^\dagger W= \sum_i \mathcal{O}(\phi_i)^\dagger \Tr \left(\mathcal{O}(\phi_j)^\dagger W \mathcal{O}(\phi_i) \right). 
    \end{equation}
    Therefore, we have 
    \begin{equation}
    \begin{aligned}
        W\sum_i \mathcal{O}(\phi_i) X\mathcal{O}(\phi_i)^\dagger &= \sum_i \sum_j\mathcal{O}(\phi_j) X \Tr \left( \mathcal{O}(\phi_j)^\dagger W \mathcal{O}(\phi_i)\right) \mathcal{O}(\phi_i)^\dagger\\
        &= \sum_j \sum_i\mathcal{O}(\phi_j) X \Tr \left( \mathcal{O}(\phi_j)^\dagger W \mathcal{O}(\phi_i)\right) \mathcal{O}(\phi_i)^\dagger\\
        &= \sum_j \mathcal{O}(\phi_j) X\mathcal{O}(\phi_j)^\dagger W.
    \end{aligned}
    \end{equation}
    This proves that 
    \begin{equation}
        \mathscr{E}(X) = \frac{1}{\Tr (\Pi)}\sum_i \mathcal{O}(\phi_i) X\mathcal{O}(\phi_i)^\dagger \in \mathrm{Sym}_{+}(\Pi)'\cap \mathfrak{A}_+. 
    \end{equation}
    By Theorem \ref{theorem:: ground state of H}, there are $W_i\in \mathrm{Sym}_{+}(\Pi)\widehat{\Pi}$ such that $\mathcal{O}(\phi_i) = \Xi W_i$. 
    Observe that $\widehat{\Pi}\Xi = \Xi \widehat{\Pi} = \Xi$, and $\widehat{\Pi}$ commutes with $W_i$ for all $i$.
    Therefore, we have  
    \begin{equation}
        \sum_i \mathcal{O}(\phi_i) X\mathcal{O}(\phi_i)^\dagger \widehat{\Pi} =  \sum_i \mathcal{O}(\phi_i) X\mathcal{O}(\phi_i)^\dagger  = \widehat{\Pi}\sum_i \mathcal{O}(\phi_i) X\mathcal{O}(\phi_i)^\dagger . 
    \end{equation}
    That is, $ \mathscr{E}(X) \in \mathrm{Sym}_{+}(\Pi)'\widehat{\Pi}$. 
\end{proof}
\begin{proposition}\label{prop:: kernel of the sesquilinear form}
    For any $Y\in \mathfrak{A}_+$, we have $\psi(Y) = 0$ if and only if $\Tr (Y^\dagger X) = 0$ for all $X\in \mathrm{Sym}_{+}(\Pi)' \widehat{\Pi}$. 
\end{proposition}
\begin{proof}
    We have $\psi(Y) = 0$ if and only if $\Tr (Y^\dagger \mathscr{E}(X)) = 0$ for all $X\in \mathfrak{A}_{+}$. 
    By Proposition \ref{prop:: image of E}, such operators span a subspace of $\mathrm{Sym}_{+}(\Pi)'\widehat{\Pi}$. 
    Therefore, it remains to show that operators of this form in fact span $\mathrm{Sym}_{+}(\Pi)'\widehat{\Pi}$. 

    To see this, note that, by Theorem \ref{theorem:: ground state of H}, there are unique $W_i\in \mathrm{Sym}_{+}(\Pi)\widehat{\Pi}$ such that $\mathcal{O}(\phi_i) = \Xi W_i$. 
    We have 
    \begin{equation}
        F = \sum_i \mathcal{O}(\phi_i) \Xi^{-2}\mathcal{O}(\phi_i)^\dagger = \sum_i W_iW^\dagger_i \in \mathrm{Sym}_{+}(\Pi)\widehat{\Pi}. 
    \end{equation}
    By the previous proposition, $F$ is a positive operator in $\mathcal{Z}(\mathrm{Sym}_{+}(\Pi)\widehat{\Pi})$. 
    We now show that $F$ has full support. 
    Suppose $K$ is a positive operator in $\mathcal{Z}(\mathrm{Sym}_{+}(\Pi)\widehat{\Pi})$ such that $\Tr (\Xi FK\Xi) = 0$. 
    This means 
    \begin{equation}
        0 = \sum_i \Tr (\Xi W_iKW^\dagger_i\Xi) = \sum_i \bra{ \phi_i} K\ket{ \phi_i} = \Tr ((\mathrm{I}_{\mathcal{H}_{-}}\otimes K)\Pi) = \Tr (K\Tr_{\mathcal{H}_-}(\Pi)). 
    \end{equation}
    By definition, the range projection of $\Tr_{\mathcal{H}_-}(\Pi)$ is $\widehat{\Pi}$, which dominates that of $K$. 
    Therefore, we have $K=0$, so $F$ is invertible in $\mathcal{Z}(\mathrm{Sym}_{+}(\Pi)\widehat{\Pi})$. 
    Let $A\in \mathrm{Sym}_{+}(\Pi)'\widehat{\Pi}$ be arbitrary.
    Then
    \begin{equation}
        \sum_i \mathcal{O}(\phi_i) (F^{-1}\Xi^{-1}A \Xi^{-1})\mathcal{O}(\phi_i)^\dagger = FF^{-1}A = A.
    \end{equation}
    This shows that operators of the form $\sum_i \mathcal{O}(\phi_i) X\mathcal{O}(\phi_i)^\dagger $ span $\mathrm{Sym}_{+}(\Pi)'\widehat{\Pi}$, and the claim follows. 
\end{proof}
\begin{remark}
    An RP projection is akin to a biprojection in the theory of subfactors \cite{bisch1994biprojection}, and the CP map $\mathscr{E}$ is analogous to the Fourier transform of a biprojection. 
    Note that for irreducible subfactors, a projection with positive Fourier transform is automatically a biprojection \cite{Liu2016ExchangeRP}. 
    The complexity here arises from reducible inclusions. 
\end{remark}
Osterwalder and Schrader use the algebra of observables in Euclidean space to define the physical observables on $\mathscr{H}$. 
Accordingly, we consider the following definition. 
\begin{definition}\label{def:: field operators and field algebra}
    Let $T\in \mathfrak{A}_+$ be an operator such that $T\ker \langle \cdot,\cdot\rangle_0\subseteq \ker \langle \cdot,\cdot\rangle_0$. 
    We define the operator $\Phi(T)$ on $\mathscr{H}$ as 
    \begin{equation}
        \Phi(T)\psi(X) = \psi(TX),\quad X\in \mathfrak{A}_+.
    \end{equation}
    We define $\mathcal{M}$ as the $C^*$-algebra generated by all operators $\Phi(T)$, where $T$ runs over all operators in $\mathfrak{A}_+$ such that $T\ker \langle \cdot,\cdot\rangle_0\subseteq \ker \langle \cdot,\cdot\rangle_0$. 
\end{definition}
We relate the $C^*$-algebra produced by Osterwalder-Schrader reconstruction from the RP projection $\Pi$ to the \emph{interaction algebras} introduced in \cite{zanardi2000StabilizingQuantumInformation}. 
The interaction algebra $\mathcal{A}_+(O)$ of an operator $O$ on $\mathcal{H}_{-}\otimes \mathcal{H}_{+}$ is the $C^*$-algebra generated by operators of the form $\Tr_{\mathcal{H}_{-}}(OQ_{-})$ for all $Q_{-}\in \mathfrak{A}_{-}$. 
If the operator $O$ has a decomposition $O = \sum_j O^j_{-}\otimes O^j_+$, where the $O^j_{-}$ are linearly independent, then $\mathcal{A}_+(O)$ is the $C^*$-algebra generated by the operators $\{O^j_+\}_j$. 
The notion of interaction algebra is related to our discussion as follows:
\begin{proposition}\cite[Lemma 7]{gross2012QCAIndex}\label{prop:: interaction algebra and local symmetry}
    For any operator $O\in\mathfrak{A}_-\otimes \mathfrak{A}_+$, we have $\mathcal{A}_{+}(O) = \mathrm{Sym}_{+}(O)'\cap \mathfrak{A}_{+}$. 
\end{proposition}
\begin{proof}
    For any $Q_-\in \mathfrak{A}_-$ and any $X\in \mathrm{Sym}_{+}(O)$, 
    \begin{equation}
        \Tr_{\mathcal{H}_-}(Q_-O)X = \Tr_{\mathcal{H}_-}(Q_-OX) = \Tr_{\mathcal{H}_-}(XQ_-O) = X \Tr_{\mathcal{H}_-}(Q_-O).
    \end{equation}
    Hence $\mathcal{A}_{+}(O)\subseteq \mathrm{Sym}_{+}(O)'\cap \mathfrak{A}_{+}$. 
    Conversely, suppose $X$ commutes with $\Tr_{\mathcal{H}_-}(Q_-O)$ for all $Q_-\in \mathfrak{A}_-$. 
    Then by expressing $O$ as a sum of operators $\sum_jO^j_{-}\otimes O^j_{+}$, we see that $X$ commutes with $O^j_{+}$ for all $j$. 
    Therefore, $\mathrm{I}_{\mathcal{H}_-}\otimes X$ also commutes with $O$, and hence $X\in \mathrm{Sym}_{+}(O)$. 
    This proves that $\mathcal{A}_{+}(O)'\cap \mathfrak{A}_{+}\subseteq \mathrm{Sym}_{+}(O)$. 
    Since $\mathfrak{A}_{+}$ is a type-I factor, the double commutant theorem implies that $\mathcal{A}_{+}(O) = \mathrm{Sym}_{+}(O)'\cap \mathfrak{A}_{+}$. 
\end{proof}

\begin{proposition}\label{prop:: criterion for field operators}
    Let $T\in \mathfrak{A}_{+}$; then left multiplication by $T$ leaves $\ker \langle \cdot,\cdot\rangle_0$ invariant if and only if $\widehat{\Pi}T\widehat{\Pi} \in \mathcal{A}_{+}(\Pi)\widehat{\Pi}$ and $\widehat{\Pi}T(\mathrm{I}-\widehat{\Pi}) = 0$.
\end{proposition}
\begin{proof}
    By Proposition \ref{prop:: kernel of the sesquilinear form}, $T\ker \langle \cdot,\cdot\rangle_0\subseteq \ker \langle \cdot,\cdot\rangle_0$ if and only if $\mathcal{A}_{+}(\Pi)\widehat{\Pi}T\subseteq \mathcal{A}_{+}(\Pi)\widehat{\Pi}$. 
    This happens if and only if $\widehat{\Pi}T\in \mathcal{A}_{+}(\Pi)\widehat{\Pi}$. 
    Inserting $\mathrm{I} = \widehat{\Pi} + (\mathrm{I}- \widehat{\Pi})$ on the right shows that $\widehat{\Pi}T\in \mathcal{A}_{+}(\Pi)\widehat{\Pi}$ if and only if $\widehat{\Pi}T\widehat{\Pi} \in \mathcal{A}_{+}(\Pi)\widehat{\Pi}$ and $\widehat{\Pi}T(\mathrm{I}-\widehat{\Pi}) = 0$. 
\end{proof}
\begin{proposition}\label{prop:: criterion of field operators}
    For any operator $T\in\mathfrak{A}_+$ such that $\Phi(T)$ is well-defined, $A = \widehat{\Pi}T\widehat{\Pi}$ is the unique operator in $\mathcal{A}_{+}(\Pi)\widehat{\Pi}$ such that $\Phi(T) = \Phi(A)$. 
    Moreover, we have $\Phi(A)^* = \Phi(\Xi^{-1}A^\dagger \Xi)$, where $\Xi\in \mathcal{A}_{+}(\Pi)\widehat{\Pi}$ is the positive square root of $\Tr_{\mathcal{H}_-}\vert \phi_{\text{PF}}\rangle\bra{\phi_{\text{PF}}}$. 
\end{proposition}
\begin{proof}
Consider the projection $\mathrm{I}-\widehat{\Pi}\in \mathfrak{A}_+$. 
Since $\mathrm{I}_{-}\otimes (\mathrm{I}_{+} - \widehat{\Pi})\Pi = 0$, we have $(\mathrm{I}-\widehat{\Pi})\mathfrak{A}_+\subseteq \ker \langle \cdot,\cdot\rangle_0$. 
Therefore $\Phi(\mathrm{I}-\widehat{\Pi}) = 0$. 
Since $\Phi(XY) = \Phi(X)\Phi(Y)$ when both $\Phi(X)$ and $\Phi(Y)$ are well-defined, we have 
\begin{equation}
    0 = \Phi(\mathrm{I}-\widehat{\Pi})\Phi(T)\Phi(\widehat{\Pi}) = \Phi((\mathrm{I}-\widehat{\Pi})T\widehat{\Pi}) = \Phi(T) - \Phi(\widehat{\Pi} T\widehat{\Pi}).
\end{equation}
By the above lemma, $\widehat{\Pi}T\widehat{\Pi}\in \mathcal{A}_{+}(\Pi)\widehat{\Pi}$. 
The uniqueness follows from the injectivity of $\Phi$ on $\mathcal{A}_{+}(\Pi)\widehat{\Pi}$. 
Suppose that $B\in\mathcal{A}_{+}(\Pi)\widehat{\Pi}$ and $\Phi(B)=0$. 
Then $\psi(B)=\Phi(B)\psi(\mathrm{I})=0$, and Proposition \ref{prop:: kernel of the sesquilinear form}, applied with $X=B$, gives $\Tr(B^\dagger B)=0$. 
Thus $B=0$, proving that $\Phi$ is injective on $\mathcal{A}_{+}(\Pi)\widehat{\Pi}$. 

To prove the second claim, note that for $X,Y\in\mathfrak{A}_{+}$, we have 
\begin{equation}
\begin{aligned}
    \langle \psi(Y),\Phi(A)\psi(X)\rangle &= \frac{1}{\Tr (\Pi)}\Tr (\Pi \Theta(Y)\otimes AX) = \frac{1}{\Tr (\Pi)}\sum_i \Tr (\mathcal{O}(\phi_i) AX\mathcal{O}(\phi_i)^\dagger Y^\dagger)\\
    &= \frac{1}{\Tr (\Pi)}\sum_i \Tr (Y^\dagger \Xi A\Xi^{-1} \mathcal{O}(\phi_i) X\mathcal{O}(\phi_i)^\dagger )\\
    &= \frac{1}{\Tr (\Pi)}\sum_i \Tr ( (\Xi^{-1} A^\dagger \Xi Y)^\dagger  \mathcal{O}(\phi_i) X\mathcal{O}(\phi_i)^\dagger )\\
    &= \langle \Phi(\Xi^{-1}A^\dagger \Xi)\psi(Y),\psi(X)\rangle,
\end{aligned}
\end{equation}
where, in the third equality, we use Theorem \ref{theorem:: ground state of H} to obtain operators $W_i\in \mathrm{Sym}_{+}(\Pi)\widehat{\Pi}$ such that $\mathcal{O}(\phi_i) = \Xi W_i$, so that for each $i$ we have $\mathcal{O}(\phi_i)A = \Xi A\Xi^{-1}\mathcal{O}(\phi_i)$. 
\end{proof}

\begin{theorem}\label{thm:: field algebra and interaction algebra}
    Let $\Pi$ be a reflection positive projection on $\mathcal{H}_-\otimes \mathcal{H}_+$. 
    Then the map 
    \begin{equation}
        \rho(A) = \Phi(\Xi^{-1/2}A\Xi^{1/2}),\quad A\in \mathcal{A}_{+}(\Pi)\widehat{\Pi},
    \end{equation}
    defines a $*$-isomorphism from $\mathcal{A}_{+}(\Pi)\widehat{\Pi}$ to $\cM$. 
\end{theorem}
\begin{proof}
    It follows directly that $\rho$ is an algebra isomorphism from $\mathcal{A}_{+}(\Pi)\widehat{\Pi}$ to $\cM$. 
    By Proposition \ref{prop:: criterion for field operators}, we have 
    \begin{equation}
        \rho(A)^* = \Phi(\Xi^{-1/2}A\Xi^{1/2})^* = \Phi(\Xi^{-1/2}A^\dagger\Xi^{1/2}) = \rho( A^\dagger),
    \end{equation}
    which shows that $\rho$ is a $*$-isomorphism. 
\end{proof}
\begin{corollary}\label{cor:: field algebra from reflection positive Hamiltonian}
    Let $H$ be a reflection positive Hamiltonian on $\mathcal{H}_-\otimes \mathcal{H}_+$ with ground state projection $\Pi$. 
    Denote by $\cM$ the $C^*$-algebra produced by OS reconstruction from the ground state correlation functions of $H$. 
    Then we have $\cM\cong \mathcal{A}_{+}(H)\widehat{\Pi}$, where $\widehat{\Pi}$ is the entanglement support of $\Pi$. 
\end{corollary}
\begin{proof}
    This follows directly from Theorem \ref{thm:: field algebra and interaction algebra} and Proposition \ref{prop:: interaction algebra and local symmetry}, as we have $\mathrm{Sym}_{+}(H)\widehat{\Pi} = \mathrm{Sym}_{+}(\Pi)\widehat{\Pi}$. 
\end{proof}

\begin{definition}
    We define the \emph{vacuum representation} $\pi_+$ of the $C^*$-algebra $\cM$ on $\mathcal{H}_+$ to be the inverse of $\rho$. 
    Written out explicitly, we have $\pi_+(\Phi(A)) = \Xi^{1/2}A \Xi^{-1/2}$, for all $A\in \mathcal{A}_{+}(\Pi)\widehat{\Pi}$. 
\end{definition}
We call $\Omega = \psi(\mathrm{I})$ the vacuum vector, which defines the vacuum state on $\cM$ as 
\begin{equation}
    \omega(\Phi(A)) = \langle \Omega,\Phi(A)\Omega\rangle = \frac{1}{\Tr (\Pi)}\Tr \left(\Pi (\mathrm{I}\otimes A) \right), \quad A\in \mathcal{A}_{+}(\Pi)\widehat{\Pi}. 
\end{equation}
Note that the set $\cM\Omega$ spans $\mathscr{H}$. 
Therefore, $(\mathscr{H},\Phi,\Omega)$ is the GNS representation of the vacuum state $\omega$. 
Using the isomorphism $\pi_+$, we can view $\omega$ as a state on $\mathcal{A}_{+}(\Pi)\widehat{\Pi}$, and we have
\begin{equation}
    \begin{aligned}
        \omega\circ \rho(A) & = \langle \psi(\mathrm{I}),\psi(\Xi^{-1/2}A\Xi^{1/2})\rangle
        = \frac{1}{\Tr (\Pi)}\Tr (\Pi (\mathrm{I}\otimes \Xi^{-1/2}A\Xi^{1/2}))\\
        &= \frac{1}{\Tr (\Pi)}\Tr (\Xi^{-1/2}A\Xi^{1/2}\Tr_{\mathcal{H}_-}(\Pi)) = \frac{1}{\Tr (\Pi)}\Tr (A\Tr_{\mathcal{H}_-}(\Pi)). 
    \end{aligned}
\end{equation}
In the last equality, we used the fact that $[\Xi^{1/2},\Tr_{\mathcal{H}_-}(\Pi)] = 0$. 
This follows from
\begin{equation*}
    \Tr_{\mathcal{H}_{-}}(\Pi)/\Tr (\Pi) = \mathscr{E}(\mathrm{I}) = \Xi^2F, 
\end{equation*}
where $F = \sum_i W_iW_i^\dagger /\Tr (\Pi)$ is a strictly positive operator in $\mathcal{Z}(\mathcal{A}_{+}(\Pi)\widehat{\Pi})$ (see the proof of Proposition \ref{prop:: kernel of the sesquilinear form}). 
Thus, $\omega\circ \rho$ is faithful, and its modular group on $\mathcal{A}_{+}(\Pi)\widehat{\Pi}$ is given by 
\begin{equation}\label{eqn:: modular group of the vacuum state}
    \sigma^{\omega}_t(A) = \Xi^{2it}A\Xi^{-2it},\quad A\in \mathcal{A}_{+}(\Pi)\widehat{\Pi}.
\end{equation}
We call $\omega\circ \rho$ the vacuum state on $\mathcal{A}_{+}(\Pi)\widehat{\Pi}$. 

\subsection{Local net of operator algebras}
Using the results of the previous sections, we construct a local net of operator algebras from the ground state correlation functions of a frustration-free interaction with RP. 
If the interaction has finite range, then the local net reduces to a boundary algebra. 

Let $\mathcal{X}$ be the set of all finite symmetric subsets of $\Gamma$ obtained by taking intersections with rectangles. 
For a reflection positive frustration-free interaction $\varPhi$, denote by $\Pi(\Lambda)\in \mathfrak{A}_\Lambda$ the ground state projection of the local Hamiltonian $H_\Lambda$. 
Then, for each $\Lambda\in \mathcal{X}$, $\Pi(\Lambda)$ is a reflection positive projection in $\mathfrak{A}_\Lambda$. 
Therefore, we can apply OS reconstruction to the linear functional $\Tr(\Pi(\Lambda)\cdot)$ on $\mathfrak{A}_\Lambda$. 
This yields a family of finite-dimensional $C^*$-algebras $\{\cM_\Lambda\}_{\Lambda\in\mathcal{X}}$. 
Denote by $\widehat{\Pi}(\Lambda)\in\mathfrak{A}_{\Lambda_+}$ the entanglement support of $\Pi(\Lambda)$. 
For an operator $O\in \mathfrak{A}_{\text{loc}}$, denote by $\mathcal{A}_{+}(O)$ the interaction algebra of $O$ in $\mathfrak{A}_{\Gamma_+}$. 
Then by Corollary \ref{cor:: field algebra from reflection positive Hamiltonian}, there is a $*$-isomorphism $\pi_{\Lambda_+}$ from $\cM_\Lambda$ to $\mathcal{A}_{+}(H_\Lambda)\widehat{\Pi}(\Lambda)$. 

We would like to assemble the family of operator algebras $\{\cM_\Lambda\}_{\Lambda\in\mathcal{X}}$ into a local net of operator algebras. 
This requires us to define injective $*$-homomorphisms $\iota_{\Lambda_2,\Lambda_1}: \cM_{\Lambda_1}\rightarrow \cM_{\Lambda_2}$ for any pair of symmetric subsets $\Lambda_1\subseteq \Lambda_2$, such that the following two conditions are satisfied:
{\rm
\begin{enumerate}
    \item Isotony: For any $\Lambda_1\subseteq \Lambda_2\subseteq \Lambda_3$, $\iota_{\Lambda_3,\Lambda_2}\circ \iota_{\Lambda_2,\Lambda_1} = \iota_{\Lambda_3,\Lambda_1}$.
    \item Locality: For any $\Lambda_1,\Lambda_2\subseteq \Lambda_3$ with $\Lambda_1\cap \Lambda_2 = \emptyset$, $[\iota_{\Lambda_3,\Lambda_1}(\cM_{\Lambda_1}),\iota_{\Lambda_3,\Lambda_2}(\cM_{\Lambda_2})] = 0$. 
\end{enumerate}
}
With these conditions satisfied, one can define the inductive limit $\displaystyle \cM_{\mathcal{X}} = \varinjlim_{\Lambda\in\mathcal X} \cM_\Lambda$, whose closure with respect to the unique $C^*$-norm is a quasi-local algebra \cite[Chap. 2.6]{bratteli1981OperatorAlgebrasQuantum}. 

\begin{lemma}\label{lemma:: nested CP idempotents}
    Let $F_1,F_2$ be symmetric completely positive idempotent maps on $\mathcal{B}(\mathcal{H})$ satisfying
    \begin{equation}
        F_2F_1 = F_2 = F_1F_2.
    \end{equation}
    Let $\Xi_i$ be the canonical Perron--Frobenius eigenvector of $F_i$, and let $p_i$ be the range projection of $\Xi_i$.
    Then there exists $D\geq 0$ in $\mathrm{Sym}(F_1)p_1$ such that $\Xi_2 = \Xi_1D$.
    Furthermore, for any operator $X$ in $p_1\mathcal{B}(\mathcal{H})p_1$ that commutes with $\mathrm{Sym}(F_1)p_1$, we have $Xp_2 = p_2X$, and
    \begin{equation}
        [Xp_2,Y] = 0,\quad Y\in \mathrm{Sym}(F_2)p_2.
    \end{equation}
\end{lemma}
\begin{proof}
    From the assumption $F_2 = F_1F_2$ and the fact that $\Xi_2 = F_2(\Xi_2)$, we have
    \begin{equation}
        F_1(\Xi_2) = F_1(F_2(\Xi_2)) = F_2(\Xi_2) = \Xi_2. 
    \end{equation}
    Therefore, by Theorem \ref{thm:: structure of maximum eigenspace of a symmetric CP map}, there exists $D\in \mathrm{Sym}(F_1)p_1$ such that $\Xi_2 = \Xi_1D$.
    Since $D\Xi_1 = \Xi_1D$ and $\Xi_2 = \Xi_1D\geq 0$, the positive operators $\Xi_1$ and $\Xi_2$ commute, and hence 
    \begin{equation*}
        D = \Xi_1^{-1/2}\Xi_2 \Xi_1^{-1/2}\geq 0,
    \end{equation*}
    where the inverse of $\Xi_1$ is taken in the corner $p_1\mathcal{B}(\mathcal{H})p_1$.
    Consequently, $p_2$ is the range projection of $D$, which implies $p_2\in \mathrm{Sym}(F_1)p_1$.
    
    Thus, for all operators $X$ in $p_1\mathcal{B}(\mathcal{H})p_1$ that commute with $\mathrm{Sym}(F_1)p_1$, we have $Xp_2 = p_2X$.
    Since the image of $F_2$ is $\Xi_2\mathrm{Sym}(F_2) = \Xi_1D\mathrm{Sym}(F_2)$, from $F_1F_2 = F_2$ we deduce
    \begin{equation}
        \Xi_1D\mathrm{Sym}(F_2)p_2\subseteq \Xi_1\mathrm{Sym}(F_1)p_1. 
    \end{equation}
    Compressing both sides by $p_2$ and then applying the inverse of $\Xi_1D$ on $p_2\mathcal{H}$ yields
    \begin{equation}
    \begin{aligned}
        \mathrm{Sym}(F_2)p_2 &= p_2\mathrm{Sym}(F_2)p_2\\
        &= (\Xi_1D)^{-1}p_2\Xi_1D\mathrm{Sym}(F_2)p_2\\
        &\subseteq (\Xi_1D)^{-1}p_2\Xi_1\mathrm{Sym}(F_1)p_1p_2\\
        &\subseteq p_2\mathrm{Sym}(F_1)p_2. 
    \end{aligned}
    \end{equation}
    Since $X$ commutes with $\mathrm{Sym}(F_1)p_1$, we know that $Xp_2$ commutes with $p_2\mathrm{Sym}(F_1)p_2$, and thus it also commutes with $\mathrm{Sym}(F_2)p_2$.
\end{proof}

\begin{proposition}\label{prop:: inclusion of interaction algebras}
Consider a family of ground state projections $\{\Pi(\Lambda)\}_{\Lambda\in \mathcal{X}}$ of a frustration-free, reflection positive interaction $\varPhi$.
Then, for any $\Lambda_1,\Lambda_2\in \mathcal{X}$ with $\Lambda_1\subseteq \Lambda_2$, we have $[\mathcal{A}_{+}(H_{\Lambda_1})\widehat{\Pi}(\Lambda_1),\widehat{\Pi}(\Lambda_2)]=0$, and the map $\alpha_{\Lambda_2,\Lambda_1}: \mathcal{A}_{+}(H_{\Lambda_1})\widehat{\Pi}(\Lambda_1) \to \mathcal{A}_{+}(H_{\Lambda_2})\widehat{\Pi}(\Lambda_2)$ defined by 
\begin{equation}
    \alpha_{\Lambda_2,\Lambda_1}(x) = (x\otimes I_{\Lambda_{2,+}\backslash \Lambda_{1,+}})\widehat{\Pi}(\Lambda_2),\quad x\in \mathcal{A}_{+}(H_{\Lambda_1})\widehat{\Pi}(\Lambda_1)
\end{equation}
is an $*$-homomorphism. 
Moreover, $\alpha_{\Lambda_2,\Lambda_1}$ is injective if  
    \begin{equation}
        \range(\Tr_{\overline{\Lambda_{1,+}}}\widehat{\Pi}(\Lambda_2)) = \widehat{\Pi}(\Lambda_1). 
    \end{equation}
    where $\range(A)$ denotes the range projection of a positive operator $A$. 
\end{proposition}
\begin{proof}
In the algebra $\mathfrak{A}_{\Lambda_{2,+}}$, set $\Lambda'_2 = \Lambda_2\backslash \Lambda_1$. 
We then have two symmetric completely positive idempotents on $\mathfrak{A}_{\Lambda_{2,+}}$:
\begin{equation}
    F_{\Lambda_1} = \mathcal{O}\bigl(\Pi(\Lambda_1)\otimes \mathrm{I}_{\Lambda'_2}\bigr)\mathcal{O}^{-1} = \mathcal{O}\Pi(\Lambda_1)\mathcal{O}^{-1}\otimes \mathrm{id}_{\Lambda'_{2,+}},\quad F_{\Lambda_2} = \mathcal{O}\Pi(\Lambda_2)\mathcal{O}^{-1},
\end{equation}
where $\mathrm{id}_{\Lambda'_{2,+}}$ is the identity map on $\mathfrak{A}_{\Lambda'_{2,+}}$. 
Since the interaction is frustration-free, we have $F_{\Lambda_1} F_{\Lambda_2} = F_{\Lambda_2} = F_{\Lambda_2} F_{\Lambda_1}$. 
For any $x\in \mathcal{A}_{+}(H_{\Lambda_1})\widehat{\Pi}(\Lambda_1)$, we have $x\otimes I_{\Lambda'_{2,+}}\in \mathrm{Sym}(F_{\Lambda_1})'\widehat{\Pi}(\Lambda_1)$. 
By Lemma \ref{lemma:: nested CP idempotents}, we conclude that $\widehat{\Pi}(\Lambda_2)$ commutes with $(x\otimes I_{\Lambda'_{2,+}})$, and $(x\otimes I_{\Lambda'_{2,+}})\widehat{\Pi}(\Lambda_2)$ is an element of $\mathrm{Sym}(F_{\Lambda_2})'\widehat{\Pi}(\Lambda_2)$. 
This implies that the map $\alpha_{\Lambda_2,\Lambda_1}$ is a well-defined $*$-homomorphism. 
The pullback of the vacuum state $\omega_{\Lambda_2}$ along $\alpha_{\Lambda_2,\Lambda_1}$ is the following state on $\mathcal{A}_{+}(H_{\Lambda_1})\widehat{\Pi}(\Lambda_1)$: 
\begin{equation}
    \alpha_{\Lambda_2,\Lambda_1}^*(\omega_{\Lambda_2}) (x) = \frac{1}{\Tr(\Pi(\Lambda_2))}\Tr(x\otimes \mathrm{I}_{\Lambda_2\backslash \Lambda_1}\Pi(\Lambda_2)) = \frac{1}{\Tr(\Pi(\Lambda_2))}\Tr(x\Tr_{\overline{\Lambda_{1,+}}}\Pi(\Lambda_2)). 
\end{equation}
Thus the pull-back state $\alpha_{\Lambda_2,\Lambda_1}^*(\omega_{\Lambda_2})$ is faithful on $\mathcal{A}_{+}(H_{\Lambda_1})\widehat{\Pi}(\Lambda_1)$, if we have 
\begin{equation}
        \range(\Tr_{\overline{\Lambda_{1,+}}}\widehat{\Pi}(\Lambda_2)) = \widehat{\Pi}(\Lambda_1). 
\end{equation}
This proves the claim. 
\end{proof}
\begin{remark}
    The entanglement supports of reflection positive projections play a special role in establishing the $*$-homomorphism between the interaction algebras. 
    In contrast, there is no apparent inclusion relation between $\mathcal{A}_{+}(H_{\Lambda_1})$ and $\mathcal{A}_{+}(H_{\Lambda_2})$. 
\end{remark}
\begin{remark}
    When $\Lambda\in \mathcal{X}$ satisfies $H_\Lambda = 0$, we have $\Pi(\Lambda) = \mathrm{I}_\Lambda$ and $\mathcal{A}_{+}(H_\Lambda)\widehat{\Pi}(\Lambda) = \mathbb{C}\mathrm{I}_{\Lambda_+}$. 
\end{remark}
\begin{definition}\label{def:: extendable family of local RP projections}
    A family of local reflection positive projections $\{\Pi(\Lambda)\}_{\Lambda\in \mathcal{X}}$ is called \emph{extendable} if, for any $\Lambda_1,\Lambda_2\in \mathcal{X}$ with $\Lambda_1\subseteq \Lambda_2$, we have 
    \begin{equation}
        \range(\Tr_{\overline{\Lambda_{1,+}}}\widehat{\Pi}(\Lambda_2)) = \widehat{\Pi}(\Lambda_1). 
    \end{equation}
\end{definition}
\begin{theorem}\label{thm:: local net of field algebras}
    Suppose that the family of local ground state projections $\{\Pi(\Lambda)\}_{\Lambda\in \mathcal{X}}$ of a frustration-free, reflection positive interaction is extendable. 
   Then for any $\Lambda_1\subseteq \Lambda_2$ in $\mathcal{X}$, there exists an injective $*$-homomorphism $\iota_{\Lambda_2,\Lambda_1}: \cM_{\Lambda_1}\rightarrow \cM_{\Lambda_2}$ such that the following properties hold: 
    \begin{enumerate}
        \item[(a)] For $\Lambda_1\subseteq \Lambda_2\subseteq \Lambda_3$ in $\mathcal{X}$, we have $\iota_{\Lambda_3,\Lambda_2} \circ \iota_{\Lambda_2,\Lambda_1} = \iota_{\Lambda_3,\Lambda_1}$. 
        \item[(b)] For $\Lambda_1,\Lambda_2,\Lambda_3\in \mathcal{X}$ with $\Lambda_1,\Lambda_2\subseteq \Lambda_3$ and $\Lambda_1\cap \Lambda_2 = \emptyset$, $\iota_{\Lambda_3,\Lambda_1}(\cM_{\Lambda_1})$ and $\iota_{\Lambda_3,\Lambda_2}(\cM_{\Lambda_2})$ commute. 
    \end{enumerate}
\end{theorem}
\begin{proof}
    By the isomorphisms $\pi_{\Lambda_{i,+}}: \cM_{\Lambda_i}\rightarrow \mathcal{A}_{+}(H_{\Lambda_i})\widehat{\Pi}(\Lambda_i)$, $i=1,2$, constructed in Theorem \ref{thm:: field algebra and interaction algebra} and Proposition \ref{prop:: inclusion of interaction algebras}, there exists a unique $*$-homomorphism $\iota_{\Lambda_2,\Lambda_1}: \cM_{\Lambda_1}\rightarrow \cM_{\Lambda_2}$ such that the following diagram commutes: 
    \begin{equation}
        \begin{tikzcd}
        \cM_{\Lambda_1} \arrow[r, "\iota_{\Lambda_2,\Lambda_1}"] \arrow[d, "\pi_{\Lambda_{1,+}}"] & \cM_{\Lambda_2} \arrow[d, "\pi_{\Lambda_{2,+}}"]\\
        \mathcal{A}_{+}(H_{\Lambda_1})\widehat{\Pi}(\Lambda_1) \arrow[r, "\alpha_{\Lambda_2,\Lambda_1}"] & \mathcal{A}_{+}(H_{\Lambda_2})\widehat{\Pi}(\Lambda_2)
        \end{tikzcd}
    \end{equation}
    Observe that (a) follows directly from the fact that $\widehat{\Pi}(\Lambda_3)\leq \widehat{\Pi}(\Lambda_2)\leq \widehat{\Pi}(\Lambda_1)$. 
    For (b), note that the two algebras $\mathcal{A}_{+}(H_{\Lambda_1})\widehat{\Pi}(\Lambda_1)$ and $\mathcal{A}_{+}(H_{\Lambda_2})\widehat{\Pi}(\Lambda_2)$ commute, since they are supported on disjoint regions. 
    In addition, $\widehat{\Pi}(\Lambda_3)$ commutes with both algebras since $\Lambda_1,\Lambda_2\subseteq \Lambda_3$. 
    Therefore, $\iota_{\Lambda_3,\Lambda_1}(\cM_{\Lambda_1})$ and $\iota_{\Lambda_3,\Lambda_2}(\cM_{\Lambda_2})$ commute as well. 
\end{proof}
\begin{remark}
For a finite subset $D\subset \Gamma_+$, set $D_s = \theta D\cup D$. 
Because $\Pi(D_s)$ is reflection positive, there is a completely positive map $\mathscr{E}_D$ on $\mathfrak{A}_{D}$ defined by 
\begin{equation}
    \mathscr{E}_D(X) = \frac{1}{\Tr(\Pi(D_s))}\Tr_{\theta D}(\Pi(D_s)\Theta(X^\dagger)\otimes \mathrm{I}),\quad X\in \mathfrak{A}_{D}.
\end{equation}
By the proof of Proposition \ref{prop:: kernel of the sesquilinear form}, the image of $\mathscr{E}_D$ is precisely the $C^*$-algebra $\mathcal{A}_+(H_{D_s})\widehat{\Pi}(D_s) \cong \cM_{D_s}$. 
Therefore, the local net of operator algebras $\{\cM_{D_s}\}_{D\subset \Gamma_+}$ can be viewed as the image of a family of local completely positive maps $\mathscr{E}_D$. 
In other words, local symmetric operators are selected by a family of completely positive maps. 

The construction in \cite{JNPW2025} uses a different selection mechanism.
There, the region of interest is enlarged by enough space to contain the support of every interaction term that meets the region, except along the cut, and one compresses by the ground state projection of the enlarged region. 
We will see later that, in the examples considered here and in \cite{JNPW2025}, this compression also selects the local symmetric operators. 
\end{remark}

Consider the extendable family of ground state projections $\{\Pi(\Lambda)\}_{\Lambda\in \mathcal{X}}$. 
For $\Lambda_1\subseteq \Lambda_2\in \mathcal{X}$, denote by $\omega_{\Lambda_1}$ and $\omega_{\Lambda_2}$ the vacuum states on $\cM_{\Lambda_1}$ and $\cM_{\Lambda_2}$, respectively. 
In general, the inclusion $\iota_{\Lambda_2,\Lambda_1}$ does \emph{not} intertwine the vacuum states; that is, $\iota^*_{\Lambda_2,\Lambda_1}(\omega_{\Lambda_2}) \neq \omega_{\Lambda_1}$. 
Thus, there is no apparent way to define a canonical vacuum state on the inductive limit. 
Nevertheless, the modular automorphism groups of the vacuum states are compatible with the inclusions $\iota_{\Lambda_2,\Lambda_1}$. 
\begin{theorem}\label{thm:: consistent dynamics in the inductive limit}
    Let $\{\Pi(\Lambda)\}_{\Lambda\in \mathcal{X}}$ be an extendable family of local ground state projections.
    For each $\Lambda\in \mathcal{X}$, let $\sigma^{\Lambda}_t$ denote the modular automorphism group associated with the vacuum state $\omega_{\Lambda}$ on $\cM_{\Lambda}$.
    Then for all $\Lambda_1,\Lambda_2\in \mathcal{X}$ satisfying $\Lambda_1\subseteq \Lambda_2$, we have 
    \begin{equation}
        \sigma^{\Lambda_2}_t\circ \iota_{\Lambda_2,\Lambda_1} = \iota_{\Lambda_2,\Lambda_1}\circ \sigma^{\Lambda_1}_t,\quad \forall t\in\mathbb{R}. 
    \end{equation}
\end{theorem}
\begin{proof}
    We use the same notation as in the proof of Proposition \ref{prop:: inclusion of interaction algebras}.
    Let $\Xi_1 = F_{\Lambda_1}(\mathrm{I}_{\Lambda_{1,+}})\otimes \mathrm{I}_{\Lambda'_{2,+}}$ and $\Xi_2 = F_{\Lambda_2}(\mathrm{I}_{\Lambda_{2,+}})$ be the corresponding PF eigenvectors.
    The modular automorphism group of $\omega_{\Lambda_1}$ is given by $\sigma^{\Lambda_1}_t(x) = \Xi_1^{2it}x \Xi_1^{-2it}$ (for $x\in \mathrm{Sym}(F_{\Lambda_1})'\widehat{\Pi}(\Lambda_1)$), while the modular automorphism group of $\omega_{\Lambda_2}$ is given by $\sigma^{\Lambda_2}_t(y) = \Xi_2^{2it}y \Xi_2^{-2it}$ (for $y\in \mathrm{Sym}(F_{\Lambda_2})'\widehat{\Pi}(\Lambda_2)$).
    By Lemma \ref{lemma:: nested CP idempotents}, there exists a positive operator $D\in\mathrm{Sym}(F_{\Lambda_1})\widehat{\Pi}(\Lambda_1)$ such that $\Xi_2 = D \Xi_1$, and $\range(D) = \range(\Xi_2) = \widehat{\Pi}(\Lambda_2)$.
    All powers of $D$ below are taken in the corner $\widehat{\Pi}(\Lambda_2)\mathcal{B}(\mathcal{H})\widehat{\Pi}(\Lambda_2)$, in which $D$ is invertible.
    Therefore,
    \begin{equation}
    \begin{aligned}
        \sigma^{\Lambda_2}_t \left( (x\otimes \mathrm{I}_{\Lambda'_{2,+}})\widehat{\Pi}(\Lambda_2)\right) &= \Xi_1^{2it}D^{2it}(x\otimes \mathrm{I}_{\Lambda'_{2,+}})\widehat{\Pi}(\Lambda_2) D^{-2it}\Xi_1^{-2it} \\
        &= \left((\Xi_1^{2it} x\Xi_1^{-2it}) \otimes \mathrm{I}_{\Lambda'_{2,+}}\right)\widehat{\Pi}(\Lambda_2),
    \end{aligned}
    \end{equation}
    because $(x\otimes \mathrm{I}_{\Lambda'_{2,+}})\widehat{\Pi}(\Lambda_2) \in \mathrm{Sym}(F_{\Lambda_2})'\widehat{\Pi}(\Lambda_2)$ commutes with $D$.
\end{proof}
\begin{remark}
    For a symmetric subset $\Lambda$, the density matrix of the state $\omega_{\Lambda}$ is $\frac{\Tr_{\mathcal{H}_-}(\Pi(\Lambda))}{\Tr(\Pi(\Lambda))}$, and $K_{\Lambda_+} = -\log \Tr_{\mathcal{H}_-}(\Pi(\Lambda))$ is an entanglement Hamiltonian up to an additive scalar.
    We note that when restricted to the boundary algebra $\cM_{\Lambda}$, the dynamics generated by $K_{\Lambda_+}$ preserves the locality of operators. 
    If the interaction satisfies the LTQO condition, then $\iota^*_{\Lambda_2,\Lambda_1}(\omega_{\Lambda_2})$ stabilizes as $\Lambda_2\nearrow \Gamma$ and defines a state on $\cM_{\mathcal{X}}$, as in \cite[Section 2.3]{JNPW2025}. 
\end{remark}
\subsection{Boundary algebra from a local interaction}
With the additional assumption that the interaction has finite range, we can further simplify the operator algebras $\cM_{\Lambda}$. 
In particular, if $\varPhi$ has interaction range $R$, then, for any finite subset $\Lambda$ with $\Lambda\cap \Gamma_+\neq \emptyset$ and $\Lambda\cap \Gamma_-\neq \emptyset$, the support of $\varPhi(\Lambda)$ is contained in the $R$-neighborhood of the reflection hyperplane. 

\begin{proposition}\label{prop:: field algebra of a local RP frustration-free Hamiltonian}
    Consider a reflection positive, frustration-free interaction $\varPhi$. 
    Given a finite symmetric subset $\Lambda$, define $H^{\Lambda}_0 = H_{\Lambda} - H_{\Lambda_+} - H_{\Lambda_-} = \sum_{\substack{\Lambda'\subseteq \Lambda,\,\Lambda'\cap \Lambda_+\neq \emptyset,\\\Lambda'\cap \Lambda_-\neq \emptyset}}\varPhi(\Lambda')$. 
    Then 
    \begin{equation}
        \mathcal{A}_{+}(H_{\Lambda})\widehat{\Pi}(\Lambda) = \mathcal{A}_{+}(H^{\Lambda}_0)\widehat{\Pi}(\Lambda). 
    \end{equation}
\end{proposition}
\begin{proof}
    By Proposition \ref{prop:: interaction algebra and local symmetry}, the two sides are the commutants, in the corner $\widehat{\Pi}(\Lambda)\mathfrak{A}_{\Lambda_+}\widehat{\Pi}(\Lambda)$, of $\mathrm{Sym}_{\Lambda_+}(H_{\Lambda})\widehat{\Pi}(\Lambda)$ and $\widehat{\Pi}(\Lambda)\mathrm{Sym}_{\Lambda_+}(H^{\Lambda}_0)\widehat{\Pi}(\Lambda)$, respectively. 
    These compressed symmetry algebras coincide by Theorem \ref{theorem:: local symmetries of a RP frustration-free Hamiltonian}, so their commutants coincide. 
\end{proof}
\begin{remark}
    Although $\widehat{\Pi}(\Lambda)$ is a central projection of the interaction algebra $\mathcal{A}_{+}(H_{\Lambda})$, in general we do not have $\widehat{\Pi}(\Lambda)\in \mathcal{A}_{+}(H^{\Lambda}_0)$, since the inclusion $\mathcal{A}_{+}(H^{\Lambda}_0)\subseteq \mathcal{A}_{+}(H_{\Lambda})$ can be strict. 
\end{remark}
\begin{theorem}\label{thm:: boundary algebra of a local RP frustration-free Hamiltonian}
    Let $\varPhi$ be a reflection positive, frustration-free interaction with interaction range $R$ such that the resulting family of ground state projections $\{\Pi(\Lambda)\}_{\Lambda\in \mathcal{X}}$ is extendable.  
    Then for any $\Lambda_1,\Lambda_2\in \mathcal{X}$ with $\Lambda_1\subseteq \Lambda_2$ such that $\Lambda_2\backslash \Lambda_1$ contains no sites within distance $R$ of the reflection hyperplane, the inclusion $\iota_{\Lambda_2,\Lambda_1}:\cM_{\Lambda_1}\rightarrow \cM_{\Lambda_2}$ is an isomorphism. 
\end{theorem}
\begin{proof}
    By Proposition \ref{prop:: field algebra of a local RP frustration-free Hamiltonian}, we have 
    \begin{equation}
        \cM_{\Lambda_1}\cong \mathcal{A}_+(H^{\Lambda_1}_0)\widehat{\Pi}(\Lambda_1),\quad \cM_{\Lambda_2}\cong \mathcal{A}_+(H^{\Lambda_2}_0)\widehat{\Pi}(\Lambda_2). 
    \end{equation}
    Since $\Lambda_2\backslash \Lambda_1$ contains no sites within distance $R$ of the reflection hyperplane, we have $H^{\Lambda_2}_0 = H^{\Lambda_1}_0$. 
    Therefore, $\mathcal{A}_+(H^{\Lambda_2}_0) = \mathcal{A}_+(H^{\Lambda_1}_0)$. 
    By the definition of $\iota_{\Lambda_2,\Lambda_1}$, we have 
    \begin{equation}
        \iota_{\Lambda_2,\Lambda_1}(\mathcal{A}_+(H^{\Lambda_1}_0)\widehat{\Pi}(\Lambda_1)) = \mathcal{A}_+(H^{\Lambda_1}_0)\widehat{\Pi}(\Lambda_2) = \mathcal{A}_+(H^{\Lambda_2}_0)\widehat{\Pi}(\Lambda_2) \cong \cM_{\Lambda_2},
    \end{equation}
    and we conclude that $\iota_{\Lambda_2,\Lambda_1}$ is surjective.
    By extendability and Proposition \ref{prop:: inclusion of interaction algebras}, $\iota_{\Lambda_2,\Lambda_1}$ is injective. 
    Therefore, $\iota_{\Lambda_2,\Lambda_1}$ is an isomorphism. 
\end{proof}
We illustrate how the above theorem can be applied to define a boundary algebra. 
Consider the hypercubic lattice $\mathbb{Z}^n$ in $\mathbb{R}^n$, and define the reflection $\theta$ by $\theta(x_1,x_2,\ldots,x_{n-1},x_n) = (x_1,x_2,\ldots,x_{n-1},-x_n)$. 
For simplicity, we take $\mathcal{X}$ to be the set of all finite symmetric rectangular regions in $\mathbb{Z}^n$. 
For $\Lambda\in \mathcal{X}$, $H^\Lambda_0$ is supported in a neighborhood of $\mathbb{R}^{n-1}\times [-R,R]$, so $\mathcal{A}_+(H^\Lambda_0)$ is supported in $\mathbb{R}^{n-1}\times[0,R]$, which is a box of the form $D\times [0,R]$. 
By Corollary \ref{cor:: field algebra from reflection positive Hamiltonian} and Proposition \ref{prop:: field algebra of a local RP frustration-free Hamiltonian}, we have $\cM_\Lambda\cong\mathcal{A}_+(H_\Lambda)\widehat{\Pi}(\Lambda)=\mathcal{A}_+(H^\Lambda_0)\widehat{\Pi}(\Lambda)$. 
Therefore, $\cM_\Lambda$ is isomorphic to a $C^*$-algebra supported in $D\times [0,R]$, where $D$ is a box in $\mathbb{Z}^{n-1}$.  

\begin{equation*}
    \vcenter{\hbox{\begin{tikzpicture}[x=0.75cm,y=0.75cm,font=\small]
        \fill[gray!20] (0,0) rectangle (4.3,6.1);
        \fill[gray!45] (0,1.83) rectangle (4.3,4.07);
        \draw[densely dashed] (-3.06,2.95) -- (7.40,2.95);
        \draw[line width=0.8pt] (0,2.95) -- (4.3,2.95);
        \node[anchor=west] at (5.65,2.30) {$\mathbb{Z}^{n-1}$};
        \node[anchor=west] at (0,2.05) {$D\times[-R,0]$};
        \node[anchor=west] at (2.95,0.25) {$\Lambda_-$};
        \node[anchor=west] at (2.88,5.55) {$\Lambda_+$};
        \node[anchor=west] at (0,3.55) {$D\times[0,R]$};
    \end{tikzpicture}}}
\end{equation*}

Thus, the structure of the net $\{\cM_\Lambda\}_{\Lambda\in \mathcal{X}}$ is entirely determined by $\{\cM_{D\times [-R,R]}\}_D$. 
The latter is manifestly a boundary algebra, as it is parameterized by $(n-1)$-dimensional rectangular regions. 

\begin{remark}
    We end this section with a remark on the extendability of local ground states.
    The extendability condition for $\{\Pi(X)\}_{X\in \mathcal{X}}$, namely that $\widehat{\Pi}(X)$ is the range projection of $\Tr_{\overline{X_+}}\widehat{\Pi}(Y)$ for all $Y\supseteq X$, is a consequence of the following stronger condition: 
    \begin{equation}\label{eqn:: extendability of ground states}
        \range( \Tr_{\overline{X}}\Pi(Y)) = \Pi(X),\quad \forall Y\supseteq X. 
    \end{equation}
    This condition can be understood as the extendability of the local ground states. 
    That is, for any ground state $\ket{\psi_X}$ of $H_X$, there exists a ground state $\ket{\psi_Y}$ of $H_Y$ such that 
    \begin{equation}
        \braket{ \psi_Y|\varphi_{Y\backslash X}\otimes \psi_X}\neq 0
    \end{equation}
    for some $\ket{\varphi_{Y\backslash X}}\in \mathcal{H}_{Y\backslash X}$. 
    Moreover, when $Y$ is taken to be the full system, so that $\Pi(Y)$ is the projection onto the global ground state subspace, Equation \eqref{eqn:: extendability of ground states} is equivalent to TQO-2 in \cite{BravyiHastingsMichalakis2010}. 
    For a commuting projector Hamiltonian with interaction range $r$, Ogata proved that the uniqueness of a frustration-free ground state in the infinite-volume limit implies the LTQO condition, provided that Equation \eqref{eqn:: extendability of ground states} is valid when $Y$ is taken to be the $r$-neighborhood of $X$ \cite[Lemma 7.2]{ogata2023BoundaryStates}. 
    We will see that the Levin--Wen string-net models \cite{LevinWen2005} satisfy this property, and we believe that the Walker-Wang models \cite{WalkerWang2012} also satisfy this property. 
\end{remark}

\section{Examples: Toric-Code and String-Net Hamiltonians}\label{sec:: examples of boundary algebras}
We compute this net for two exactly solvable models: the Kitaev toric code model \cite{Kitaev2003} and the Levin--Wen string-net models \cite{LevinWen2005}. 
Our computations explicitly show that the net of local operator algebras constructed by OS reconstruction is the algebra of local symmetric operators. 
\subsection{The toric code model}
In this section, we perform OS reconstruction for the $(2+1)$-dimensional toric code model \cite{Kitaev2003}. 
Spins are placed on the edges of the lattice, each of which carries a qubit $\mathbb{C}^2$. 
Vertex and plaquette operators are of the form $A_v = \bigotimes_{e:\,v\in \partial e}\sigma^x_e$ and $B_p = \bigotimes_{e\in \partial p}\sigma^z_e$. 
The local Hamiltonian on a finite region $\Lambda$ is given by $H_\Lambda = -\sum_{v\in \Lambda} A_v - \sum_{p\in \Lambda} B_p$.
After a shift by a constant, the Hamiltonian is a sum of commuting local projections and is therefore frustration-free. 

We first show that the Hamiltonian is reflection positive with respect to vertical reflection. 
Consider the antiunitary on $\mathbb{C}^2$ defined by its action on the computational basis: $\theta_0:\vert 0\rangle\mapsto \vert 0\rangle$ and $\theta_0:\vert 1\rangle\mapsto \vert 1\rangle$. 
Note that we have $\theta_0 \sigma^x\theta_0^{-1} = \sigma^x$ and $\theta_0 \sigma^z\theta_0^{-1} = \sigma^z$. 
Consider a vertex $v$ on the x-axis.
Then $A_v = \sigma^x_1\sigma^x_2\otimes \sigma^x_3\sigma^x_4$, where edges $1,2$ are in the upper half-plane and $3,4$ are in the lower half-plane. 
Thus, we can write $A_v = \Theta(\sigma^x_1\sigma^x_2)\otimes \sigma^x_1\sigma^x_2$. 
Similarly, for a plaquette $p$ that crosses the x-axis, we have $B_p = \sigma^z_1\sigma^z_2\otimes \sigma^z_3\sigma^z_4 = \Theta(\sigma^z_1\sigma^z_2)\otimes \sigma^z_1\sigma^z_2$. 
Thus the interactions across the x-axis are of the form required in Theorem \ref{thm:: structure of RP Hamiltonian}. 
The interactions away from the x-axis are manifestly reflection-invariant, so the toric code Hamiltonian has reflection positivity. 

We now compute the boundary net of the toric code. 
Consider a symmetric rectangular region $\Lambda$ in the plane. 
Then, by locality and the frustration-free property of $H$, we have $\mathcal{A}_+(H_\Lambda)\widehat{\Pi}(\Lambda) = \mathcal{A}_+(H^\Lambda_0)\widehat{\Pi}(\Lambda)$, where $H^\Lambda_0$ is supported in a slab around the part of the x-axis in $\Lambda$. 
Our first task is to compute the interaction algebra $\mathcal{A}_+(H^\Lambda_0)$. 
Since each vertex and plaquette operator acts on four neighboring edges, $H^\Lambda_0$ is supported in a tubular neighborhood of the x-axis, which contains all vertices in $\Lambda$ that are on the x-axis and the edges that connect to them. 
Denote the positive half of the tubular neighborhood by $\mathcal{I}$. 
\begin{equation*}
    \begin{tikzpicture}[scale = 0.75]
        \begin{scope}[rotate = -45]
        \draw (-3,2) -- (3,2);
        \draw (-3,1) -- (3,1);
        \draw (-3,0) -- (3,0);
        \draw (-3,-1) -- (3,-1);
        \draw (-3,-2) -- (3,-2);
        \draw (-2,3) -- (-2,-3);
        \draw (-1,3) -- (-1,-3);
        \draw (0,3) -- (0,-3);
        \draw (1,3) -- (1,-3);
        \draw (2,3) -- (2,-3);
        \foreach \x in {-2,-1,0,1,2}{
            \foreach \y in {-2,-1,0,1,2}{
                \filldraw (\x,\y) circle (1pt);
            }
        }
        \draw[thick,densely dotted] (-3,-3) -- (3,3);
        \draw[color = blue!75, opacity = 0.5, line cap=round, line join=round,line width=2pt] (-1,-1) -- (-1,0) -- (0,0);
        \draw[color = red!75, opacity = 0.5, line cap=round, line join=round,line width=2pt] (0,1) -- (1,1) -- (1,2);
        \filldraw[color = gray!40, opacity = 0.25] (2,3) -- (-3,-2) -- (-3,-3) -- (3,3) -- cycle;
        \node at (2.5,3.00) {$\mathcal{I}$};
        \end{scope}
        \node at (3,2) {$\Lambda_+$};
        \node at (3,-2) {$\Lambda_-$};
    \end{tikzpicture}
\end{equation*}
Suppose that there are $L$ edges in the region $\Lambda_+$ with one endpoint on the x-axis. 
We number these edges from left to right by $1,2,\ldots,L$. 
Then $\mathcal{A}_L:= \mathcal{A}_+(H^\Lambda_0)$ is generated by Pauli operators 
\begin{equation*}
    \{\sigma^x_{2k+1}\sigma^x_{2k+2}\}^{\left\lfloor \frac{L-2}{2} \right\rfloor}_{k=0}\cup \{\sigma^z_{2k}\sigma^z_{2k+1}\}^{ \left\lfloor \frac{L-1}{2} \right\rfloor}_{k=1},
\end{equation*} 
as indicated in the above figure. 
When $L = 2$, we have $\mathcal{A}_{+}(H^\Lambda_0) = \langle \sigma^x_1\sigma^x_2\rangle \cong \mathbb{C}^2$. 
Then $\mathcal{A}_3$ is generated by $\mathcal{A}_2$ together with an additional projection $\displaystyle\frac{1 + \sigma^z_2\sigma^z_3}{2}$. 
The structure of $\mathcal{A}_3$ is determined by the algebraic relation  
\begin{equation}
    \frac{1 + \sigma^z_2\sigma^z_3}{2}\sigma^x_1\sigma^x_2\frac{1 + \sigma^z_2\sigma^z_3}{2} = 0 = \mathbb{E}_{\mathbb{C}}(\sigma^x_1\sigma^x_2)\frac{1 + \sigma^z_2\sigma^z_3}{2}. 
\end{equation}
That is, $\langle\mathcal{A}_2,\frac{1 + \sigma^z_2\sigma^z_3}{2}\rangle$ forms a basic construction of $\mathbb{C}\subset \mathcal{A}_2$, so that $\mathcal{A}_3\cong M_2(\mathbb{C})$ \cite{Jones1983}. 
This pattern continues: $\mathcal{A}_{L+1}$ is the basic construction of $\mathcal{A}_{L-1}\subset \mathcal{A}_L$, with Jones projection $\displaystyle\frac{1 + \sigma^z_{L}\sigma^z_{L+1}}{2}$ for even $L$ and $\displaystyle\frac{1 + \sigma^x_{L}\sigma^x_{L+1}}{2}$ for odd $L$. 
Thus, $\mathcal{A}_L\cong \underbrace{M_2(\mathbb{C})\otimes \cdots \otimes M_2(\mathbb{C})}_{(L-1)/2}$ for odd $L$ and $\mathcal{A}_L\cong \underbrace{M_2(\mathbb{C})\otimes \cdots \otimes M_2(\mathbb{C})}_{L/2-1}\otimes \mathbb{C}^2$ for even $L$. 
In particular, the center of $\mathcal{A}_L$ is trivial when $L$ is odd and is generated by the Pauli string $S_L  =\sigma^x_1\sigma^x_2\cdots\sigma^x_L$ when $L$ is even. 
The inclusion $\mathcal{A}_L\subset \mathcal{A}_{L+1}$ is given by tensoring with the identity on the $(L+1)$-th qubit. 

Next, we compute the entanglement support $\widehat{\Pi}(\Lambda)$. 
Denote by $v_1,v_2,\ldots,v_{L/2}$ the vertices in $\Lambda$ that are on the x-axis, numbered from left to right. 
Denote by $p_j$ the plaquette between $v_{j}$ and $v_{j+1}$, $1\leq j\leq L/2-1$. 
Then $H^\Lambda_0 = -\sum^{L/2}_{k = 1} A_{v_k} - \sum^{L/2-1}_{j=1} B_{p_j}$. 
By the unitarity of the Pauli operators, $-\mathcal{O}H^\Lambda_0\mathcal{O}^{-1}(\mathrm{I}) = (L-1)\mathrm{I}$ is a scalar multiple of the identity. 
Since $H^\Lambda_0$ commutes with $H_{\Lambda_+}$ and $H_{\Lambda_-}$, we have $\widehat{\Pi}(\Lambda) = \Pi(\Lambda_+)$. 
Thus, the boundary algebra obtained from the ground states of $H_\Lambda$ is isomorphic to $\mathcal{A}_+(H^\Lambda_0)\Pi(\Lambda_+) = \mathcal{A}_L\Pi(\Lambda_+)$. 
Finally, multiplication by $\Pi(\Lambda_+)$ is injective on $\mathcal{A}_L$, as $\mathcal{A}_L$ is simple for odd $L$, whereas for even $L$ the $\sigma^z$-string $\prod^L_{k=1} \sigma^z_k$ acts nontrivially on the ground state subspace of $H_{\Lambda_+}$. 
In this case, the action of the one-parameter automorphism group $\sigma_t$ is also trivial. 
The above local net is identical, up to locality-preserving isomorphisms, to the local $\mathbb{Z}_2$-symmetric operator algebra considered in \cite[Section IV]{ChatterjeeWen2023} and \cite[Remark 3.13]{JNPW2025}. 

\subsection{The Levin--Wen string-net models}
We now show that the pattern observed in the toric code model extends to Levin--Wen string-net models by explicitly computing the boundary algebras. 
First, we recall the categorical ingredients underlying reflection positivity for Levin--Wen string-net models from \cite{JL2020}. 
Let $\mathscr{C}$ be a spherical unitary fusion category. 
We denote by $x,y,z,i,j,k,\dots$ the objects in $\mathscr{C}$ and adopt the following notation:\\

\noindent
$\Irr$: the set of simple objects in $\mathscr{C}$.\\
$\mathbb{1}$: the tensor unit of $\mathscr{C}$.\\
$x^{-}$: the dual of an object $x^{+} = x$.\\
$\hom_{\mathscr{C}}(x,y)$: the space of morphisms from $x$ to $y$.\\
$\ev_x\in \hom_{\mathscr{C}}(x^{-}\otimes x,\mathbb{1})$: the evaluation map of $x$.\\
$\coev_x\in \hom_{\mathscr{C}}(\mathbb{1},x\otimes x^{-})$: the coevaluation map of $x$.\\
$\tr_{\mathscr{C}}$: the categorical trace.\\
$\mathrm{1}_x$: the identity morphism of an object $x$.\\
$d(x) = \tr_{\mathscr{C}}(\mathrm{1}_x)$: the quantum dimension of $x$.\\
$\mu = \sum_{j\in \Irr}d(j)^2$: the global dimension of $\mathscr{C}$.\\

\noindent
For simple objects $i,j,k$, we define the inner product on $\hom_{\mathscr{C}}(\mathbb{1},i\otimes j\otimes k)$ as 
\begin{equation}
    \langle \eta,\xi\rangle = \frac{1}{\sqrt{d(i)d(j)d(k)}}\tr_{\mathscr{C}}(\eta^*\xi),\quad \forall \xi,\eta\in \hom_{\mathscr{C}}(\mathbb{1},i\otimes j\otimes k). 
\end{equation}
We denote by $\ONB(i\otimes j\otimes k)$ an orthonormal basis of $\hom_{\mathscr{C}}(\mathbb{1},i\otimes j\otimes k)$.  
We will use the graphical calculus freely. 
For an object $x$, we represent the $1_x$ by $\vcenter{\hbox{\begin{tikzpicture}[scale=0.5]
    \draw[thick, postaction={decorate}, decoration={markings, mark=at position 0.5 with {\arrow{stealth}}}] (0,1) --(0,-1);
    \node at (0,1.25) {$x$};
\end{tikzpicture}}}$ and $1_{x^{-}}$ by $\vcenter{\hbox{\begin{tikzpicture}[scale=0.5]
    \draw[thick, postaction={decorate}, decoration={markings, mark=at position 0.5 with {\arrow{stealth}}}] (0,-1) --(0,1);
    \node at (0,1.25) {$x$};
\end{tikzpicture}}}$. 
The choice of the inner product implies the following partition of unity: 
\begin{equation}
    \vcenter{\hbox{\begin{tikzpicture}[scale=0.75]
        \draw[thick, postaction={decorate}, decoration={markings, mark=at position 0.5 with {\arrow{stealth}}}] (-0.5,1) -- (-0.5,-1);
        \draw[thick, postaction={decorate}, decoration={markings, mark=at position 0.5 with {\arrow{stealth}}}] (0.5,1) -- (0.5,-1);
        \node at (-0.5,1.25) {$i$};
        \node at (0.5,1.25) {$j$};
    \end{tikzpicture}}} = \sum_{k\in \Irr} \sum_{\alpha\in \ONB(i\otimes j\otimes k^{-})}\frac{\sqrt{d(k)}}{\sqrt{d(i)d(j)}}\vcenter{\hbox{\begin{tikzpicture}[scale=0.75]
        \draw[thick, postaction={decorate}, decoration={markings, mark=at position 0.5 with {\arrow{stealth}}}] (0,-0.5) --(0,-1);
        \draw[thick, postaction={decorate}, decoration={markings, mark=at position 0.5 with {\arrow{stealth}}}] (0.5,-1) arc (0:90:0.5); 
        \draw[thick, postaction={decorate}, decoration={markings, mark=at position 0.75 with {\arrow{stealth}}}] (0,-0.5) arc (90:180:0.5);
        \node at (0,-0.25) {$\alpha$};
        \fill (0,-0.5) circle (1.5pt); 
        \draw[thick, postaction={decorate}, decoration={markings, mark=at position 0.5 with {\arrow{stealth}}}] (0,1) --(0,0.5);
        \draw[thick, postaction={decorate}, decoration={markings, mark=at position 0.5 with {\arrow{stealth}}}] (-0.5,1) arc (180:270:0.5);
        \draw[thick, postaction={decorate}, decoration={markings, mark=at position 0.75 with {\arrow{stealth}}}] (0,0.5) arc (270:360:0.5);
        \node at (0,0.25) {$\alpha^*$};
        \fill (0,0.5) circle (1.5pt); 
        \draw[thick] (0.5,1) arc (180:0:0.25);
        \draw[thick, postaction={decorate}, decoration={markings, mark=at position 0.5 with {\arrow{stealth}}}] (1,1) -- (1,-1);
        \draw[thick] (1,-1) arc (0:-180:0.25);
        \node at (1.25,0) {$k$};
        \node at (-0.5,1.25) {$i$};
        \node at (0,1.25) {$j$};
    \end{tikzpicture}}}. 
\end{equation}
Define $A = \bigoplus_{j\in \Irr}j$. 
We use a red line without an arrow to represent $1_A$ in the graphical calculus (which is legitimate since $A$ is self-dual). 
Define the rotation $\rho$ on $\hom_{\mathscr{C}}(\mathbb{1},A^3)$ as 
\begin{equation}\label{eqn:: rotation on hom(1,A^n)}
    \rho \left( \vcenter{\hbox{\begin{tikzpicture}[scale=0.75]
        \draw[thick, color= red] (0,0) -- (-0.5,-1);
        \draw[thick, color= red] (0,0) -- (0,-1);
        \draw[thick, color= red] (0,0) -- (0.65,-1);
        \fill (0,0) circle (1.5pt); 
        \node at (0,0.25) {$\xi$};
    \end{tikzpicture}}} \right) = 
    \vcenter{\hbox{\begin{tikzpicture}
        \draw[thick, color= red] (0,0) -- (-0.5,-1);
        \draw[thick, color= red] (0,0) -- (0,-1);
        \draw[thick, color= red] (0,0) -- (0.65,-1);
        \fill (0,0) circle (1.5pt); 
        \node at (0,0.25) {$\xi$};
        \draw[thick, color= red] (-0.5,-1) arc (0:-180:0.25) -- +(0,1) arc (180:0:1);
        \draw[thick, color= red] (1,0) -- (1,-1);
    \end{tikzpicture}}}
\end{equation}
For objects $x,y,z$, we define the modular conjugation $\theta_{\mathscr{C}}: \hom_{\mathscr{C}}(x\otimes y,z)\rightarrow \hom_{\mathscr{C}}(y^{-}\otimes x^{-}\otimes z^{-})$ by 
\begin{equation}\label{eqn:: modular conjugation on hom(xy,z)}
    \theta_{\mathscr{C}}\left( \vcenter{\hbox{\begin{tikzpicture}[scale=0.75]
        \draw[thick, postaction={decorate}, decoration={markings, mark=at position 0.65 with {\arrow{stealth}}}] (150:1) --(0,0);
        \draw[thick, postaction={decorate}, decoration={markings, mark=at position 0.65 with {\arrow{stealth}}}] (30:1) --(0,0);
        \draw[thick, postaction={decorate}, decoration={markings, mark=at position 0.65 with {\arrow{stealth}}}] (0,0) --(-90:1);
        \node at (210:0.25) {$\xi$};
        \node at (150:1.25) {$x$};
        \node at (30:1.25) {$y$};
        \node at (-90:1.25) {$z$};
        \fill (0,0) circle (1.5pt); 
    \end{tikzpicture}}} \right) = 
    \vcenter{\hbox{\begin{tikzpicture}
        \draw[thick] (0,0) -- (-150:0.5);
        \draw[thick] (0,0) -- (-30:0.5);
        \draw[thick] (0,0) --(90:0.5);
        \draw[thick,postaction={decorate}, decoration={markings, mark=at position 0.75 with {\arrow{stealth}}}] (-150:0.5) arc (0:-180:0.25) -- +(0,1);
        \draw[thick,postaction={decorate}, decoration={markings, mark=at position 0.75 with {\arrow{stealth}}}] (-30:0.5) arc (0:-180:1) -- +(0,1);
        \draw[thick,postaction={decorate}, decoration={markings, mark=at position 0.25 with {\arrow{stealth}}}] (1,-0.5) -- +(0,1) arc (0:180:0.5);
        \node at (150:0.25) {$\xi^*$};
        \node at (-0.9,1) {$x$};
        \node at (-1.8,1) {$y$};
        \node at (0.9,-1) {$z$};
        \fill (0,0) circle (1.5pt); 
    \end{tikzpicture}}}. 
\end{equation}
In particular, $\theta_{\mathscr{C}}$ defines an antiunitary on $\hom_{\mathscr{C}}(\mathbb{1},A^3)$. 

We now consider the Levin--Wen string-net models on the honeycomb lattice $\Gamma$. 
We choose the reflection hyperplane to be the x-axis, and we assume that the vertical reflection $\theta$ interchanges $\Gamma_\pm$. 
Let $(V_\pm,E_\pm)$ be the sets of vertices and edges above and below the x-axis, and denote by $E_0$ the set of edges that cross the x-axis. 
Denote by $s, t: E\rightarrow V$ the source and target functions. 
We require that $\theta$ reverses the orientation of each edge, meaning that, for any edge $e\in E_+$, 
\begin{equation}
    s(\theta (e)) = \theta (t(e)) \text{ and } t(\theta (e)) = \theta (s(e)).
\end{equation}
For each vertex $v\in V$, denote by $E(v)$ the set of edges that are incident to $v$. 
We fix $\kappa_v: \{1,2,3\}\rightarrow E(v)$ to be a counterclockwise ordering of $E(v)$. 
Note that $\kappa_v$ is determined by $\kappa_v(1)$. 
Define $\epsilon_v(e) = +$ if $t(e) = v$ and $\epsilon_v(e) = -$ if $s(e) = v$. 
Following the convention in \cite{Kong2014Universal}, we assign the degrees of freedom only to the vertices. 
For each vertex $v$, we associate the Hilbert space $\mathcal{H}_v = \hom_{\mathscr{C}}(\mathbb{1},A^3)$. 
For a vertex $v\in V_+$, the antiunitary $\widehat{\theta}_v: \mathcal{H}_v\rightarrow \mathcal{H}_{\theta(v)}$ is given by modular conjugation: 
\begin{equation}
    \widehat{\theta}_v\ket{\xi} = \ket{\theta_{\mathscr{C}}(\xi)},\quad \ket{\xi} \in \mathcal{H}_v. 
\end{equation}

The string-net Hamiltonian consists of two types of local interactions. 
For $\overrightarrow{j}\in \Irr^3$, denote by $P_{v,\overrightarrow{j}}$ the orthogonal projection from $\mathcal{H}_v$ onto $\hom_{\mathscr{C}}(\mathbb{1},\overrightarrow{j})$. 
For each edge $e\in E$ and each simple object $i$, define $Q^i_e$ on $\mathcal{H}_{t(e)}\otimes \mathcal{H}_{s(e)}$ to be the projection onto the subspace spanned by configurations where the label on $e$ from both vertices is $i$. 
Define $Q_e = \sum_{i\in \Irr} Q^i_e$. 
The operator $Q_e$ projects $\mathcal{H}_{t(e)}\otimes \mathcal{H}_{s(e)}$ onto the subspace where the labels on $e$ from both vertices match. 
The plaquette term is more involved. 
For an object $x$ and morphisms $\xi\in\hom_{\mathscr{C}}(x^{-}\otimes A,A)$ and $\eta\in\hom_{\mathscr{C}}(A\otimes x,A)$, define the operator $C_{\xi,\eta}$ on $\hom_{\mathscr{C}}(\mathbb{1},A^3)$ by
\begin{equation}\label{eqn:: operator C_xi,eta}
    C_{\xi,\eta}\left( \vcenter{\hbox{\begin{tikzpicture}
        \draw[thick, color= red] (0,0) -- (-0.5,-1);
        \draw[thick, color= red] (0,0) -- (0,-1);
        \draw[thick, color= red] (0,0) -- (0.5,-1);
        \fill (0,0) circle (1.5pt); 
        \node at (0,0.25) {$\zeta$};
    \end{tikzpicture}}} \right) = (\xi\otimes 1_{A}\otimes \eta)(1_A\otimes \zeta\otimes 1_A)\coev_{A} =
    \vcenter{\hbox{\begin{tikzpicture}
        \draw[thick, color= red] (0,0) -- (-0.3,-0.5);
        \draw[thick, color= red] (0,0) -- (0,-1);
        \draw[thick, color= red] (0,0) -- (0.3,-0.5);
        \draw[thick] (-0.3,-0.5) -- (-0.6,-0.25) -- (-0.6,0);
        \draw[thick] (0.3,-0.5) -- (0.6,-0.25) -- (0.6,0);
        \draw[thick, postaction={decorate}, decoration={markings, mark=at position 0.65 with {\arrow{stealth}}}] (-0.6,0) arc (180:0:0.6);
        \draw[thick, color= red] (-0.3,-0.5) -- (-0.3,-1);
        \draw[thick, color= red] (0.3,-0.5) -- (0.3,-1);
        \fill (0,0) circle (1.5pt); 
        \fill (-0.3,-0.5) circle (1.5pt); 
        \fill (0.3,-0.5) circle (1.5pt); 
        \node at (0,0.25) {$\zeta$};
        \node at (-0.6,-0.5) {$\xi$};
        \node at (0.15,-0.5) {$\eta$};
        \node at (0,0.85) {$x$};
    \end{tikzpicture}}}. 
\end{equation}
With a suitable normalization, the matrix units of $C_{\xi,\eta}$ produce the F-symbols of $\mathscr{C}$. 
Let $p$ be a plaquette with boundary vertices and edges $v_1,e_1,v_2,e_2,\ldots,v_6,e_6$ ordered clockwise, and let $j$ be a simple object. 
Intuitively, the operator $B^j_p$ on $\displaystyle \bigotimes_{m=1}^6\mathcal{H}_{v_m}$ is defined by inserting a loop $\vcenter{\hbox{\begin{tikzpicture}[scale = 0.5]
    \draw[thick, postaction={decorate}, decoration={markings, mark=at position 0.5 with {\arrow{stealth}}}] (0,0) circle (0.75);
    \node at (0,0) {$j$};
\end{tikzpicture}}}$ labeled by $j$ in the plaquette and then resolving the double line $\vcenter{\hbox{\begin{tikzpicture}[scale = 0.75]
    \draw[thick, color= red] (-0.25,0.5) -- (-0.25,-0.5);
        \draw[thick, postaction={decorate}, decoration={markings, mark=at position 0.5 with {\arrow{stealth}}}] (0.25,0.5) -- (0.25,-0.5);
        \node at (0.5,0) {$j$};
\end{tikzpicture}}}$ using the partition of unity. 
Explicitly, $B^j_p$ can be expressed as the following matrix-product operator:  
\begin{equation}
\begin{aligned}
    &B^j_{p} = \sum_{i_1,\ldots,i_6\in \Irr}\sum_{j_1,\ldots,j_6\in \Irr}\sum_{y_m\in \ONB(j^{-}\otimes i_m,j_m)} \\ 
    &\prod^6_{m=1} \left( \frac{d(j_m)}{d(j)d(i_m)} \right)^{1/4} \left( \frac{d(j_{m-1})}{d(j)d(i_{m-1})} \right)^{1/4} \mathrm{Ad}^{1-\kappa^{-1}_{v_m}(e_m)}_\rho \left[ C_{y_m,\theta_{\mathscr{C}}(y_{m-1})}\right]\prod^6_{m=1}Q^{i_m}_{e_m}
\end{aligned}
\end{equation}
Here the indices are cyclic: $i_0=i_6$ and $j_0=j_6$. 
Each factor indexed by $m$ acts on $\mathcal{H}_{v_m}$. 
Note that we have $\Theta(B^j_p) = B^{j^{-}}_{\theta (p)}$.
The plaquette operator is then defined as $\displaystyle B_p = \frac{1}{\mu}\sum_{j\in \Irr}d(j) B^j_p$. 
It is well known that the edge terms and plaquette terms are mutually commuting projections. 
The string-net Hamiltonian is defined as
\begin{equation}
    H_{\Lambda} = -\sum_{e\subset \Lambda}Q_e - \sum_{p\subset \Lambda}B_p, 
\end{equation}
for a finite subset $\Lambda$ of $\Gamma$. 
Since $B_p$ evaluates to $0$ on configurations with mismatched edge labels, the Hamiltonian $H_{\Lambda}$ is frustration-free. 

The result of \cite{JL2020} can be used to show that $H_{\Lambda}$ is reflection positive for every symmetric region $\Lambda$. 
First, consider a plaquette $p$ that is contained in the upper-half lattice $\Lambda_+$. 
Then $\Theta(B^j_p) = B^{j^{-}}_{\theta (p)}$; hence, $\Theta(B_p) = B_{\theta (p)}$. 
For an edge $e\in E_+$, we have $\Theta(Q_e) = Q_{\theta (e)}$ as the reflection reverses the orientation of $e$. 
Therefore these interactions are reflection symmetric. 
Now, for a plaquette $p$ that crosses the reflection axis, label the vertices and the edges on the boundary of $p$ that are contained in $\Lambda_+$ by $f_0,v_1,f_1,v_2,f_2,v_3,f_3$ ordered clockwise, with $f_0$ and $f_3$ being the two edges that cross the x-axis. 
\begin{equation*}
    \vcenter{\hbox{\begin{tikzpicture}[x=0.55cm,y=0.55cm,font=\small,line cap=round,line join=bevel]
        \draw (0,-0.8) -- (1.4,-1.6) -- (2.8,-0.8) -- (2.8,0.8) -- (1.4,1.6) -- (0,0.8) -- cycle;
        \draw (0,0.8) -- (-1.4,1.6);
        \draw (1.4,1.6) -- (1.4,3.2);
        \draw (2.8,0.8) -- (4.2,1.6);
        \draw (0,-0.8) -- (-1.4,-1.6);
        \draw (1.4,-1.6) -- (1.4,-3.2);
        \draw (2.8,-0.8) -- (4.2,-1.6);
        \draw[densely dashed] (-2.1,0) -- (5.4,0);
        \node[left] at (0,0.4) {$f_0$};
        \node[below] at (0.7,1.2) {$f_1$};
        \node[below] at (2.1,1.2) {$f_2$};
        \node[right] at (2.8,0.4) {$f_3$};
        \node[above left] at (0,0.8) {$v_1$};
        \node[above] at (1.4,1.6) {$v_2$};
        \node[above right] at (2.8,0.8) {$v_3$};
        \node[anchor=west] at (4.55,1.1) {$\Lambda_+$};
        \node[anchor=west] at (4.55,-1.1) {$\Lambda_-$};
    \end{tikzpicture}}}
\end{equation*}
Then we can decompose $B^j_p$ as
\begin{equation}
    B^j_p = \sum_{i_0,i_3\in \Irr}\sum_{j_0,j_3\in \Irr}\sum_{y_0\in\ONB(j^{-}\otimes i_0,j_0)}\sum_{y_3\in \ONB(j^{-}\otimes i_3,j_3)} \Theta(T_{\theta_{\mathscr{C}}(y_0),y_3})\otimes T_{\theta_{\mathscr{C}}(y_0),y_3}, 
\end{equation}
where $T_{\theta_{\mathscr{C}}(y_0),y_3}$ is the operator on $\mathcal{H}_{v_1}\otimes \mathcal{H}_{v_2}\otimes \mathcal{H}_{v_3}$ defined as
\begin{equation}\label{eqn:: partial plaquette operator}
\begin{aligned}
    T_{\theta_{\mathscr{C}}(y_0),y_3} &= \sum_{i_1,i_2\in \Irr}\sum_{j_1,j_2\in \Irr}\sum_{y_k\in \ONB(j^{-}\otimes i_k,j_k),k=1,2} \\
    &\prod^3_{m=1}\left( \frac{d(j_m)}{d(j)d(i_m)} \right)^{1/4} \left( \frac{d(j_{m-1})}{d(j)d(i_{m-1})} \right)^{1/4}\mathrm{Ad}^{1-\kappa^{-1}_{v_m}(f_m)}_\rho \left[ C_{y_m,\theta_{\mathscr{C}}(y_{m-1})}\right] \prod^3_{m=0}Q^{i_m}_{f_m}. 
\end{aligned}
\end{equation}
These operators are called partial plaquette operators in \cite{kitaev_kong_2012,Kong2014Universal}, and they play a major role in classifying gapped boundaries of Levin--Wen string-net models. 

For a simply connected region $D$ with boundary $\partial D$ meeting the lattice at $m$ edges, one defines the map $\mathrm{eval}:\bigotimes_{v\in D} \mathcal{H}_{v}\rightarrow \Hom_{\mathscr{C}}(1,A^m)$ by evaluating the string-net configurations in $D$ with compatible edge labels \cite{kitaev_kong_2012}. 
It was shown in \cite[Theorem 2.9]{Green2024enrichedstringnet} that the ground state projection of $H_D$ is given by 
\begin{equation}\label{eqn:: evaluation map and ground state projection}
    \Pi(D) = \mu^{-\vert p(D)\vert}\mathrm{eval}^\dagger\mathrm{eval},
\end{equation}
where $p(D)$ is the set of plaquettes in $D$. 
\begin{lemma}
    Let $D_1\subseteq D_2$ be two rectangular regions, and let $\Pi(D_i)$ be the ground state projection of $H_{D_i}$ for $i=1,2$. 
    Then we have $\range\bigl(\Tr_{\overline{D_1}}\Pi(D_2)\bigr) = \Pi(D_1)$. 
\end{lemma}
\begin{proof}
    Consider a nonzero ground state $\ket{\psi_1}\in\Pi(D_1)\mathcal{H}_{D_1}$, and set $\eta=\mathrm{eval}_{D_1}\ket{\psi_1}$. 
    We need to find $\ket{\chi}\in \mathcal{H}_{D_2\backslash D_1}$ and $\ket{\psi_2}\in \Pi(D_2)\mathcal{H}_{D_2}$ such that 
    \begin{equation}
        \braket{\psi_2|\chi\otimes \psi_1}\neq 0. 
    \end{equation}
    Write $m=\lvert\partial D_1\rvert$. 
    By linearity, we may assume $\eta \in \hom_{\mathscr{C}}\left(\mathbb{1},i_1\otimes\cdots\otimes i_m\right)$. 
    For each boundary edge $e$ of $D_1$, there is a simple path in $D_2\backslash D_1$ that begins at $e$ and ends at an edge on $\partial D_2$. 
    Along the path starting at the $k$-th edge, label each edge by $i_k$ or $i_k^{-}$ according to its orientation; then label every remaining edge of $D_2\backslash D_1$ by $\mathbb{1}$ and choose the canonical evaluation or coevaluation morphism at every vertex on a path.
    This determines a configuration $\ket{\chi}\in\mathcal{H}_{D_2\backslash D_1}$. 
    Now take $\ket{\psi_2} = \Pi(D_2)\ket{\chi\otimes \psi_1}$; then Equation \eqref{eqn:: evaluation map and ground state projection} gives
    \begin{equation}
        \braket{\psi_2|\chi \otimes \psi_1}=\mu^{-\lvert p(D_2)\rvert}\left\lVert\mathrm{eval}_{D_2}\ket{\chi \otimes \psi_1}\right\rVert^2 \propto \tr_{\mathscr{C}}(\eta^*\eta)>0,
    \end{equation}
    proving the lemma. 
\end{proof}
\noindent As an immediate consequence, we have $\widehat{\Pi}(\Lambda) = \Pi(\Lambda_+)$ for any reflection-symmetric rectangular region $\Lambda$. 

\begin{figure}[htb]
    \centering
    \def\svgwidth{0.6\textwidth}
    \begingroup%
  \makeatletter%
  \providecommand\color[2][]{%
    \errmessage{(Inkscape) Color is used for the text in Inkscape, but the package 'color.sty' is not loaded}%
    \renewcommand\color[2][]{}%
  }%
  \providecommand\transparent[1]{%
    \errmessage{(Inkscape) Transparency is used (non-zero) for the text in Inkscape, but the package 'transparent.sty' is not loaded}%
    \renewcommand\transparent[1]{}%
  }%
  \providecommand\rotatebox[2]{#2}%
  \newcommand*\fsize{\dimexpr\f@size pt\relax}%
  \newcommand*\lineheight[1]{\fontsize{\fsize}{#1\fsize}\selectfont}%
  \ifx\svgwidth\undefined%
    \setlength{\unitlength}{472.93917463bp}%
    \ifx\svgscale\undefined%
      \relax%
    \else%
      \setlength{\unitlength}{\unitlength * \real{\svgscale}}%
    \fi%
  \else%
    \setlength{\unitlength}{\svgwidth}%
  \fi%
  \global\let\svgwidth\undefined%
  \global\let\svgscale\undefined%
  \makeatother%
  \begin{picture}(1,0.4824754)%
    \lineheight{1}%
    \setlength\tabcolsep{0pt}%
    \put(0,0){\includegraphics[width=\unitlength,page=1]{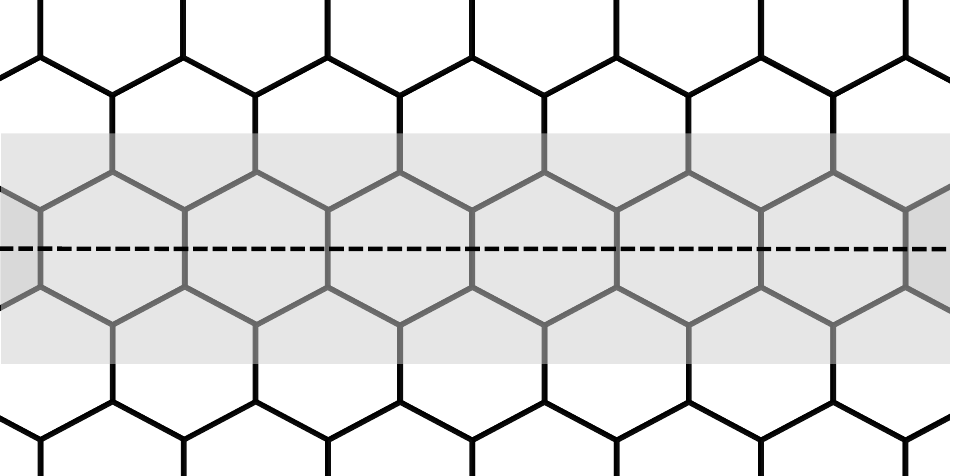}}%
    \put(0.01192677,0.12536789){\color[rgb]{0,0,0}\makebox(0,0)[lt]{\lineheight{1.25}\smash{\begin{tabular}[t]{l}$X_{-}$\end{tabular}}}}%
    \put(0.00413502,0.28868293){\color[rgb]{0,0,0}\makebox(0,0)[lt]{\lineheight{1.25}\smash{\begin{tabular}[t]{l}$X_{+}$\end{tabular}}}}%
    \put(0.78797006,0.44409826){\color[rgb]{0,0,0}\makebox(0,0)[lt]{\lineheight{1.25}\smash{\begin{tabular}[t]{l}$\Lambda_{+}$\end{tabular}}}}%
    \put(0.78797006,0.014741){\color[rgb]{0,0,0}\makebox(0,0)[lt]{\lineheight{1.25}\smash{\begin{tabular}[t]{l}$\Lambda_{-}$\end{tabular}}}}%
    \put(0.07449332,0.20337562){\color[rgb]{0,0,0}\makebox(0,0)[lt]{\lineheight{1.25}\smash{\begin{tabular}[t]{l}$p_1$\end{tabular}}}}%
    \put(0.21620973,0.20337562){\color[rgb]{0,0,0}\makebox(0,0)[lt]{\lineheight{1.25}\smash{\begin{tabular}[t]{l}$p_2$\end{tabular}}}}%
    \put(0.3630033,0.20313431){\color[rgb]{0,0,0}\makebox(0,0)[lt]{\lineheight{1.25}\smash{\begin{tabular}[t]{l}$p_3$\end{tabular}}}}%
    \put(0.51226682,0.20546778){\color[rgb]{0,0,0}\makebox(0,0)[lt]{\lineheight{1.25}\smash{\begin{tabular}[t]{l}$p_4$\end{tabular}}}}%
    \put(0.65891665,0.20546778){\color[rgb]{0,0,0}\makebox(0,0)[lt]{\lineheight{1.25}\smash{\begin{tabular}[t]{l}$p_5$\end{tabular}}}}%
    \put(0.8025214,0.20546778){\color[rgb]{0,0,0}\makebox(0,0)[lt]{\lineheight{1.25}\smash{\begin{tabular}[t]{l}$p_6$\end{tabular}}}}%
  \end{picture}%
\endgroup%
    \caption{A symmetric tubular neighborhood of the reflection hyperplane that contains $6$ plaquettes across the reflection axis (the dashed line).}
    \label{fig:: tubular neighborhood}
\end{figure}

We now compute the boundary algebra derived from the ground states of the Levin--Wen string-net model by OS reconstruction. 
Since the Hamiltonian is local and frustration-free, Theorem \ref{thm:: boundary algebra of a local RP frustration-free Hamiltonian} allows us to restrict attention to symmetric subsets that do not include plaquettes in the interior of $\Gamma_{\pm}$. 
These subsets are tubular neighborhoods of the x-axis that contain only the plaquettes that cross the axis (Fig.~\ref{fig:: tubular neighborhood}).
Let $\Lambda$ be such a subset, and number the plaquettes contained in $\Lambda$ by $p_1,p_2,\ldots,p_L$. 
Denote by $e_1,e_2,\dots,e_L,e_{L+1}$ the edges that cross the x-axis and are contained in $\Lambda$, so that $e_{l},e_{l+1}$ lie on the boundary of $p_l$ for $1\leq l\leq L$. 
Then the local Hamiltonian $H_\Lambda$ is 
\begin{equation}
    H_\Lambda = -\sum^{L}_{k=1}B_{p_k} - \sum^{L+1}_{k=1}Q_{e_k}.
\end{equation}
For simple objects $j,i,i',k,k'$ and morphisms $\xi\in \hom_{\mathscr{C}}(j^{-}\otimes i,i')$ and $\eta\in \hom_{\mathscr{C}}(j^{-}\otimes k,k')$, denote by $T^l_{\theta_{\mathscr{C}}(\eta),\xi}$ the partial plaquette operators appearing in the decomposition of $B_{p_l}$. 
Then $\mathcal{A}_+(H_\Lambda)\widehat{\Pi}(\Lambda)$ is the $C^*$-algebra generated by operators of the form 
\begin{equation}
    T^l_{\theta_{\mathscr{C}}(\eta),\xi}\widehat{\Pi}(\Lambda), \quad 1\leq l\leq L, \xi,\eta\in \hom_{\mathscr{C}}(A\otimes A,A). 
\end{equation}
Denote by $\mathrm{eval}_{\Lambda_+}$ the evaluation map from $\mathcal{H}_{\Lambda_+}$ to $\hom_{\mathscr{C}}(A^{L+2},A^{L+1})$. 
A configuration in $\Lambda_+$ is evaluated to a path of length $L+2$ consisting of trivalent vertices labeled by $\xi_1,\xi_2,\ldots,\xi_{L+2}$:
\begin{equation}
    \vcenter{\hbox{\begin{tikzpicture}
        \begin{scope}
            \VeryThinLine (0:0) -- (-90:0.5);
            \VeryThinLine (0:0) -- (150:0.5);
            \VeryThinLine (0:0) -- (30:1);
            \draw[fill=black] (0:0) ellipse (0.05 and 0.05);
            \node at (-0.25,-0.125) {$\xi_1$};
        \end{scope}
        \begin{scope}[shift={(1.3,0.75)}]
            \VeryThinLine (0:0) -- (90:0.5);
            \VeryThinLine (0:0) -- (210:0.5);
            \VeryThinLine (0:0) -- (-30:0.5);
            \draw[fill=black] (0:0) ellipse (0.05 and 0.05);
            \node at (-0.25,0.125) {$\xi_2$};
        \end{scope}
        \begin{scope}[shift={(2.6,0)}]
            \VeryThinLine (0:0) -- (-90:0.5);
            \VeryThinLine (0:0) -- (150:1);
            \VeryThinLine (0:0) -- (30:1);
            \draw[fill=black] (0:0) ellipse (0.05 and 0.05);
            \node at (-0.25,-0.125) {$\xi_3$};
        \end{scope}
        \begin{scope}[shift={(3.9,0.75)}]
            \VeryThinLine (0:0) -- (90:0.5);
            \VeryThinLine (0:0) -- (210:0.5);
            \VeryThinLine (0:0) -- (-30:0.5);
            \draw[fill=black] (0:0) ellipse (0.05 and 0.05);
            \node at (-0.25,0.125) {$\xi_4$};
        \end{scope}
        \node at (5.2,0.375) {$\cdots$};
        \begin{scope}[shift = {(6.5,0)}]
            \VeryThinLine (0:0) -- (-90:0.5);
            \VeryThinLine (0:0) -- (150:0.5);
            \VeryThinLine (0:0) -- (30:0.5);
            \draw[fill=black] (0:0) ellipse (0.05 and 0.05);
            \node at (-0.45,-0.125) {$\xi_{L+2}$};
        \end{scope}
    \end{tikzpicture}}}
\end{equation}
Then the operators $T^l_{\theta_{\mathscr{C}}(\eta),\xi}\widehat{\Pi}(\Lambda)$ satisfy the following relation with $\mathrm{eval}_{\Lambda_+}$:
\begin{equation}
    \mathrm{eval}_{\Lambda_+}T^l_{\theta_{\mathscr{C}}(\eta),\xi} \ket{\xi_1\otimes \xi_2\otimes \cdots \otimes \xi_{L+2}} = \left(\mathrm{1}_{A^{l-1}}\otimes F(\eta,\xi)\otimes \mathrm{1}_{A^{L-l-1}}\right) \mathrm{eval}_{\Lambda_+}\ket{\xi_1\otimes \xi_2\otimes \cdots \otimes \xi_{L+2}},
\end{equation}
where $F(\eta,\xi)\in \hom_{\mathscr{C}}(k^{-}\otimes i,k'^{-}\otimes i')$ is defined as 
\begin{equation}
    F(\eta,\xi) = \left( \frac{d(k')}{d(j)d(k)} \right)^{1/4} \left( \frac{d(i')}{d(j)d(i)} \right)^{1/4}\vcenter{\hbox{\begin{tikzpicture}
        \draw[thick, postaction={decorate}, decoration={markings, mark=at position 0.75 with {\arrow{stealth}}}] (-0.5,0) -- (-0.5,0.75);
        \draw[thick, postaction={decorate}, decoration={markings, mark=at position 0.5 with {\arrow{stealth}}}] (-0.5,-0.75) -- (-0.5,0);
        \draw[thick, postaction={decorate}, decoration={markings, mark=at position 0.5 with {\arrow{stealth}}}] (0.5,0.75) -- (0.5,0);
        \draw[thick, postaction={decorate}, decoration={markings, mark=at position 0.75 with {\arrow{stealth}}}] (0.5,0) -- (0.5,-0.75);
        \draw[thick, postaction={decorate}, decoration={markings, mark=at position 0.75 with {\arrow{stealth}}}] (0.5,0) arc (0:180:0.5);
        \fill (-0.5,0) circle (1.5pt); 
        \fill (0.5,0) circle (1.5pt); 
        \node at (-1.15,0) {$\theta_{\mathscr{C}}(\eta)$};
        \node at (0.75,0) {$\xi$};
        \node at (0,0.75) {$j$};
        \node at (-0.5,1) {$k$};
        \node at (0.5,1) {$i$};
        \node at (-0.5,-1) {$k'$};
        \node at (0.5,-1) {$i'$};
    \end{tikzpicture}}}
\end{equation}
This follows from summing over $y_1$ and $y_2$ in Equation \eqref{eqn:: partial plaquette operator} and the partition of unity.  
As $j,i,i',k,k'$ range over all simple objects, these morphisms span $\hom_{\mathscr{C}}(A^2,A^2)$. 
In general, $\mathcal{A}_+(H_\Lambda)\Pi(\Lambda_+)$ is isomorphic to $\hom_{\mathscr{C}}(A^{L+1},A^{L+1})$. 

Finally, we compute the action of the modular group $\sigma^\Lambda_t$ on $\cM_\Lambda$. 
To this end, we take advantage of the fact that $\mathcal{A}_+(H_\Lambda)\widehat{\Pi}(\Lambda)$ is generated by the subalgebras $\mathcal{A}_+(B_{p_l})\widehat{\Pi}(\Lambda)$. 
If $x = x_1x_2\cdots x_L$, where $x_l\in \mathcal{A}_+(B_{p_l})\Pi(\Lambda_+)$ for $1\leq l\leq L$, then, by Theorem \ref{thm:: consistent dynamics in the inductive limit}, 
\begin{equation}\label{eqn:: modular group on the tubular neighborhood}
    \sigma^\Lambda_t(x) = \sigma^{p_1}_t(x_1)\sigma^{p_2}_t(x_2)\cdots \sigma^{p_L}_t(x_L), 
\end{equation}
where $\sigma^{p_l}_t$ is the modular group of $\omega_{p_l}$ on $\mathcal{A}_+(B_{p_l})\Pi(p_{l,+})$. 
Therefore, it is enough to compute the modular group of the state $\omega_p$ on $\mathcal{A}_+(B_p)\Pi(p_+)$ for a single plaquette $p$. 
The state $\omega_p$ is computed by the intertwining relation between $T_{\theta_{\mathscr{C}}(\eta),\xi}$ and $F(\eta,\xi)$. 
Indeed, we have
\begin{equation}
    \omega_p(T_{\theta_{\mathscr{C}}(\eta),\xi}) = \frac{1}{\mu}d(k)d(i)\tr_{\mathscr{C}}(F(\eta,\xi)). 
\end{equation}
Therefore, the modular group of $\omega_p$ is given by 
\begin{equation}
    \sigma^{p}_t(T_{\theta_{\mathscr{C}}(\eta),\xi}) = \left( \frac{d(k)}{d(k')} \right)^{it} \left( \frac{d(i)}{d(i')} \right)^{it} T_{\theta_{\mathscr{C}}(\eta),\xi},\quad \xi\in \hom_{\mathscr{C}}(j^{-}\otimes i,i'), \eta\in \hom_{\mathscr{C}}(j^{-}\otimes k,k').
\end{equation}
Now consider a monomial $\prod^L_{l=1}T^l_{\theta_{\mathscr{C}}(\eta_l),\xi_l}$, where $\xi_l\in \hom_{\mathscr{C}}(j^{-}_l\otimes i_l,i'_l)$, $\eta_l\in \hom_{\mathscr{C}}(j^{-}_l\otimes k_l,k'_l)$. 
In order for this monomial to be nonzero, we must have $k'_l = i_{l-1}$ for $2\leq l\leq L$. 
Then by Equation \eqref{eqn:: modular group on the tubular neighborhood}, we have 
\begin{equation}
\begin{aligned}
    \sigma^\Lambda_t\left(\prod^L_{l=1}T^l_{\theta_{\mathscr{C}}(\eta_l),\xi_l}\right) &= \prod_{l=1}^{L}\sigma^{p_l}_t\left(T^l_{\theta_{\mathscr{C}}(\eta_l),\xi_l}\right) = \prod^L_{l=1}\left( \frac{d(k_l)}{d(k'_l)} \right)^{it} \left( \frac{d(i_l)}{d(i'_l)} \right)^{it}\prod^L_{l=1}T^l_{\theta_{\mathscr{C}}(\eta_l),\xi_l}\\
    &= \left( \frac{d(i_L)\prod^{L}_{l=1}d(k_l)}{d(k'_1)\prod^{L}_{l=1}d(i'_l)} \right)^{it}\prod^L_{l=1}T^l_{\theta_{\mathscr{C}}(\eta_l),\xi_l}.
\end{aligned}
\end{equation}
The boundary algebra of Levin--Wen string-net models constructed by OS reconstruction is isomorphic to that derived in \cite{JNPW2025} from the LTO axioms. 
Furthermore, the one-parameter automorphism computed using Theorem \ref{thm:: consistent dynamics in the inductive limit} is the modular automorphism group of the canonical state constructed from the same set of axioms \cite[Sect. 5.3]{JNPW2025}. 

\begin{conflictofinterest}
There is no conflict of interest. 
\end{conflictofinterest}

\begin{dataavailability}
Data sharing is not applicable to this article as no new data were created or analyzed in this study. 
\end{dataavailability}

\bibliographystyle{alpha}
\bibliography{reference}
\end{document}